\documentclass[fleqn,usenatbib,onecolumn]{mnras}

\usepackage{newtxtext,newtxmath}

\usepackage[T1]{fontenc}

\DeclareRobustCommand{\VAN}[3]{#2}
\let\VANthebibliography\thebibliography
\def\thebibliography{\DeclareRobustCommand{\VAN}[3]{##3}\VANthebibliography}

\usepackage{graphicx}	
\usepackage{amsmath}	
\usepackage{multirow} 

\usepackage{longtable}
\usepackage{array}
\usepackage{footnote}
\usepackage{stfloats}
\usepackage{natbib}
\usepackage{lineno}

\def\apjl{ApJ}
\def\apjs{ApJS}
\def\ao{Appl.~Opt.}
\def\aap{A\&A}

\newcommand{\be}{\begin{equation}}
\newcommand{\ee}{\end{equation}}
\newcommand{\bary}{\begin{eqnarray}}
\newcommand{\eary}{\end{eqnarray}}

\title[Identifying the poynting-flux-dominated outflow of Fermi GRBs]{Identifying the poynting-flux-dominated outflow of Fermi GRBs with non-thermal spectrum and its energy-resolved light curve fitting}

\author[Chang et al.]{
Xue-Zhao Chang,$^{1,2}$
HouJun L\"{u},$^{1}$\thanks{E-mail: lhj@gxu.edu.cn}
Zhao-Wei Du$^{1}$
Xing Yang$^{1}$
and En-Wei Liang$^{1}$
\\
$^{1}$Guangxi Key Laboratory for Relativistic Astrophysics, School of Physical Science and Technology, Guangxi University, Nanning 530004, China\\
$^{2}$Institute of Astrophysics, Chuxiong Normal University, Chuxiong 675000, China\\
}

\date{Accepted XXX. Received YYY; in original form ZZZ}

\pubyear{2025}

\begin{document}
\label{firstpage}
\pagerange{\pageref{firstpage}--\pageref{lastpage}}
\maketitle

\begin{abstract}
The jet compositions of gamma-ray bursts (GRBs) are very important to understand the energy dissipation and radiation mechanisms, but it remains an open question in GRB physics. In this paper, we present a systematic analysis to search for 88 bright GRBs that include a total of 129 pulses observed by Fermi/GBM with redshift measured, and extract the spectra of each pulse with Band function (Band), cutoff power-law (CPL), blackbody (BB), non-dissipative photospheric (NDP), Band+BB, as well as CPL+BB. We find that 80 pulses, 35 pulses, and 14 pulses present purely non-thermal, hybrid, and thermal spectra, respectively. By focusing on those 80 pulses with purely non-thermal spectra, one can estimate the lower limits of magnetization factor ($\sigma$) via suppressing the pseudo-thermal component. It is found that 30 pulses in 21 GRBs are the lower limit of $\sigma>5$ at the photosphere by adopting $R_{0}=10^{10}$ cm. It suggests that at least the outflow of those GRB jets with high $\sigma$ is dominated by Poynting-flux. On the other hand, we also perform the light curve fitting with a fast-rise-exponential-decay (FRED) model for 15 bright GRBs with a high magnetization factor in our sample, and find that a correlation between pulse width ($w$) and energy of 13 GRBs really exists in their energy-resolved light curves. It is also a piece of independent evidence for those GRBs with a high value $\sigma$ to support the origin of the Poynting flux outflow.
\end{abstract}


\section{Introduction} 
\label{intro}
Gamma-ray bursts (GRBs) are extremely luminous with a total energy of $\sim 10^{51}-10^{53}$ ergs in a few or hundreds of seconds, making them one of the most luminous electromagnetic (EM) phenomena in the universe since they were first discovered in the 1967 (see \citealt{2015PhR...561....1K} for a review). Although the field of GRBs has rapidly advanced and made great breakthroughs in both observations and theory in the past 50 years, there still exist several open questions in GRB physics, such as what is the composition of a GRB jet? How energy is dissipated to give rise to prompt emission? What is the radiation mechanism of GRB prompt emission \citep{2011CRPhy..12..206Z}?

In the classical “fireball” model of GRB, the GRB jet was launched from an initially hot fireball which is composed of photons, electron/positron pairs, and a small amount of baryons \citep{1986ApJ...308L..43P,1986ApJ...308L..47G,1990ApJ...365L..55S}. After an initially rapid acceleration phase under fireball thermal pressure, a fraction of the thermal energy is converted to the kinetic energy of the outflow \citep{1993ApJ...415..181M}. The hot photons escape and a quasi-thermal component of the fireball photosphere should be powered when the jet expands to the radius of the photosphere (photosphere model; \citealt{1986ApJ...308L..43P,1986ApJ...308L..47G}). Moreover, the baryonic matter continues to expand to even greater distances, reconverting the kinetic energy into energetic particles through internal collisional dissipation, then the synchrotron (or synchrotron self-Compton) radiation by these particles gives rise to the observed non-thermal gamma-ray emission (internal shock model; \citealt{1994ApJ...430L..93R,1997ApJ...490...92K,1998MNRAS.296..275D,2014AcASn..55..354Z}). Alternatively, another scenario invokes a non-thermal component from the Poynting-flux-dominated outflow where most of the energy is stored in the magnetic field \citep{2011ApJ...726...90Z}. The magnetic energy can be dissipated through magnetic reconnection or current instability to power the observed prompt emission of GRB \citep{2003astro.ph.12347L,2009ApJ...700L..65Z}, or the internal-collision-induced magnetic reconnection and turbulence (ICMART) model \citep{2011ApJ...726...90Z}.

From the observational point of view, the spectrum of several GRBs are dominated by quasi-thermal component, such as GRB 090902B \citep{2009ApJ...706L.138A,2010ApJ...709L.172R,2012MNRAS.420..468P}, and it implies that at least the outflow of GRB 090902B is the matter-dominated. Other GRBs of thermal dominated outflow are also discussed in the literatures \citep{2020ApJ...893..128A}. On the contrary, a good fraction of GRBs are consistent with not having a thermal component, such as GRB 080916C \citep{2009Sci...323.1688A,2009ApJ...700L..65Z}, GRB 130606B \citep{2016ApJ...816...72Z,2019A&A...625A..60R,2020NatAs...4..174B} and GRB 230307A \citep{2024MNRAS.529L..67D,2025ApJ...985..239Y}, and it suggests that the outflow of those GRBs are Poynting-flux-dominated. Moreover, a dominated (or sub-dominated) non-thermal component have been discovered in a hybrid GRB jet with both quasi-thermal component and Poynting-flux component, e.g., GRB 100724B \citep{2011ApJ...727L..33G}, GRB 110721A \citep{2012ApJ...757L..31A}, GBR 160625B \citep{2017ApJ...849...71L,2017Natur.547..425T,2018NatAs...2...69Z}, GRB 081221 \citep{2018ApJ...866...13H}, and 211211A \citep{2023ApJ...943..146C}. Moreover, a systematic search for such hybrid GRB jet by using the Fermi/GBM data is also performed by \cite{2020ApJ...894..100L} and \cite{2024ApJ...972....1L}.

From the theoretical point of view, the prediction of the photosphere model has a magnetization parameter ($\sigma$) with $\sigma \ll$ 1, and the internal shock model has a magnetization parameter $\sigma$ less than unity at the GRB emission site. On the contrary, the ICMART model has a large $\sigma \gg$ 1 at the emission site with the GRB emission powered by directly dissipating the magnetic energy to radiation \citep{2011ApJ...726...90Z,2015ApJ...801...2,2024ApJS..275....9C}. Here, the magnetization parameter $\sigma$ can be defined as the ratio between Poynting flux and baryon flux. On the other hand, the light curves of prompt emission predicted by different models also exhibit different properties \citep{2018pgrb.book.....Z,2025ApJ...985..239Y}. For example, the predictions of the photosphere and ICMART models show that the shape of the light curve is energy-dependent, including spectral lags and global hard-to-soft evolution \citep{1986ApJ...308L..43P,1986ApJ...308L..47G,2000ApJ...530..292M,2016ApJ...825...97U,2018ApJ...869..100U}, while the predicted light curve of prompt emission from internal shock model do not show any shape-energy dependence \citep{2025ApJ...985..239Y}. On the other hand, the jet composition of GRB is also constrained via independent method with non-detected high-energy neutrinos \citep{2024ApJ...976..174O}.

Recently, an interesting GRB 230307A, called “short-duration with extended emission” which is originated from a binary star merger, is quite similar to that of GRBs 060614 \citep{2006Natur.444.1044G}, 211211A \citep{2022Natur.612..223R,2022Natur.612..232Y,2025ApJ...988L..46L}, 211227A \citep{2022ApJ...931L..23L,2023A&A...678A.142F}, and it was attracted attention with redshift $z=0.065$ \citep{2023ApJ...954L..29D,2024Natur.626..737L,2024Natur.626..742Y,2025NSRev..12..401S}. Despite the long duration of the prompt emission, it is associated with kilonovae emission instead of supernovae and suggests that the progenitor of GRB 230307A is likely to originate from binary compact object system \citep{2024Natur.626..737L,2024Natur.626..742Y,2024ApJ...962L..27D,2025NSRev..12..401S}. More interestingly, many rapidly variable short pulses of light curve and without thermal emission of the spectrum in the prompt emission, the plateau emission in the X-ray afterglow, the high magnetization parameter, together with the high radiation efficiency of prompt emission, suggest that the outflow of GRB 230307A originated from merger of binary neutron stars should be poynting-flux-dominated and is consistent with the ICMART model \citep{2025ApJ...985..239Y,2024MNRAS.529L..67D,2025NSRev..12..401S}. Motivated by analysis of the jet composition of GRB 230307A, several questions emerge. What is the fraction of observed non-thermal spectra that are truly dominated by Poynting-flux? To answer this question is very important to understand the physical process and mechanism of GRBs.

In this paper, we intend to systematically apply the methodology mentioned in \citet{2009ApJ...700L..65Z} to a large sample of GRBs observed by Fermi/GBM, and estimate the lower limit of $\sigma$ for our sample with a featureless non-thermal spectral to present the fraction of GRBs with  Poynting flux-dominated outflows. On the other hand, we also adopt an independent method to analyse the light curves of our sample with high-magnetization parameter and cross-check the results based on those two independent methods. The methodology of the data analysis and sample selection are shown in \S 2. In \S 3, we present the the constraints of magnetization parameter from the suppressed thermal component. The details of our temporal analysis approach are described in \S 4. Conclusions are drawn in \S 5 with some additional discussions. Throughout the paper, we adopt a concordance cosmology with parameters $H_{0}=70 \mathrm{~km} \mathrm{~s}^{-1} \mathrm{Mpc}^{-1}$, $\Omega_{\mathrm{M}}=0.30$ and $\Omega_{\Lambda}=0.70$, respectively.

\section {Data Reduction and Sample Selection}
The Fermi satellite has operated for more than 15 years since it was successfully launched in 2008. It has a very wide energy band from 8 keV to 300 GeV with coverage over seven orders of magnitude \citep{2009ApJ...702..791M}, and includes two kinds of main scientific instruments, such as Gamma-ray Burst Monitor (GBM) and Large Area Telescope (LAT). The GBM has 12 sodium iodide (NaI) detectors with an energy range spanning from 8 keV to 1 MeV and two bismuth germanate (BGO) scintillation detectors within the range of 200 keV to 40 MeV. In this work, we only focus on GBM data
for the temporal and spectral analysis, and ignore the contributions of LAT data if it was triggered by LAT\footnote{The reason is that the physical origin of high-energy photons remains in high debate, such as originating from internal or external dissipations, remain unknown.}. 

\subsection{Light-curve extraction}
We download the original GBM data from the Fermi Science Support Center’s FTP site \footnote{https://heasarc.gsfc.nasa.gov/FTP/fermi/}. The GBM detectors include three distinct data modes: Continuous Time (CTIME), Continuous Spectroscopy (CSPEC), and Time-Tagged Event (TTE). The CTIME and CSPEC modes offer a temporal resolution of 64 milliseconds and 1.024 seconds, respectively, whereas the TTE mode is composed of individually digitized-pulse height-resolution events with the type of an unbinned data \citep{2012ApJS..199...18P}. The TTE mode provides an energy resolution spanning 128 channels and documents the temporal intervals of photons extending from -20 seconds to 300 seconds post-trigger. The TTE data enables the flexibility to employ any bin size for temporal or spectral analysis and offers a versatile platform for comprehensive studies. We select the data from all the bright NaI detectors (one to three NaI detectors) and the brightest BGO detector to do the analysis. The standard response files (RSP or RSP2) provided by the GBM team are adopted in this work. Moreover, we develop a {\em Python} code to extract the energy-dependent light curves by invoking the spectrum source package {\em Fermi ScienceTools}. For more details refer to our previous paper \citep{2018ApJ...862..155L,2024ApJS..275....9C}.

\subsection{Sample selection}
As of January 2024, we have extracted the light curves of more than 3400 GRBs that were detected by Fermi/GBM. The motivation of this study is to identify Poynting-flux-dominated GRBs generated by the ICMART model. There are three criteria adopted for our sample selection: (1) in order to calculate the magnetization factor $\sigma$, the redshift of a GRB needs to be measured; (2) the signal-to-noise ratio (S/N; \citealt{2018ApJS..236...17V}) is required to be higher than 20 during the range of $T_{90}$, and make sure that enough photons can be to extract the spectrum; (3) if the light curve of a GRB is composed by multiple pulses (or called sub-pulse), the time-integrated spectrum can not fully reflect the properties of prompt emission because of the existence of spectral evolution during the multiple pulses. Thus, it requires that the S/N of each sub-pulse is also higher than 20. 

It is worth noting to identify the pulse mentioned above. Traditionally, an individual pulse is described as asymmetric "fast-rise-exponential-decay" (FRED) profiles \citep{1996ApJ...459..393N,2005ApJ...627..324N,2018ApJ...863...77H}. In this work, a pulse is required to exhibit a continuous "rise-peak-decay" profile: the start time is defined as the moment when the count rate begins to rise continuously from the background level, while the end time corresponds to when the count rate stabilizes back to the background or enters a flat plateau \citep{2025ApJ...991..230G}. A pulse should possess a unified global evolution trend, even if local fluctuations (e.g., secondary peaks and deep dips) are present. The global pulse is classified as a single pulse as long as these fluctuations are enclosed within a unified pulse envelope (e.g., GRB 230307A; \citealt{2025ApJ...985..239Y}), as such emission likely originates from a single radiative process.

By satisfying with above three criteria, we obtain 88 bursts out of more than 3400 GRBs detected by Fermi GBM until 2024 January 1. It is found that only one clear pulse is included in the light curve for 60 bursts out of 88 GRBs, 19 bursts having two distinct pulses, 6 bursts exhibiting three different pulses, 2 bursts having four different pulses, and 1 burst displaying five pulses. In total, 129 pulses are obtained from these 88 GRBs (see Table \ref{Table1} for details).


\subsection{Spectral model selected and spectral fitting}
By selecting the 88 GRBs which include 129 pulses, we then fit the spectra of each pulse by invoking several spectral models, such as (1) Band function (Band), (2) cutoff power-law (CPL), (3) blackbody (BB), (4) non-dissipative photospheric (NDP) model, or (5) Band (or CPL) $+$ BB. The functions of those models can be expressed as follows,\\
(1) Band function (Band; \citealt{1993ApJ...413..281B}):
\begin{eqnarray}
N_{\textrm{Band}}(E)=A_{\rm Band}\left\{\begin{array}{clcc}
(\frac{E}{100~\mathrm{keV}})^{\alpha }\mathrm{exp}\left[-\frac{(\alpha+2)E}{E_{p}} \right ], E \leq (\alpha-\beta) E_{c}, \\
(\frac{E}{100~\mathrm{keV}})^{\beta }\mathrm{exp}(\beta -\alpha )(\frac{(\alpha-\beta)E_{c}}{100~\mathrm{keV}})^{\alpha-\beta }, E > (\alpha-\beta) E_{c}
\end{array}\right.
\end{eqnarray}
where the $A_{\rm Band}$ is the normalization of the Band function. $\alpha$ and $\beta$ are the low and high-energy photon spectral indices, respectively. $E_{\rm p}=(2+\alpha)E_{\rm c}$ is the peak energy, and $E_{\rm c}$ is the cut-off energy.\\
(2) Cutoff power-law:
\begin{eqnarray}
N_{\rm CPL}(E) = A_{\rm CPL}\cdot (\frac{E}{100~\mathrm{keV}})^{\alpha} \mathrm{exp}(-\frac{E}{E_{c}}).
\end{eqnarray}
where $A_{\rm CPL}$ is the normalization of the CPL model.\\
(3) Blackbody (BB) emission of the quasi-thermal component with Planck function \citep{2010ApJ...709L.172R},
\begin{eqnarray}
N_{\textrm{BB}}(E)=A_{\rm BB}\cdot \frac{E^{2}}{\exp [E / k T]-1}
\end{eqnarray}
where $k$ and $T$ are Boltzmann constant and temperature, respectively. \\
(4) The thermal emission may not look like a simple Planck spectrum, but rather has a more complicated shape \citep{2020ApJ...893..128A}, we also adopt the NDP model to describe the pure thermal spectrum. This model was initially proposed by \citep{2019MNRAS.487.5508A} and was discussed in the details in their paper. The mathematical function can be expressed as 
\begin{eqnarray}
N_{\mathrm{NDP}}(E)=A_{\mathrm{NDP}}\left(\frac{E}{E_{\text {pivot }}}\right)^{0.4} \mathrm{e}^{-\left(\frac{E}{E_{\mathrm{cut}}}\right)^{0.65}}
\end{eqnarray}
where $E_{pivot}$ (set to 100 keV) and $E_{cut}$ represent the pivot energy
and cutoff energy, respectively.

In our analysis, we employ a Bayesian methodology by utilizing the Multi-Mission Maximum Likelihood Framework (3ML; \citealt{2015arXiv150708343V}) to extract and fit the spectral data. The first step is to extract the raw photon count spectrum directly from the TTE data. Then, by selecting two temporal intervals before and after the prompt emission phase, we model these intervals with polynomials of orders ranging from 0 to 4, and the background model is subtracted from the observed data. Next, we apply a fitting model that is convolved with the standard response files (RSP or RSP2), and it generates a predicted count spectrum for comparison with the processed observational data. The goodness of the fit is evaluated using the PGstat statistic which is a maximum-likelihood-based metric including the Poisson statistics for the observed counts and Gaussian statistics for the background profile. The spectral analysis spans distinct energy ranges tailored to the detector types: for NaI detectors, we consider energies from 8 keV to 900 keV, while for BGO detectors, the range extends from 250 keV to 40 MeV. Moreover, in order to mitigate the influence of the instrument's K-edge at 33.17 keV, we exclude the adjacent energy band from 30 keV to 40 keV. Finally, by adopting the capabilities of 3ML, we conduct a Bayesian analysis to determine the optimal set of fit parameters, along with their uncertainties.

Bayesian analysis is indeed to be conducted with the consideration of prior assumptions. In this task, the prior distributions are defined as follows.\\
(1) For Band model,
\begin{eqnarray}
\left\{\begin{array}{l}
A_{\rm Band} \sim \log \mathcal{N}(\mu=0, \sigma=2) \quad \mathrm{cm}^{-2}~\mathrm{keV}^{-1} \mathrm{~s}^{-1} \\
\alpha \sim {TG}(\mu=-1, \delta=0.5, LB=-1.5, UB=1) \\
\beta \sim {TG}(\mu=-2, \delta=0.5, LB=-5, UB=-1.8) \\
E_{\mathrm{p}} \sim \log \mathcal{N}(\mu=2, \delta=1)  ~\mathrm{keV}
\end{array}\right.
\end{eqnarray}
where $\log \mathcal{N}$ is a prior of a log-normal distribution, $\mu$ and $\delta$ are the mean value and standard deviation of the distribution, respectively. $TG$ is a prior of a Truncated-Gaussian distribution, and the $LB$ and $UB$ are the lower bound and the upper bound of the distribution, respectively.\\
(2) For CPL model,
\begin{eqnarray}
\left\{\begin{array}{l}
A_{\rm CPL} \sim \log \mathcal{N}(\mu=0, \delta=2) \quad \mathrm{cm}^{-2}~\mathrm{keV}^{-1} \mathrm{~s}^{-1} \\
\alpha \sim {TG}(\mu=-1, \delta=0.5, LB=-5, UB=1)  \\
E_{\mathrm{c}} \sim \log \mathcal{N}(\mu=2, \delta=1) ~\mathrm{keV}
\end{array}\right.
\end{eqnarray}
\\
(3) For BB model,
\begin{eqnarray}
\left\{\begin{array}{l}
A_{\rm BB} \sim \log \mathcal{U}(LB=10^{-10}, UB=10^{3}) \quad \mathrm{cm}^{-2}~\mathrm{keV}^{-1} \mathrm{~s}^{-1} \\
kT \sim \log \mathcal{U}(LB=10^{0}, UB=10^{3})\quad \mathrm{keV}\\
\end{array}\right.
\end{eqnarray}
where $\log \mathcal{U}$ is a prior of a log-uniform distribution.\\
(4) For NDP model,
\begin{eqnarray}
\left\{\begin{array}{l}
A_{\rm NDP} \sim \log \mathcal{U}(LB= 10^{-11}, UB=10^{1}) \quad \mathrm{cm}^{-2}~\mathrm{keV}^{-1} \mathrm{~s}^{-1} \\
E_{\mathrm{cut}} \sim \mathcal{U}(LB=10^{0}, UB=10^{4})\quad \mathrm{keV} \\
\end{array}\right.
\end{eqnarray}
where $ \mathcal{U}$ is a prior of a uniform distribution.

\begin{figure}
\centering
 \includegraphics [angle=0,scale=0.5] {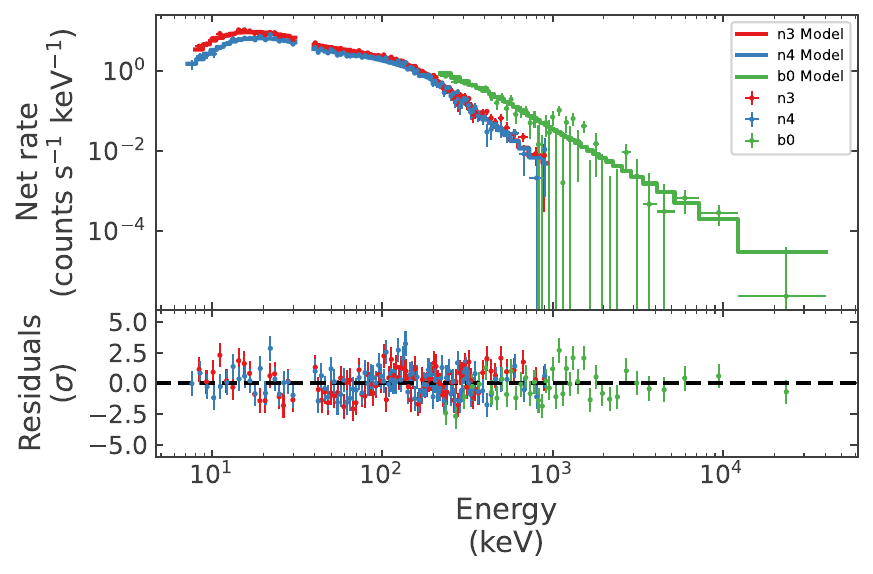}
\includegraphics [angle=0,scale=0.3] {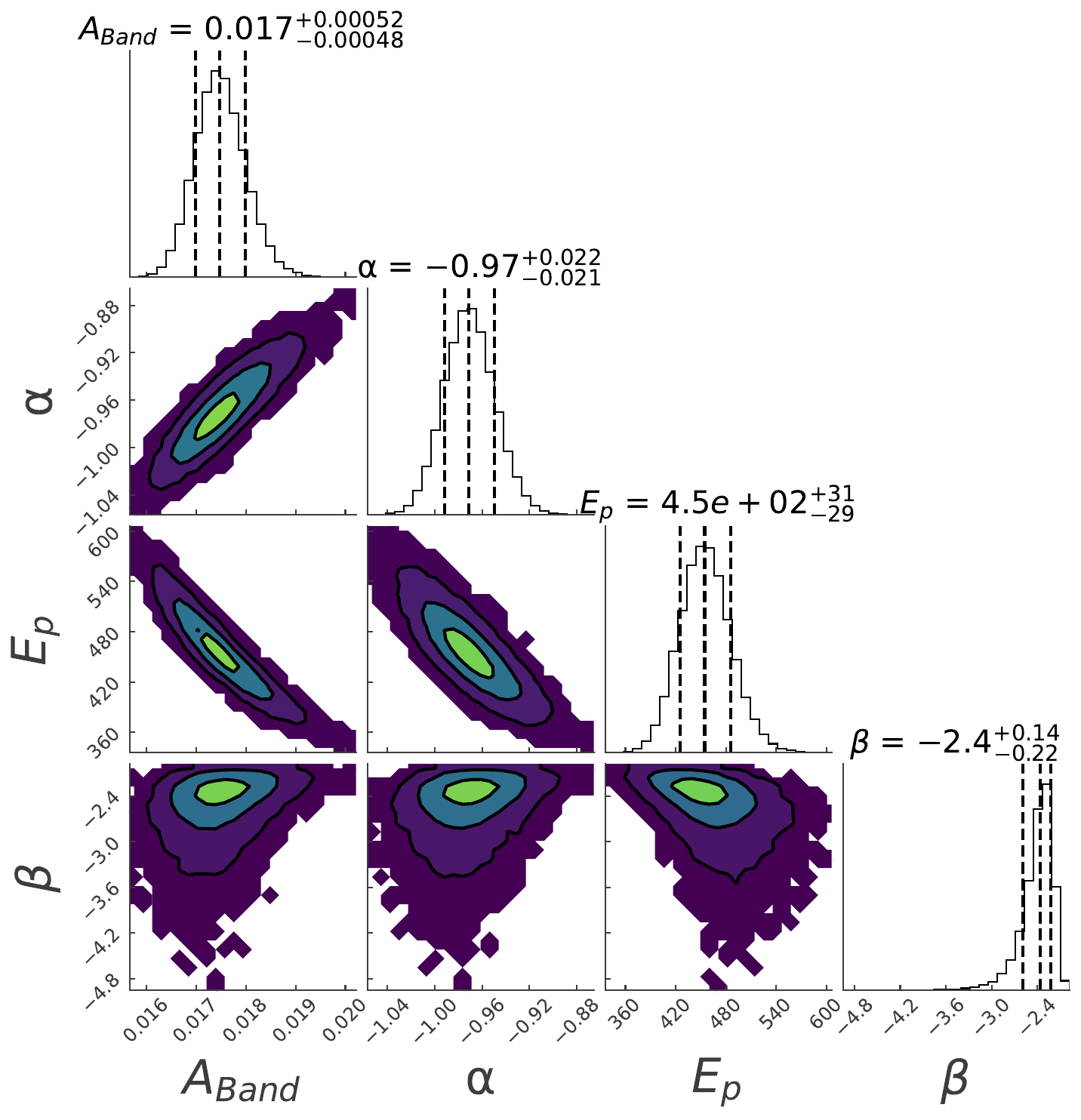}
 \caption{An example of time-integrated spectral fitting of GRB 080916C with Band function model. Left: Observed and modeled photon count spectra. Right: the parameter constraints of the spectral fits.}
 \label{fig:1}
\end{figure}

After setting the prior assumptions, we apply the Markov Chain Monte Carlo (MCMC) method from $\mathit{emcee}$ package to sample the posterior and obtain the best fitting parameters \citep{2010CAMCS...5...65G}. The uncertainties of fitting are typically provided at the 68$\%$ Bayesian confidence level and then calculated by adopting the last 75$\%$ of the MCMC chain with 10,000 iterations. It is to ensure that the uncertainties are estimated from a sufficiently converged portion of the chain. Here, we adopt the Bayesian information criterion (BIC) to determine the best-fitting model. The definition of BIC can be expressed as BIC=-2ln $L+k\cdot$ ln($n$), where $L$ is the maximum value of the likelihood function of the estimated model. $k$ and $n$ are the number of model parameters data points, respectively. The BIC is a criterion used to evaluate the best-fitting model from a finite set of models. The model with the lowest BIC value is considered as the preferred model. The BIC information of our sample is shown in Table \ref{Table1}. One example of GRB 080916C whose time-integrated spectrum is fitted by Band function and the parameter constraints are shown in Figure \ref{fig:1}.

Finally, we perform a preliminary analysis of the 129 pulses from 88 GRBs and employ distinct spectral models which include Band, CPL, Band+BB, CPL+BB, BB, and NDP models to fit the spectra of 129 pulses. The results of model comparisons based on the BIC are summarized in Table \ref{Table1}. Based on the composition of spectral components, it can be roughly divided into seven categories: (1) 47 GRBs (including 34 single-pulse bursts and 13 multiple-pulse bursts) have spectra with purely non-thermal component during the burst; (2) the spectral of 9 GRBs (including 8 single-pulse bursts and 1 multiple-pulse burst) have fully thermal component during the burst; (3) 22 GRBs (including 18 single-pulse bursts and 4 multiple-pulse burst) show a hybrid spectrum during the burst; (4) one GRB 160625B with multiple-pulse emission, shows a transition from a thermal to a hybrid spectrum; (5) two GRBs (170705A and 230204B) shows a transition from a thermal to a non-thermal spectrum; (6) 6 GRBs present a transformation from a hybrid to a non-thermal spectrum; (7) one GRB 201020B presents a unique feature, displaying an evolution from a non-thermal to a thermal spectrum. If we only focus on the 129 pulses from 88 GRBs, it is found that 80 pulses (in 56 GRBs), 35 pulses (in 30 GRBs), and 14 pulses (in 12 GRBs) present purely non-thermal spectra, hybrid spectra, and thermal spectra, respectively.


\section{Estimated the Magnetization Parameter}
A good fraction of our sample whose spectra lack detected thermal emission (47 GRBs out of 88), are dominated by non-thermal component. One can estimate the lower limits of the magnetization factor of the non-thermal spectra which is used to suppress the thermal component. More details can be referred to \citep{2009ApJ...700L..65Z} and \citep{2024MNRAS.529L..67D}.

In the framework of the fireball model, the point of emission where the fireball reaches transparency is denominated as the photosphere radius, namely, the electron scattering optical depth $\left(\tau_{\gamma e}^{\prime}=n^{\prime} \sigma_{T} \Delta^{\prime}\right)$ is close to 1. Here, $\sigma_{T}$ is the Thomson cross section, $n^{\prime}$ and $\Delta^{\prime}$ are electron number density and width of the ejecta shell in the rest frame comoving with the ejecta, respectively \citep{2000ApJ...530..292M,2005ApJ...628..847R,2007ApJ...666.1012T,2007MNRAS.382L..72G,2009ApJ...700L..47L}. By assuming a pure baryonic flux, we derive a thermal component spectrum that can be emitted from the photosphere with a total wind luminosity of $L_{w}$ \citep{2018pgrb.book.....Z}, and the photosphere radius can be written as \citep{2000ApJ...530..292M,2008ApJ...682..463P,2009ApJ...700L..65Z,2015ApJ...801...2,2018pgrb.book.....Z,2024MNRAS.529L..67D}
\begin{eqnarray}
R_{\mathrm{ph}}=\left\{\begin{array}{ll}
\left(\frac{L_{\mathrm{w}} \sigma_{\mathrm{T}} R_{0}^{2}}{8 \pi m_{\mathrm{p}} c^{3} \eta}\right)^{1 / 3}, & R_{\mathrm{ph}}<R_{\mathrm{c}} \\
\frac{L_{\mathrm{w}} \sigma_{\mathrm{T}}}{8 \pi m_{\mathrm{p}} c^{3} \Gamma^{2} \eta}, & R_{\mathrm{ph}}>R_{\mathrm{c}}
\end{array}\right.
\end{eqnarray}
where $\eta = L_{\mathbf{w}}/\dot{M}c^{2}$ is dimensionless entropy of baryonic flow, $R_{\mathrm{c}} \sim R_{0}\times\min(\eta,\eta_{*})$ is the radius where ejecta enter the ’coasting’ phase, $R_{0}=c\Delta t_{\mathrm{obs}}$ is the radius at which the ejecta is emitted from central engine, and $\Delta t_\mathrm{obs}$ is the variability time scale of the central engine. In this work, we fixed the $\Delta t_\mathrm{obs}=\frac{1}{3}$ s which is corresponding $R_{0}=10^{10}$ cm. $\eta_{*}=(L_{\mathrm{w}}\sigma_{T}/8\pi m_{\mathrm{p}}c^{3}R_{0})^{1/4}$ is critical dimensionless entropy, $m_{p}$ and $c$ are the fundamental constants proton mass and speed of light, respectively. The coasting Lorentz factor is $\Gamma = \eta$ for $R_{\mathrm{ph}}>R_{\mathrm{c}}$ and $\Gamma=\eta_{*}$ for $R_{\mathrm{ph}}\leq R_{\mathrm{c}}$, respectively. The initial total wind luminosity ($L_{w}$) of the fireball is at least larger than the observed gamma-ray luminosity ($L_{\gamma}$) because the radiation efficiency is not as high as 100\%, i.e.,  $L_{w}\geqslant L_{\gamma}$. As the outflow expands with time, such an outflow is postulated to have emitted residual thermal radiation at the photospheric radius. The luminosity of the thermal component can be written as \citep{2000ApJ...530..292M}
\begin{eqnarray}
L_{\mathrm{th}}=\left\{\begin{array}{ll}
L_{\mathrm{w}}, & \eta>\eta_{*}, R_{\mathrm{ph}}<R_{\mathrm{c}} \\
L_{\mathrm{w}}\left(\eta / \eta_{*}\right)^{8 / 3}, & \eta<\eta_{*}, R_{\mathrm{ph}}>R_{\mathrm{c}}.
\end{array}\right.
\end{eqnarray}
One can calculate the temperature of the blackbody component which is produced from the photosphere \citep{2000ApJ...530..292M,2008ApJ...682..463P},
\begin{eqnarray}
T_{\mathrm{ph}}^{\mathrm{ob}}=\left\{\begin{array}{ll}
\left(\frac{L_{\mathrm{w}}}{4 \pi R_{0}^{2} a}\right)^{1 / 4}(1+z)^{-1}, & R_{\mathrm{ph}}<R_{\mathrm{c}} \\
\left(\frac{L_{\mathrm{w}}}{4 \pi R_{0}^{2} a}\right)^{1 / 4}\left(\frac{R_{\mathrm{ph}}}{R_{\mathrm{c}}}\right)^{-2 / 3}(1+z)^{-1}, & R_{\mathrm{ph}}>R_{\mathrm{c}}
\end{array}\right.
\end{eqnarray}
where $a$ is the Stefan-Boltzman’s constant.

In our calculations, we only focus on the 80 pulses that are from 56 GRBs in our sample. By assuming that a pseudo blackbody spectrum is produced by the photosphere of GRB, we can calculate and plot the lower limits of the expected photosphere spectrum for the internal shock model ($L_{\mathrm{w}}=L_{\gamma}$) in Figure \ref{fig:2}(a)-Figure \ref{fig:4}(a) (dashed lines). In order to suppress the pseudo thermal emission, one can infer a lower limit on the magnetization parameter ($\sigma=L_{\mathrm{p}}/L_{\mathrm{b}}$), which is defined as the ratio between the Poynting flux ($L_{\mathrm{p}}$) and the baryonic flux ($L_{\mathrm{b}}$), and the wind luminosity can be rewritten as $L_{\mathrm{w}}=L_{\mathrm{p}}+L_{\mathrm{b}}=(1+\sigma)L_{\mathrm{b}}$. Based on the derived Eqs (9), (10), and (11), the $L_{\mathrm{w}}$ can be replaced with $L_{\mathrm{w}}/(1+\sigma)$ by assuming no dissipation of the Poynting flux below $R_{\mathrm{ph}}$. The precise value of $\sigma$ is difficult to obtain from observational data with the lack of thermal emission, but one can infer the minimum value of $\sigma$ which can be used to suppress (or hide) the expected thermal component from photosphere emission. For given $R_{0} = 10^{10}$ cm, one can calculate the maximum temperature ($kT_{\rm max}$) of pseudo-thermal emission in units of keV. Then, we simulate the spectrum of pseudo-thermal radiation by assuming $\sigma=0$, and compare the thermal spectrum with that of the observed non-thermal spectrum. If the pseudo-thermal spectrum exceeds the spectral lines of the non-thermal spectrum, we re-scaled up for increasing $\sigma$ by a step of 0.01 until the pseudo-thermal spectrum is entirely below the non-thermal spectrum. Finally, we set the the value of calculated $\sigma$ above as the lower limit of the magnetization factor for the pulse.

\begin{figure}
{ \centering
\resizebox*{\textwidth}{0.4\textheight}
{\includegraphics{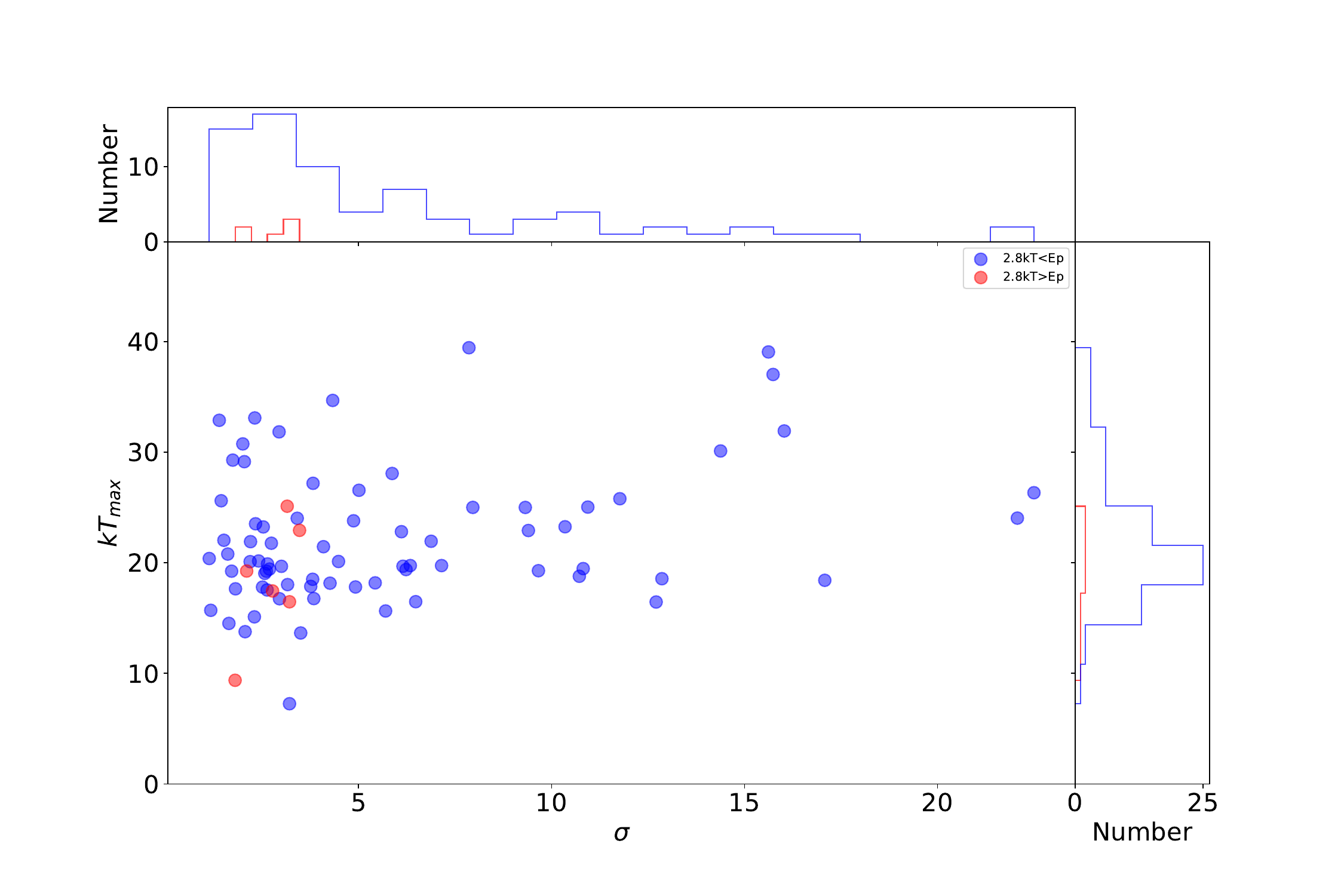}
}}
\caption{2-D distributions of $\sigma$ and $kT_{\rm max}$. The solid circles are the lower limit of estimated $\sigma$ for pulses with non-thermal spectrum.}
 \label{fig:5}
\end{figure}

In Figure \ref{fig:5}, we show 1D distribution ($kT_{\rm max}$ and lower limits of $\sigma$), and 2D ($kT_{\rm max}-\sigma$) diagram for 80 pulses with non-thermal spectra. The spectral fitting results with Band and CPL models, as well as the lower limit of $\sigma$, are shown in Table \ref{Table2}. It is found that 6 GRBs (e.g., 091127, 091208B, 111228A, 121128A, 190829A, and 211023A) whose pseudo-thermal spectra are located after the peak of non-thermal spectrum. It results in the calculated magnetization factor not being the minimum value. Therefore, we exclude these six pulses from 80 pulses in our calculations. As shown in Figure \ref{fig:5}, we find that the values of $kT_{\rm max}$ for 74 pulses with the non-thermal spectrum are distributed in the range of 7.25 keV - 39.45 keV, and the peak value of Gaussian distribution is located at 20.5 $\pm$ 0.4 keV. On the other hand, the lower limits of $\sigma$ are distributed in the range of 1.13-22.5, about $\sim$ 40.5\% (30 out of 74) with lower limits of $\sigma$ are larger than 5, and $\sim$ 18.9\% (14 out of 74) with lower limits of $\sigma$ are larger than 10. The high magnetization factor of GRBs is used to suppress the absence of observed pseudo-thermal components, which is consistent with the presence of observed non-thermal components. The high magnetization factors suggest that the baryonic model at least does not work for those GRBs, and the initial wind luminosity is not stored in the form of fireball. A possible way to solve the above contradiction is to invoke a Poynting-flux-dominated outflow of GRBs which requires a relatively high magnetization factor. 

\section{Energy-resolved light curve fitting}
Excepting the different magnetization factor $\sigma$ predicted from different models (e.g., photosphere, internal shock, and  ICMART), the light curves of prompt emission predicted by different models also exhibit different shape-energy dependence \citep{2018pgrb.book.....Z,2025ApJ...985..239Y}. Depending on the unknown jet composition, three emission sites are commonly discussed in GRB prompt emission model. One is a matter-dominated fireball, the observed emission is likely a superposition of a thermal component which is originated from the fireball photosphere at $R_{\mathrm{ph}} \sim 10^{11}-10^{12}$ cm. In this model, rapid variability of light curves in GRB prompt emission is attributed to erratic central engine activities \citep{1986ApJ...308L..43P,1986ApJ...308L..47G,2000ApJ...530..292M}. The activity of the central engine or energy injection can lead to a shape-energy dependence in the light curve, such as spectral lags and a global hard-to-soft evolution \citep{2016A&A...592A..95M,2016ApJ...825...97U,2018ApJ...869..100U,2021ApJ...922...34C}. According to the results of simulation from \citet{2025ApJ...985..239Y}, the internal shock model, the spectral feature exhibit a non-thermal spectrum with a relatively small magnetization factor, the typical emission region is $R_{\rm IS} \sim \Gamma^{2}c\Delta t \sim (3\times10^{14} \mathrm{cm})(\Gamma/100)^{2}(\delta t/1 \mathrm{s})$ \citep{1994ApJ...430L..93R,1997ApJ...490...92K} from the central engine, where $\Gamma \sim 100$ and $\delta t \sim 1$ are the bulk Lorentz factor and typical variability time scale of the fast component, respectively. The light curve also shows significant differences compared with that of the photosphere model. The broad pulse of the light curve from internal shock is composed of numerous short-duration pulses that are caused by the shell collisions. The broad pulse is caused by the history of central engine activity, and the light curves should share the same behavior in different wavebands rather than displaying a global energy-dependent behavior \citep{2025ApJ...985..239Y}. 

Alternatively, another one is a Poynting-flux-dominated outflow, such as the ICMART model. The spectrum exhibits a non-thermal with a high magnetization factor, and the collision site between two magnetically-dominated shells is at $R_{\mathrm{ICMART}} \sim (1.2\times10^{16} \mathrm{cm})(\Gamma/100)^{2}(\Delta t/40 \mathrm{s})$. The collisions between a pair of highly magnetized shells with different Lorentz factors would distort the ordered magnetic field and trigger fast magnetic seeds, which will induce relativistic magnetohydrodynamic (MHD) turbulence. Numerical simulations have demonstrated that MHD turbulence has the ability to generate random local magnetic reconnection events \citep{2015ApJ...815...16T,2017ApJ...838...91K,2018MNRAS.476.4263T}. The magnetic reconnection events effectively convert the stored magnetic energy within the fluid medium into particle energy, a process that inherently perturbs and further distorts the magnetic field structure, triggering a cascade of successive reconnection events. Subsequently, the particle energy is transformed into the kinetic energy of the bulk via adiabatic cooling mechanisms and then converted into radiation photons through synchrotron radiation. If this is the case, the broad pulse should be the simple superposition of emissions from many mini-jets. Each mini-jet emission is used to contribute to the fast pulse, and the random latitudes and orientations can cause the high variability of the light curve. As the outflow expands, the global magnetic field strength in the emission region naturally decreases, and it can lead to a well-defined shape-energy dependence behavior, as well as the spectral lags and a global hard-to-soft evolution \citep{2016ApJ...825...97U,2018ApJ...869..100U}.

In Section 3, we have identified 30 non-thermal spectral pulses from 21 GRBs with magnetization factors larger than 5, we adopt an independent method which is energy-resolved light curve fitting to double-check the jet composition of these pulses with high magnetization factor (with a lower limit of at least 5). We divide the data that are from TTE data into eight energy bands (1st to 8th channels), namely, 8-30 keV, 30-70 keV, 70-100 keV, 100-150 keV, 150-200 keV, 200-300 keV, 300-500 keV, and 500-900 keV. The peak energy ($E_{\rm p}$) of most GRBs is distributed in the range of 100 keV to 200 keV \citep{2025ApJS..276...62C}, therefore we adopt a finer and broader energy binning in the low- and high-energy bands, respectively. Moreover, the method of dividing energy bands has been widely used in studies of energy-dependent of light curve and spectral lag \citep{2007ApJS..169...62H,2018ApJ...865..153L,2025ApJ...985..239Y}. The counts are binned with 0.128 and 1.024 seconds, respectively. By subtracting the effect from the background, one can obtain the light curves without a background. Then, we adopt the empirical function which is from \citet{2005ApJ...627..324N} to fit energy-resolved pulses, and one can obtain the pulse width of each channel, the empirical function is the fast-rise-exponential-decay profile, which is defined as,
\begin{eqnarray}
I(t)=A\lambda\exp[-\tau_1/(t-t_s)-(t-t_s)/\tau_2]
\end{eqnarray}
where $t$ is the time since the trigger, $A$ is the pulse amplitude, $t_{s}$ is the pulse start time, $\tau_{1}$ and $\tau_{2}$ are the time scales of the rise and decay, respectively. $\lambda$ is a constant which is related to $\tau_{1}$ and $\tau_{2}$, namely, $\lambda=\exp[2(\tau_{1}/\tau_{2})^{1/2}]$.
The width $w$ of the pulse is therefore defined as $w=\tau_2(1+2\ln\lambda)^{1/2}$, the peak time, rise time, and decay time of pulse can be expressed as $t_{\text {p}}=t_{s}+\left(\tau_{1} / \tau_{2}\right)^{1 / 2}$, $t_{\text {r}}=\frac{1}{2} w(1+\frac{\tau_2}{w})$, $t_{\text {d}}=\frac{1}{2} w(1-\frac{\tau_2}{w})$, respectively. In this work, a 0.128 s time-bin of light curves is used to do the fitting process. The temporal resolution we adopted is approximately 0.1 seconds to mitigate statistical variations in the light curve, and it can arise from background noise, thereby ensuring the high quality of individual pulse within a burst. This approach aligns with standard practices established in prior studies \citep{2021ApJS..254...35L,2025ApJ...985..239Y,2025ApJ...991..230G}.

\begin{figure} 
    \centering %
    \begin{minipage}{0.45\textwidth} %
        \centering %
        \includegraphics[width=\textwidth]{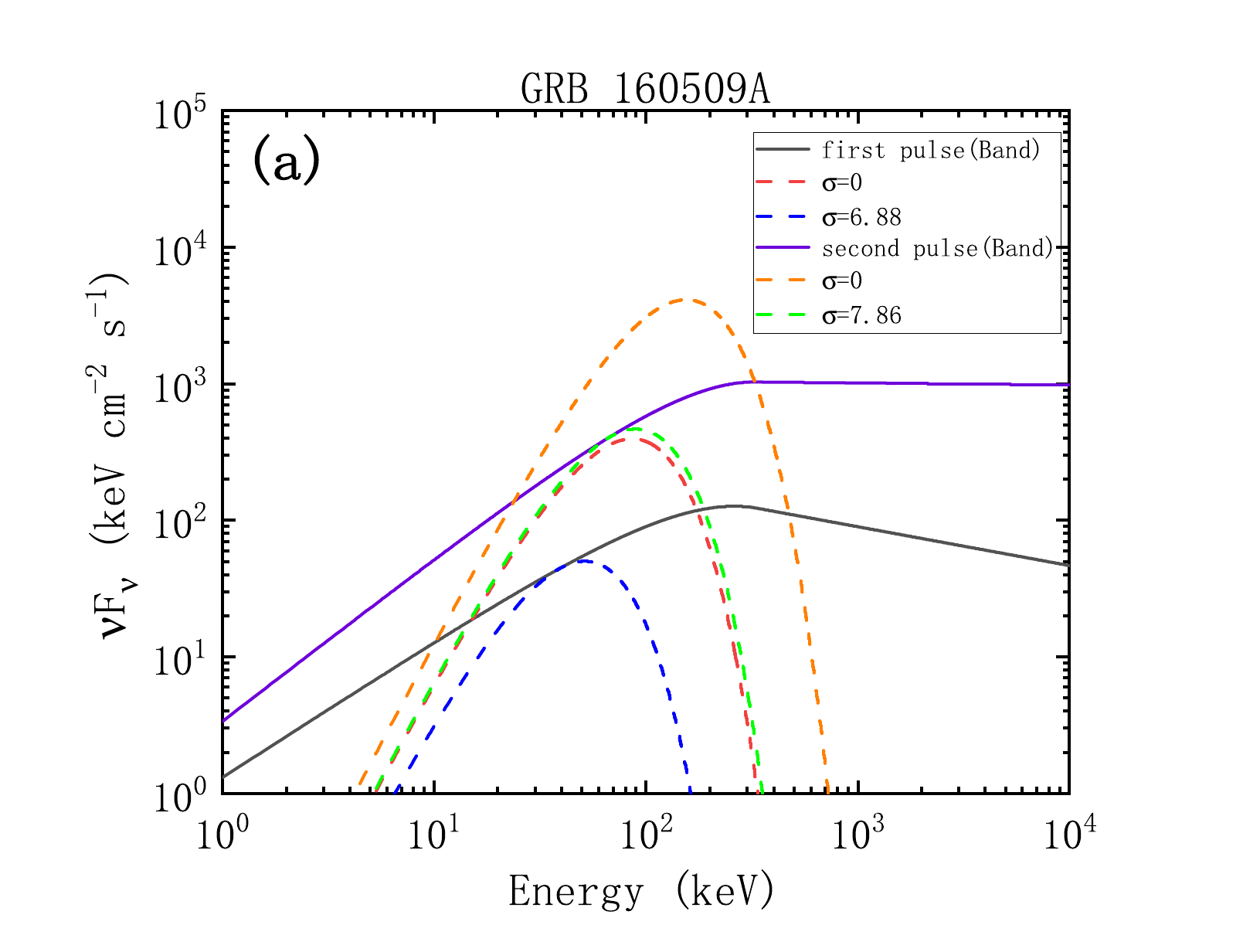} %
        \vskip\baselineskip %
        \includegraphics[width=\textwidth]{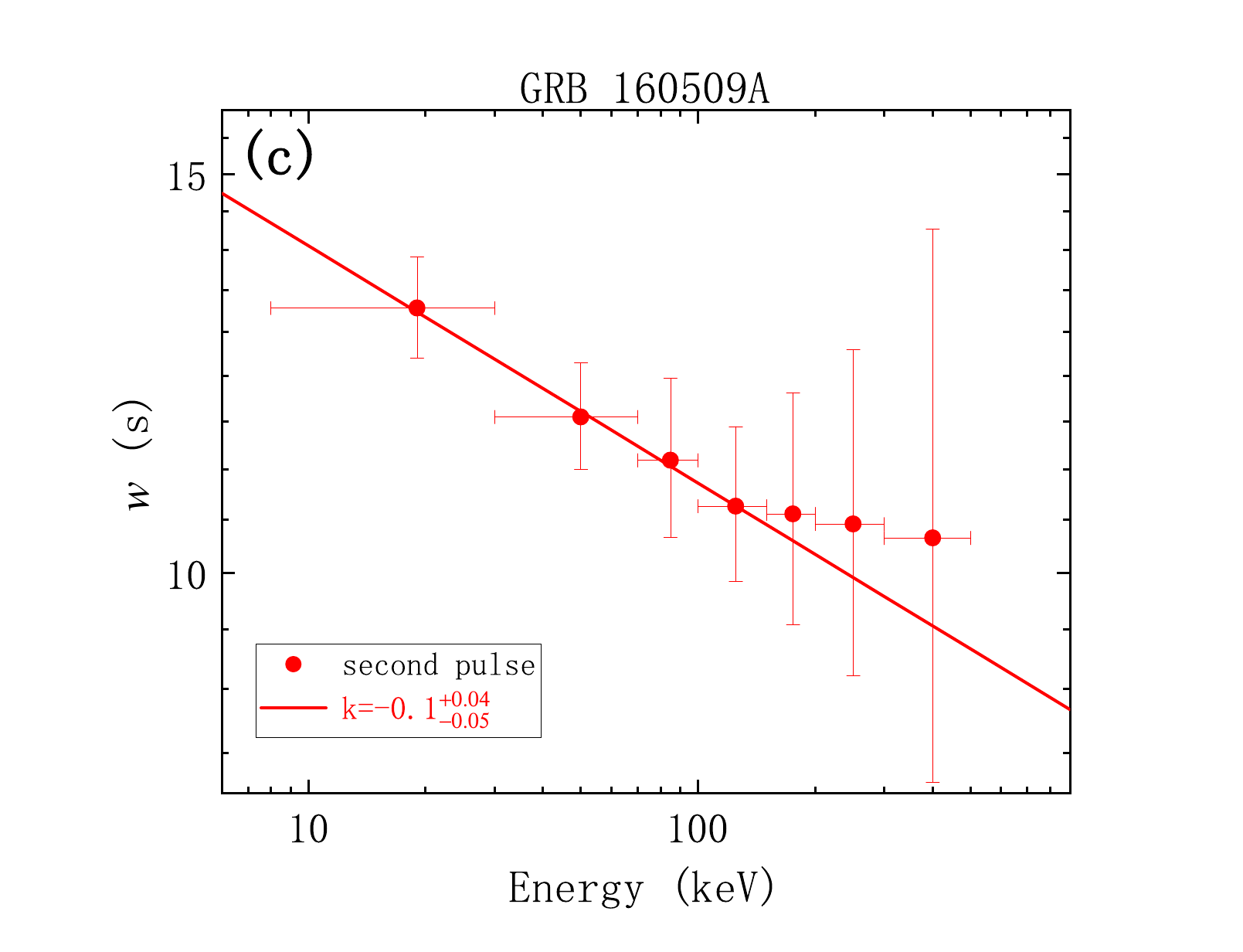} %
    \end{minipage}
    \hfill %
    \begin{minipage}{0.45\textwidth} %
        \centering %
        \includegraphics[width=\textwidth]{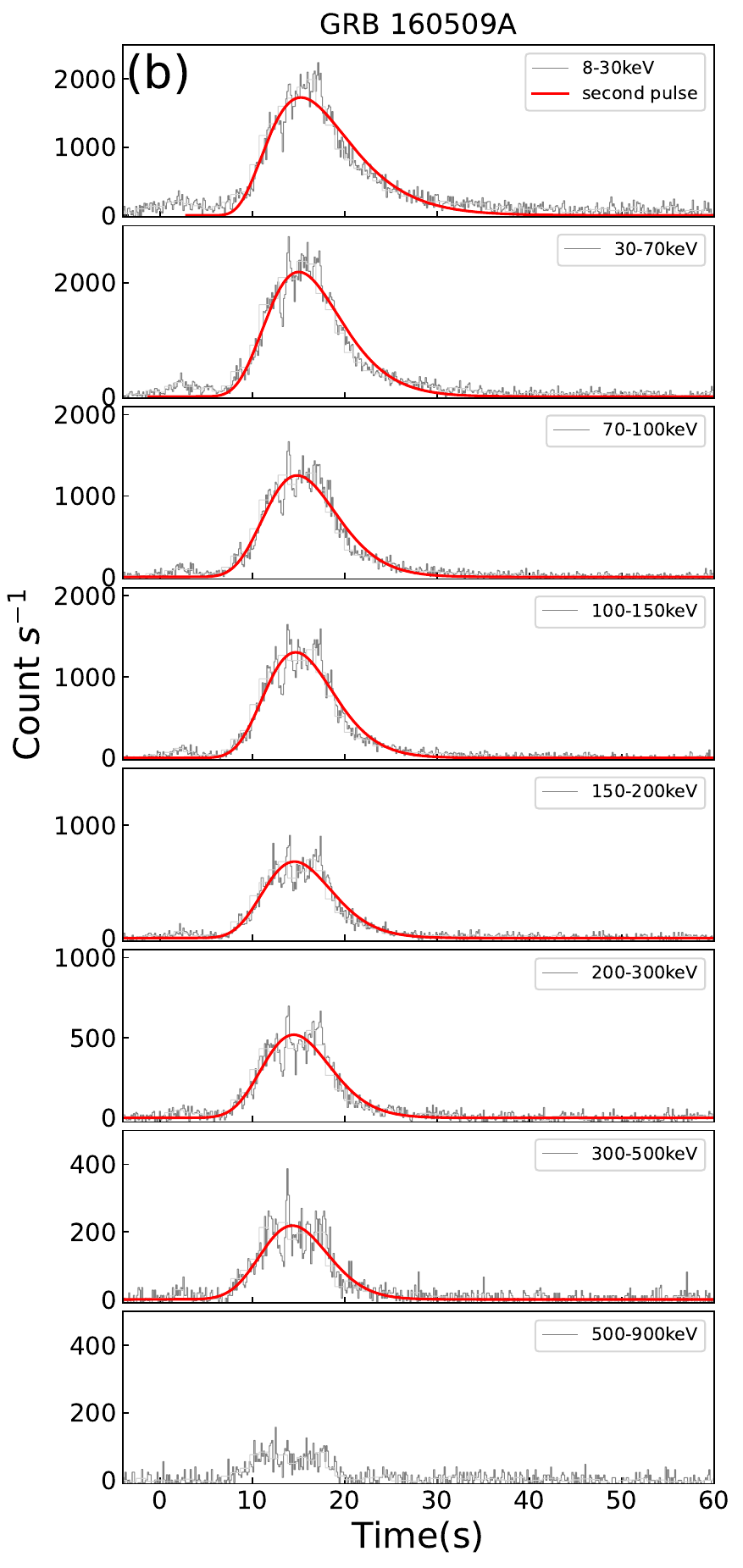} %
    \end{minipage}
    \caption{Spectral and temporal analysis of an example GRB 160519A for group (I). (a‌). The predicted lower limits of the photosphere spectra (dashed lines) with $R_{0}=10^{10}$ cm and observed non-thermal spectrum (solid line). (b). The light curves of the prompt emission of GRB 160519A (gray) in the different energy ranges and the FRED model fitting (red solid lines). (c). The pulse width ($w$) is derived from FRED model fitting as a function of energy and the power-law fitting (solid red line).}  
     \label{fig:2}
\end{figure}

\begin{figure} 
    \centering %
    \begin{minipage}{0.45\textwidth} %
        \centering %
        \includegraphics[width=\textwidth]{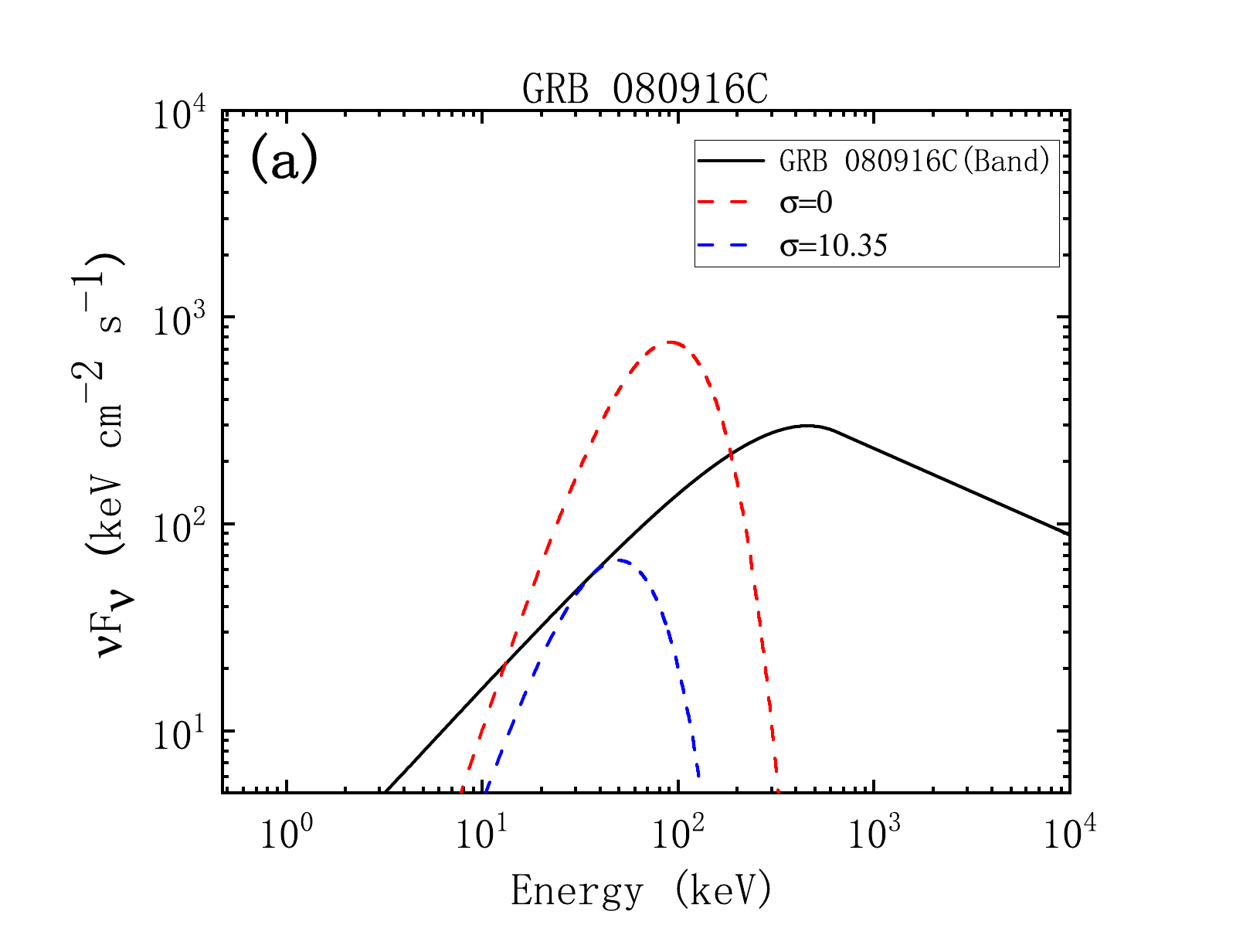} %
        \vskip\baselineskip %
        \includegraphics[width=\textwidth]{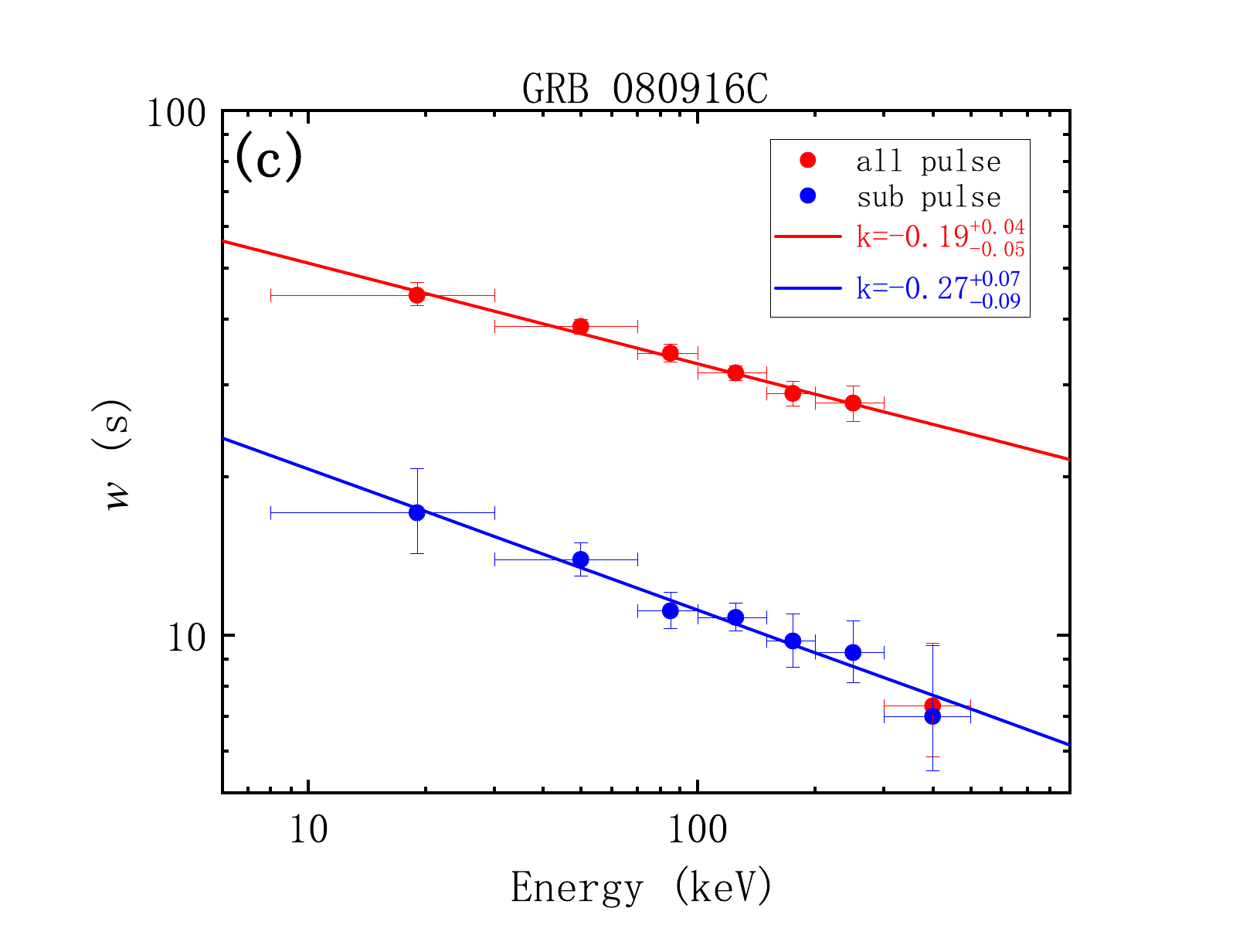} %
    \end{minipage}
    \hfill %
    \begin{minipage}{0.45\textwidth} %
        \centering %
        \includegraphics[width=\textwidth]{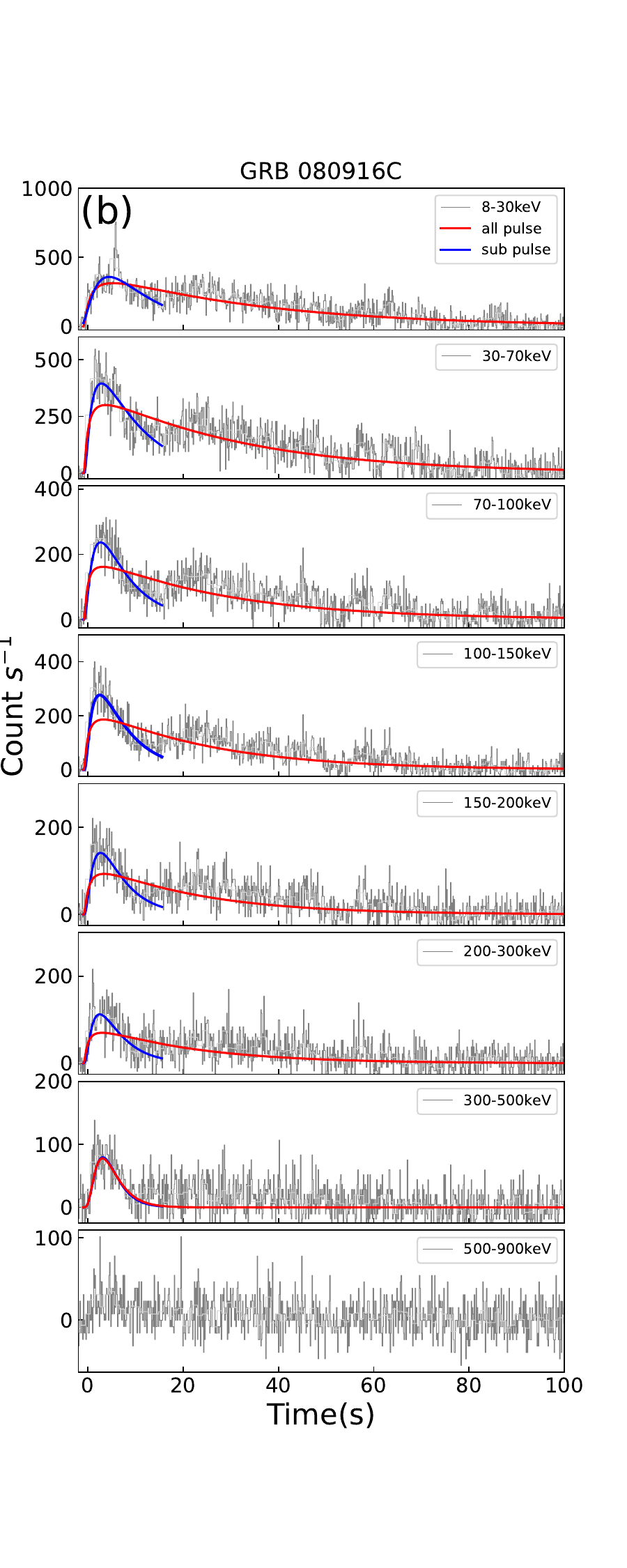} %
    \end{minipage}
    \caption{Spectral and temporal analysis of an example GRB 080916C for group (II). (a‌). The predicted lower limits of the photosphere spectra (dashed lines) with $R_{0}=10^{10}$ cm and observed non-thermal spectrum (solid line). (b). The light curves of the prompt emission of GRB 080916C (gray) in the different energy ranges and the FRED model fitting (red and blue solid lines). (c). The pulse width ($w$) is derived from the FRED model fitting for sub-pulse and whole pulse as a function of energy and the power-law fitting (red and blue solid lines).}  
     \label{fig:3}
\end{figure}

\begin{figure} 
    \centering %
    \begin{minipage}{0.45\textwidth} %
        \centering %
        \includegraphics[width=\textwidth]{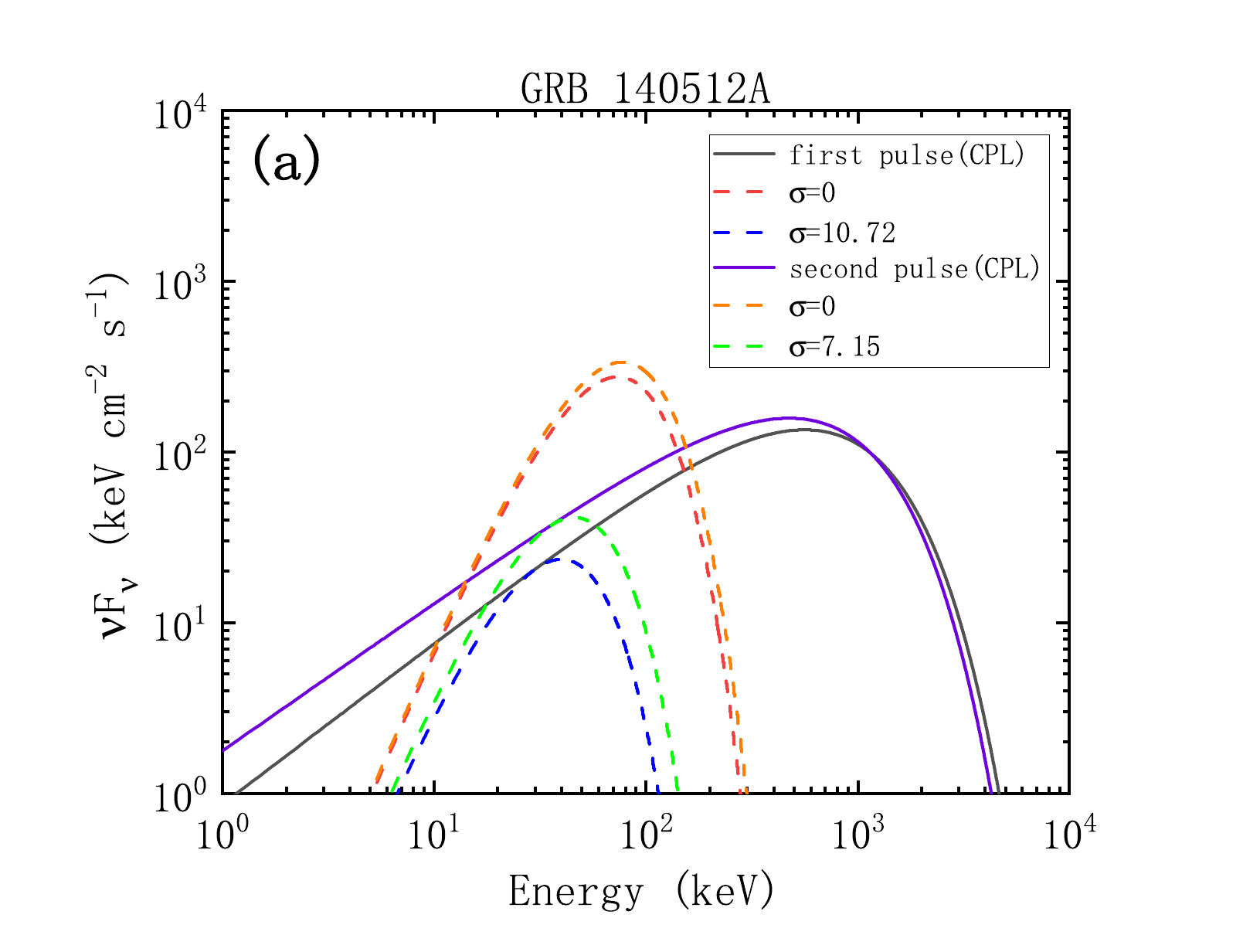} %
        \vskip\baselineskip %
        \includegraphics[width=\textwidth]{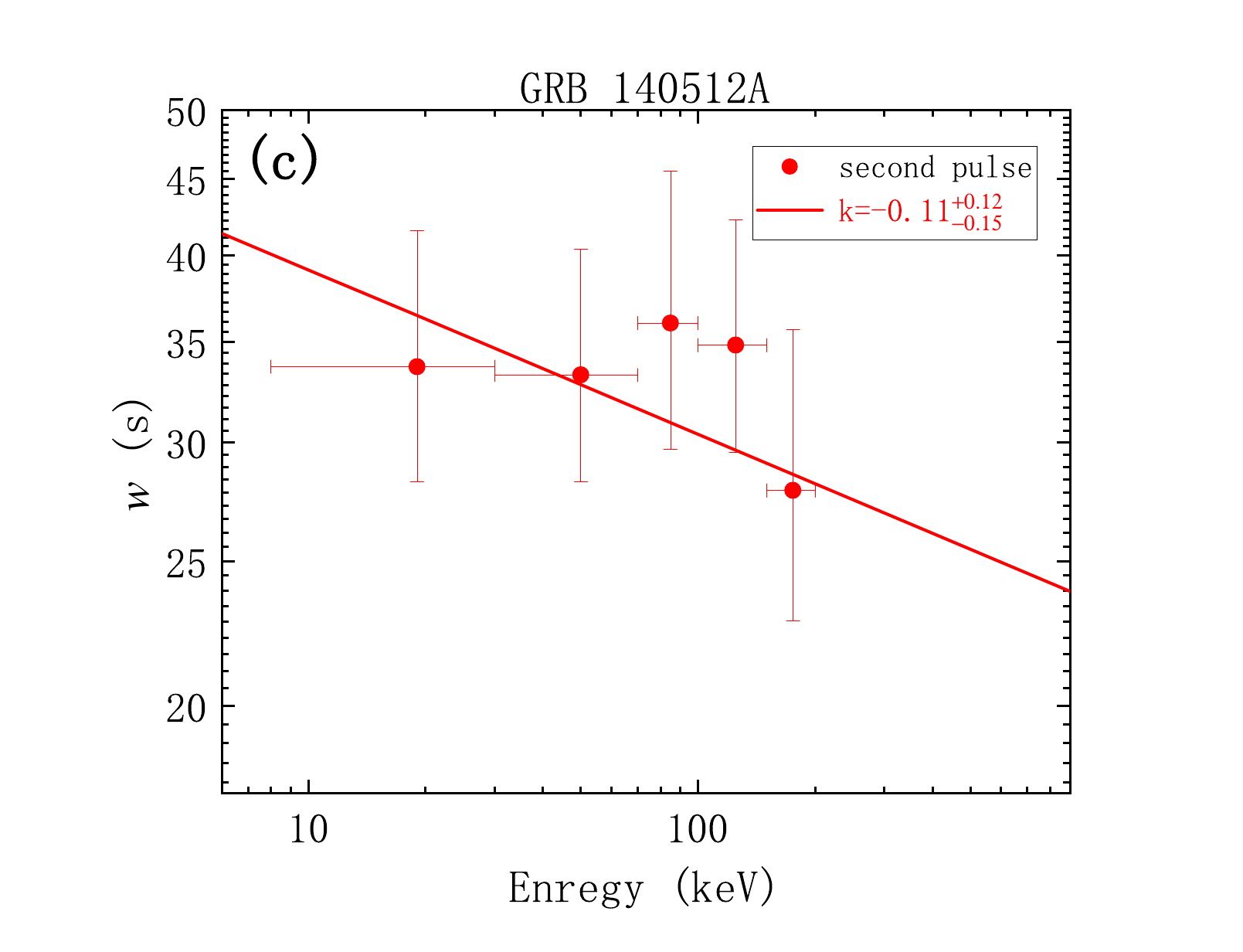} %
    \end{minipage}
    \hfill %
    \begin{minipage}{0.45\textwidth} %
        \centering %
        \includegraphics[width=\textwidth]{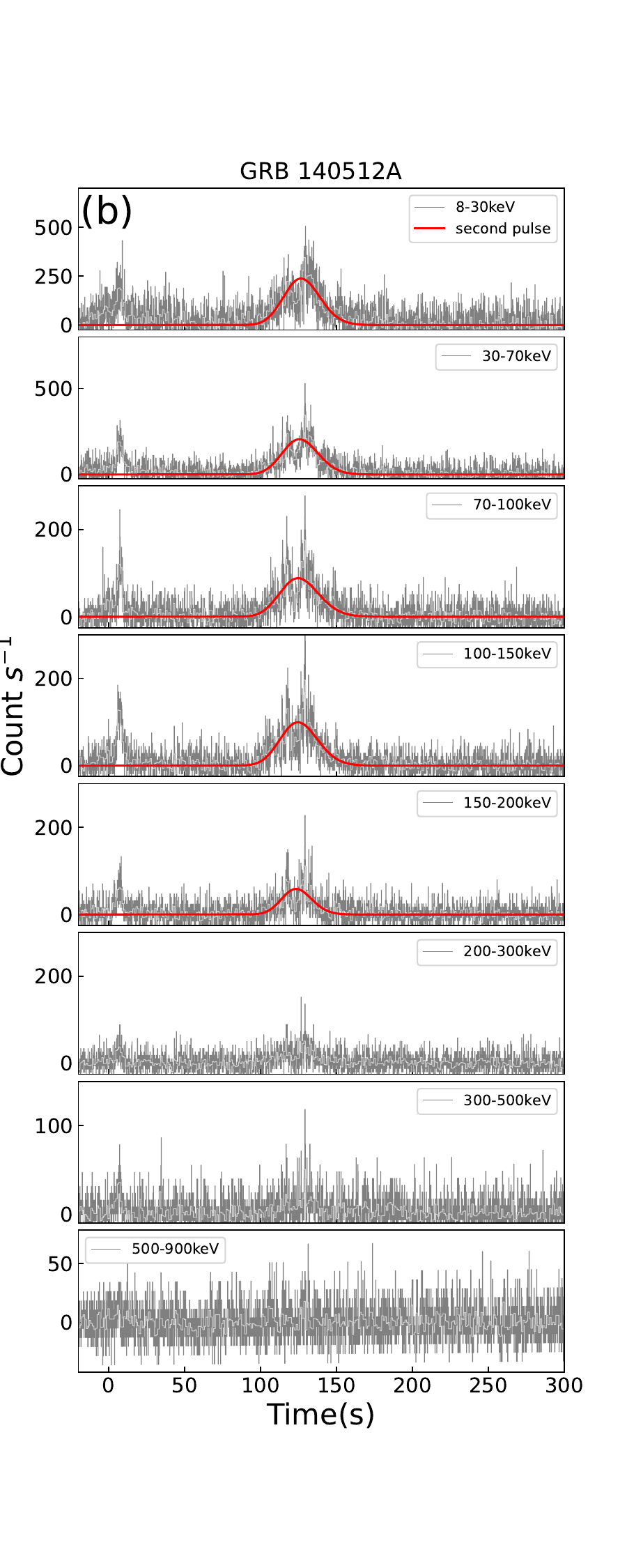} %
    \end{minipage}
    \caption{Spectral and temporal analysis of an example GRB 140512A for group (III). (a‌). The predicted lower limits of the photosphere spectra (dashed lines) with $R_{0}=10^{10}$ cm and observed non-thermal spectrum (solid lines). (b). The light curves of prompt emission of GRB 140512A (gray) in the different energy ranges and the FRED model fitting (red solid lines). (c). The pulse width ($w$) is derived from the FRED model fitting for the bright pulse as a function of energy and the power-law fitting (red solid lines).‌‌} %
     \label{fig:4}
\end{figure}

Pulses with S/N less than 35 are excluded due to the inability to obtain distinct energy-resolved pulses. The total of 15 GRBs (e,g., GRBs 080916C, 100414A, 120624B, 120711A, 131108A, 140512A, 150821A, 160509A, 170405A, 180703A, 181020A, 200524A, 220107A, 220627A, 230204B) which are including 20 pulses are analyzed in details\footnote{We find that it is difficult to proceed with the subsequent energy-resolved light curve fitting when the S/N is less than 35. So, 6 single-pulse GRBs (e.g., GRBs 091020, 130215A, 131011A, 140606B, 160629A, 231118A), together with other four pulses (the second pulse of GRB 131108A, the first pulse of GRB 140512A, the first pulse of GRB 160509A, and the first pulse of GRB 220627A), are excluded from our analysis (see Table \ref{Table2}).}, and the energy-resolved light curve fitting results are presented in Table \ref{Table3}. In Figures (\ref{fig:2}-\ref{fig:4}), we show an example of the observed non-thermal spectral (solid lines) and the pseudo-thermal spectra (dashed lines) which are suppressed with a high magnetization factor (a), the light curves of prompt emission in different energy bands (b), and the relationship between the pulse width and energy of the case (c). The results of other GRBs can be found in Appendix.

Based on the fitting results from the prompt emission of our sample, one needs to find out the relationship between pulse width ($w$) and energy ($E$). So that, we adopt a power-law function to fit the $w$ and $E$ (e.g., $w\propto E^{k}$) by employing Bayesian inference framework, and one can obtain the slope $k$ and its uncertainty from the posterior distribution. Specifically, the posterior distribution of the power-law index $k$ is sampled by using the $emcee$ package. During the fitting process, the likelihood function is constructed based on a skew-normal distribution to account for asymmetric uncertainties in $w$. The median value of the posterior distribution is adopted as the best-fit parameter, and the 68\% credible interval is reported as the uncertainty range, based on 10,000 MCMC samples with the first 2,000 steps discarded as burn-in. Moreover, we also try to find out the correlation coefficient between $w$ and $E$, we adopt a Monte Carlo sampling approach that accounts for measurement uncertainties in both variables. Specifically, we generate 50000 realizations of the dataset by drawing from normal distributions for $E$ (considering symmetric errors) and skew-normal distributions for $w$ (to reflect asymmetric uncertainties). For each dataset, we compute the Spearman rank correlation coefficient ($\rho_{\rm s}$), thereby constructing its sampling distribution. The median value of this distribution is taken as the best estimated of the correlation, with the 68\% credible interval representing its uncertainty range. In addition, we also calculate the point-estimate Spearman correlation coefficient $\rho_{\rm p}$ without accounting for measurement uncertainties. The $k$ values, $\rho_{\rm s}$, and $\rho_{\rm p}$ are reported in Table \ref{Table4}, and the power-law fitting are also shown in Figures \ref{fig:2}-\ref{fig:4} as an example. One needs to note that we exclude the last two data points of GRB 220104A and the last data point of the global pulse of GRB 080916C due to their failure to adequately capture the overall pulse structure in the fitting and correlation analysis. The results of other GRBs can be found in Appendix.

Based on the results of light curve fitting and the evolution behavior of the relationship between the pulse width and energy of the 15 GRBs, there are two criteria adopted for our sample categorizing. First, how many pulses of the light curve are used to fit. Second, whether the relationship between the pulse width and energy is existence. If this is the case, one can roughly categorize them into three distinct groups: 
\begin{itemize}	
\item[$\bullet$] (I): Only a single pulse is used for fitting, and its pulse width ($w$) is inversely  proportional to the energy (7 GRBs; 100414A, 131108A, 160509A, 180703A, 181020A, 220107A, and 220627A). GRB 160509A is shown as an example in Figure \ref{fig:2}, and the results of the other six GRBs can be found in the Appendix (Figures \ref{fig:A1}-\ref{fig:A6}). From the perspective of power-law fitting, we find that the slope $k$ remains consistently negative even when measurement uncertainties are fully incorporated, indicating an inverse correlation between pulse width ($w$) and energy ($E$). Similarly, point estimates of the Spearman correlation coefficient ($\rho_{\rm p}$) also reveal a strong negative correlation in most cases. By accounting for the full error propagation (e.g., $\rho_s$), GRBs 131108A and 220627A still exhibit highly significant negative correlations. GRBs 100414A, 160509A, and 180703A show moderately strong negative trends, while GRBs 181020A and 220107A display no statistically significant correlation. The lack of significance in the latter two cases may be attributed to the limited photon counts in some energy bands, which result in large measurement uncertainties. Taking into account both the $k$ value and the correlation coefficient, the correlations between $w$ and energy for this group of GRBs are consistent with that of the prediction of the ICMART model and independently support its origin from the ICMART model with Poynting-flux-dominated outflow.

\item[$\bullet$] (II): Two or more pulses are used to fit, and the pulse width ($w$) is inversely proportional to the energy at least in two pulses. There are six GRBs (080916C, 120624B, 120711A, 150821A, 170405A, and 230204B) which are part of this group. GRB 080916C is shown as an example in Figure \ref{fig:3}, and the results of the other five GRBs can be found in the Appendix (Figures \ref{fig:B1}-\ref{fig:B5}). The inverse correlation between $w$ and $E$ for those four GRBs (GRBs 080916C, 120624B, 120711A, 150821A) are indeed existent with power-law index $k$ which are range from -0.16 to -0.37, and the correlation coefficients of Spearman also support the inverse correlations (see Table \ref{Table4}). Even after fully accounting for measurement uncertainties, these four GRBs still exhibit strong or moderate negative correlations. Based on above results, it also independently supports its origin from the ICMART model with Poynting-flux-dominated outflow, and the results are consistent with that of a high$-\sigma$ factor derived from spectra with no-thermal emission. One needs to note that the first sub-pulse of GRB 170405A which exhibits a low correlation in the $w$-energy relation may be attributed to the presence of two outliers above 150 keV. Another noteworthy observation is that the fourth sub-pulse k value and correlation coefficient of GRB 230204B are close to zero with a large uncertainty, and it may be caused by the '$w$-$E$' relation which exhibit a positive correlation below 100 keV and a negative correlation between 100 keV and 300 keV. Excluding these two pulses, the remaining four pulses of GRB 170404A and GRB 230204B yield negative values of the power-law index $k$, indicating an inverse $w$–$E$ relationship. The point-estimate Spearman correlation coefficients ($\rho_{\rm p}$) also suggest strong or moderate negative correlations. However, when measurement uncertainties are fully taken into account, the resulting $\rho_{\rm s}$ values are not statistically significant, likely due to the limited number of data points and relatively large uncertainties in some energy bands.

\item[$\bullet$] (III) Only one single pulse is used to fit, yet its width does not exhibit a notable dependency on energy (GRB 140512A and GRB 200524A). We do not obtain a convincing correlation between $w$ and energy for those two GRBs. The slopes for the two GRBs are approximately zero, and the correlation coefficients are also low enough and Table \ref{Table4}). GRB 140512A is shown as an example in Figure \ref{fig:4}, and the result of GRB 200524A can be found in the Appendix (Figure \ref{fig:C1}). The possible reason for this phenomenon may be due to the excessive uncertainty in the fitting, leading to the error bars of $w$ being relatively large. It can result in a low significant $w$-energy relation. Of course, the other reason is that its origin may be from the internal shock model rather than the ICMART model.
\end{itemize}
In any case, one can roughly obtain that 87\% GRBs (13 out of 15) in our sample with high-$\sigma$ (e.g., $\sigma>5$) are also supported by the light curves analysis with an inverse correlation between $w$ and energy. It means that the jet composition of GRBs with high-$\sigma$ in our sample should be intrinsically dominated by Poynting-flux, and are consistent with that of the prediction of the ICMART model. For example, GRB 080916C exhibits a long-duration broad pulse spanning over 70 seconds, with a prominent dip around 15 seconds that divides the broad pulse into two distinct parts. Based on the ICMART scenario, the broad pulse is comprised of numerous mini-jets and the intensity of the pulse is determined by the number of mini-jet events. The dip is attributed to the waiting time between mini-jet events. If this is the case, then the shape-energy dependence characteristic of the global broad pulse should be present in any individual sub-pulse. It is consistent with the result from \citet{2009ApJ...700L..65Z}.

\section{Conclusion and discussion}
In this paper, We have presented a comprehensive temporal and spectral analysis of prompt emission of GRB data observed by Fermi/GBM during 15 years of operation, and focus on the bright GRBs (total 88 GRBs) which are redshift measured with $S/N>20$. The light curves of prompt emission for 60 out of 88 GRBs show one clear pulse, 19 out of 88 GRBs have two distinct pulses, 6 out of 88 GRBs exhibit three different pulses, 2 out of 88 GRBs have four different pulses, as well as 1 out of 88 GRBs has five pulses. In total, 129 pulses are obtained from these 88 GRBs. Then, we do the spectral analysis for those 129 pulses by adopting different models to fit and present the best model for each pulse based on the value of BIC. It is found that 80 pulses (in 56 GRBs), 35 pulses (in 30 GRBs), and 14 pulses (in 12 GRBs) present purely non-thermal spectra, hybrid spectra, and thermal spectra, respectively.

Our aim is to identify the jet composition of those GRBs whose spectra are without thermal component by employing two independent methods, one is the observed non-thermal spectra to estimate the magnetization factor ($\sigma$), and the other one is to find out the possible relationship between pulse width ($w$) and energy in the light curve of prompt emission. Our results are summarized as follows:
\begin{itemize}	
\item[$\bullet$] By adopting $R_{0}=10^{10}$ cm, one can roughly estimate the lower limits of $\sigma$ for 80 pulses (in 56 GRBs) \footnote{Since there are 6 GRBs (e.g., 091127, 091208B, 111228A, 121128A, 190829A and 211023A) whose pseudo-thermal spectra are located after the peak of the non-thermal spectrum, so we exclude these six pulses from 80 pulses in our analysis.}. We find that the estimated lower limits of $\sigma$ for 74 pulses by adopting the observed non-thermal spectrum to suppress the possible thermal emission are distributed in the range of 1.13-22.5, and there are 40.5\% (30/74) of the pulses to be lower limits of  $\sigma>5$. It suggests that at least more than 40\% of non-thermal pulses in the prompt emission mainly arise from magnetized dissipation in such highly magnetized environments rather than particle acceleration via internal shocks. Also, the high magnetization parameter ($\sigma >$ 5) of pulses is also consistent with the prediction of Poynting-flux-dominated outflow in the ICMART scenario. Moreover, those 30 pulses from 21 GRBs in our sample, suggest that over 23.8\% (21/88) of GRBs originated from the ICMART scenario.
\item[$\bullet$] On the other hand, by fitting those 30 pulses from 21 GRBs with $\sigma >$ 5 adopting the FRED model, it is found that 6 GRBs are not bright enough to do energy-resolved light curve fitting. So, we conduct an energy-resolved light curve fitting with the FRED model for those 20 pulses in 15 bright GRBs. We find that 13 out of 15 GRBs exhibit an inverse correlation between $w$ and energy in global pulse or sub-pulse, and it is also consistent with the predicted characteristics of mini-jets from the ICMART model. In other words, it is independent evidence to support those 13 GRBs with highly magnetized factor originating from the ICMART model. Furthermore, two GRBs (e,g., GRB 140512A and GRB 200524A) do not exhibit $w$ - Energy dependent relationships, which may be attributed to the excessive uncertainty in the fitting, leading to a large error in $w$, or its origin may be associated with internal shock model rather than ICMART model.
\end{itemize}

The estimated lower limit of the magnetization factor $\sigma$ for our sample is dependent on the $R_0$, and we only adopt $R_{0}=10^{10}$ cm to do the calculations. We find that the peak value of Gaussian distribution of $kT_{\rm max}$ is about 20.5 keV, and this value is very close to that of observations with $kT=28$ keV obtained from \citet{2024ApJ...972....1L}. Based on Eq. (11), by adopting a smaller value of $R_{0}$ which can result in a larger value of pseudo $kT_{\rm max}$, can exceed the peak of non-thermal spectrum, such as $R_{0}=10^{9}$ cm and $10^{8}$ cm. So that, we also adopt $R_{0}=10^{9}$ cm and $10^{8}$ cm to estimate the pseudo $kT_{\rm max}$ for our sample, and obtain the Gaussian distribution of $kT_{\rm max}$ peaked at 59.9 keV and 204.8 keV for $R_{0}=10^{9}$ cm and $10^{8}$ cm, respectively. They are larger than that of observations with $kT=28$ keV obtained from \citet{2024ApJ...972....1L}. It means that adopting $R_{0}=10^{10}$ cm to do the calculations is reasonable.

Moreover, we note that our estimated distribution of magnetization factors is a little bit smaller than that inferred by \citet{2024ApJ...972....1L}. It may be attributed to adopting different radius \citep{2015ApJ...801...2,2020ApJ...894..100L,2022ApJ...932...25C}, such as what we adopt the size is at the photosphere to constrain the magnetization factor values, whereas \citet{2024ApJ...972....1L} adopt the size is at the central engine to constrain the magnetization factors. Our analysis also poses one curious question in this study. We focus on the spectral evolution between pulses rather than possible spectral evolution within an individual pulse, and extract full-duration time-integrated spectra for each identified complete pulse. We do not consider the effect of spectral evolution within individual pulses, and such time-resolved spectra of individual pulses may affect that of the time-integrated spectra \citep{2025ApJ...991..230G}.

It is worth noting that the ICMART model is as one of the candidate models to explain the prompt emission of GRBs. Within this scenario, the magnetic reconnection events in strong magnetic field environments can trigger plasma turbulence, which in turn converts magnetic energy into particle kinetic energy and ultimately generates gamma-ray radiation through synchrotron radiation. It provides a new perspective for understanding GRB radiation mechanisms under extreme magnetic field conditions \citep{2011ApJ...726...90Z}. However, it is important to objectively point out that the ICMART model still has significant theoretical limitations, such as the physical process of magnetic energy dissipation has not been fully clarified. The limitation prevents the ICMART model from being on an equal footing with well-established models such as internal shocks or photospheric emission—both of which have clear physical bases for their radiation origins. The results of this study (e.g., high magnetization factor and the inverse correlation between the width and energy) are consistent with the predictions of the ICMART model, but are not the sole explanation. Further verification with higher-energy observations will be required in the future, and additional observational support is still needed to validate the physical mechanism of the ICMART model.

\section{acknowledgements} 
We acknowledge the use of the public data from the Fermi/GBM data archive. This work is supported by the Guangxi Science Foundation (grant Nos. 2023GXNSFDA026007 and 2025GXNSFDA02850010), the National Natural Science Foundation of China (grant Nos. 12494574 and 12133003), the Program of Bagui Scholars Program (LHJ), and the Guangxi Talent Program (“Highland of Innovation Talents”).

\section*{Data Availability}
The majority of the data what we adopt is publicly available data from Fermi/GBM. If one needs to adopt the calculated data in this
article, it should be to cite this reference paper and request to the authors.


\bibliographystyle{mnras}
\bibliography{MS} 



\clearpage


\clearpage
\onecolumn
\begin{longtable}{p{1.5cm}<{\centering}p{0.3cm}<{\centering}p{1.5cm}<{\centering}p{0.5cm}<{\centering}p{0.5cm}<{\centering}p{0.5cm}<{\centering}p{1.5cm}<{\centering}p{1.5cm}<{\centering}p{0.5cm}<{\centering}p{0.5cm}<{\centering}p{0.5cm}<{\centering}}
    \caption{Comparisons of BIC for different models of 129 pulses form 88 GRBs. Note: ’Non-Con.’ means that the posterior parameters of fits are not constrained.} 
    \label{Table1} \\
    \hline
    \hline
    \multicolumn{1}{c}{GRB} & 
    \multicolumn{1}{c}{redshift} & 
    \multicolumn{1}{c}{$\rm t_1\sim t_2$} & 
    \multicolumn{1}{c}{S/N} & 
    \multicolumn{1}{c}{$\quad$} & 
    \multicolumn{1}{c}{$\quad$} & 
    \multicolumn{1}{c}{BIC} & 
    \multicolumn{1}{c}{$\quad$} & 
    \multicolumn{1}{c}{$\quad$} & 
    \multicolumn{1}{c}{$\quad$} & 
    \multicolumn{1}{c}{Best} \\
    \multicolumn{1}{c}{$\quad$} & 
    \multicolumn{1}{c}{$\quad$} & 
    \multicolumn{1}{c}{(s)} & 
    \multicolumn{1}{c}{$\quad$} & 
    \multicolumn{1}{c}{CPL} &  
    \multicolumn{1}{c}{Band} &
    \multicolumn{1}{c}{CPL+BB} &
    \multicolumn{1}{c}{Band+BB} &
    \multicolumn{1}{c}{BB} &
    \multicolumn{1}{c}{NDP} &
    \multicolumn{1}{c}{Model} \\
    \hline
    \endfirsthead
    \multicolumn{11}{c}{{\bfseries \tablename\ \thetable{} -- continued from previous page}} \\
    \hline
   \multicolumn{1}{c}{GRB} & 
    \multicolumn{1}{c}{redshift} & 
    \multicolumn{1}{c}{$\rm t_1\sim t_2$} & 
    \multicolumn{1}{c}{S/N} & 
    \multicolumn{1}{c}{$\quad$} & 
    \multicolumn{1}{c}{$\quad$} & 
    \multicolumn{1}{c}{BIC} & 
    \multicolumn{1}{c}{$\quad$} & 
    \multicolumn{1}{c}{$\quad$} & 
    \multicolumn{1}{c}{$\quad$} & 
    \multicolumn{1}{c}{Best} \\
    \multicolumn{1}{c}{$\quad$} & 
    \multicolumn{1}{c}{$\quad$} & 
    \multicolumn{1}{c}{(s)} & 
    \multicolumn{1}{c}{$\quad$} & 
    \multicolumn{1}{c}{CPL} &  
    \multicolumn{1}{c}{Band} &
    \multicolumn{1}{c}{CPL+BB} &
    \multicolumn{1}{c}{Band+BB} &
    \multicolumn{1}{c}{BB} &
    \multicolumn{1}{c}{NDP} &
    \multicolumn{1}{c}{Model} \\
    \hline
    \endhead
    \hline
    \endfoot
    \hline
    \endlastfoot
080916A	&	0.689	&	0	$\sim$	40	&	25.5 	&	5137 	&	5148 	&	Non-Con.	&	Non-Con.	&	5448 	&	5161 	&	CPL	\\
080916C	&	4.35	&	0	$\sim$	70	&	94.5 	&	6267 	&	6264 	&	Non-Con.	&	Non-Con.	&	9570 	&	7321 	&	Band	\\
081222	&	2.77	&	0	$\sim$	20	&	56.4 	&	6023 	&	6030 	&	Non-Con.	&	Non-Con.	&	7109 	&	6203 	&	CPL	\\
090424	&	0.544	&	0	$\sim$	2	&	182.9 	&	4015 	&	4023 	&	4008 	&	Non-Con.	&	13064 	&	5401 	&	CPL+BB	\\
$\quad$	&	0.544	&	2	$\sim$	3.9	&	127.9 	&	3816 	&	3819 	&	3792 	&	Non-Con.	&	8253 	&	4427 	&	CPL+BB	\\
$\quad$	&	0.544	&	3.9	$\sim$	5.5	&	145.5 	&	3684 	&	3683 	&	3656 	&	3666 	&	8751 	&	4256 	&	CPL+BB	\\
090618 	&	0.54	&	-6	$\sim$	50	&	45.6 	&	3911 	&	3917 	&	Non-Con.	&	Non-Con.	&	4383 	&	3976 	&	CPL	\\
$\quad$	&	0.54	&	50	$\sim$	75	&	191.3 	&	3668 	&	3647 	&	Non-Con.	&	Non-Con.	&	14499 	&	6708 	&	Band	\\
\end{longtable}


\clearpage
\onecolumn
\begin{longtable}{p{2.5cm}<{\centering}p{1.5cm}<{\centering}p{0.5cm}<{\centering}p{0.5cm}<{\centering}p{1.5cm}<{\centering}p{1.5cm}<{\centering}p{1cm}<{\centering}p{0.5cm}<{\centering}p{0.5cm}<{\centering}}
    \caption{The parameters of spectral fitting and derived the value of $\sigma$ with no-thermal spectral.} 
    \label{Table2} \\
    \hline
    \hline
    \multicolumn{1}{c}{GRB} & 
    \multicolumn{1}{c}{$\rm t_1\sim t_2$} & 
    \multicolumn{1}{c}{S/N} & 
    \multicolumn{1}{c}{Model} & 
    \multicolumn{1}{c}{$\alpha$} & 
    \multicolumn{1}{c}{$E_{p}\backslash E_{c}$} & 
    \multicolumn{1}{c}{$\beta$} & 
    \multicolumn{1}{c}{$\sigma$} & 
    \multicolumn{1}{c}{$kT_{max}$} \\
    \multicolumn{1}{c}{$\quad$} & 
    \multicolumn{1}{c}{(s)} & 
    \multicolumn{1}{c}{$\quad$} & 
    \multicolumn{1}{c}{$\quad$} & 
    \multicolumn{1}{c}{$\quad$} & 
    \multicolumn{1}{c}{(keV)} & 
    \multicolumn{1}{c}{$\quad$} & 
    \multicolumn{1}{c}{$\quad$} & 
    \multicolumn{1}{c}{(keV)} 
   \\
    \hline
    \endfirsthead
    \multicolumn{9}{c}{{\bfseries \tablename\ \thetable{} -- continued from previous page}} \\
    \hline
     \multicolumn{1}{c}{GRB} & 
    \multicolumn{1}{c}{$\rm t_1\sim t_2$} & 
    \multicolumn{1}{c}{S/N} & 
    \multicolumn{1}{c}{Model} & 
    \multicolumn{1}{c}{$\alpha$} & 
    \multicolumn{1}{c}{$E_{p}\backslash E_{c}$} & 
    \multicolumn{1}{c}{$\beta$} & 
    \multicolumn{1}{c}{$\sigma$} & 
    \multicolumn{1}{c}{$kT_{max}$} \\
    \multicolumn{1}{c}{$\quad$} & 
    \multicolumn{1}{c}{(s)} & 
    \multicolumn{1}{c}{$\quad$} & 
    \multicolumn{1}{c}{$\quad$} & 
    \multicolumn{1}{c}{$\quad$} & 
    \multicolumn{1}{c}{(keV)} & 
    \multicolumn{1}{c}{$\quad$} & 
    \multicolumn{1}{c}{$\quad$} & 
    \multicolumn{1}{c}{(keV)} \\
    \hline
    \endhead
    \hline
    \endfoot
    \hline
    \endlastfoot
$2.8kT_{max}<E_{p}$	&				&		&		&						&						&						&		&		\\
\hline																															
080916A	&	0	$\sim$	40	&	25.5 	&	CPL	&$	-0.93 	_{	-0.08 	}^{+	0.08 	}$&$	107.1 	_{	-14.9 	}^{+	15.2 	}$&						&	2.06 	&	13.77 	\\
080916C	&	0	$\sim$	70	&	94.5 	&	Band	&$	-0.97 	_{	-0.02 	}^{+	0.02 	}$&$	456.9 	_{	-29.4 	}^{+	30.8 	}$&$	-2.42 	_{	-0.23 	}^{+	0.15 	}$&	10.35 	&	23.25 	\\
081222	&	0	$\sim$	20	&	56.4 	&	CPL	&$	-0.90 	_{	-0.08 	}^{+	0.07 	}$&$	166.2 	_{	-13.3 	}^{+	13.3 	}$&						&	2.61 	&	19.23 	\\
090618 	&	-6	$\sim$	50	&	45.6 	&	CPL	&$	-0.86 	_{	-0.06 	}^{+	0.06 	}$&$	168.8 	_{	-15.6 	}^{+	15.6 	}$&						&	3.16 	&	18.02 	\\
$\quad$	&	50	$\sim$	75	&	191.3 	&	Band	&$	-1.03 	_{	-0.08 	}^{+	0.07 	}$&$	262.8 	_{	-6.6 	}^{+	6.7 	}$&$	-2.72 	_{	-0.09 	}^{+	0.09 	}$&	2.94 	&	31.83 	\\
$\quad$	&	75	$\sim$	103	&	143.4 	&	CPL	&$	-1.16 	_{	-0.08 	}^{+	0.07 	}$&$	174.9 	_{	-6.6 	}^{+	6.5 	}$&						&	1.44 	&	25.60 	\\
$\quad$	&	103	$\sim$	116	&	71.5 	&	CPL	&$	-1.29 	_{	-0.06 	}^{+	0.05 	}$&$	100.1 	_{	-10.2 	}^{+	10.3 	}$&						&	1.13 	&	20.38 	\\
\end{longtable}


\clearpage
\onecolumn
\begin{longtable}{p{1cm}<{\centering}p{1.5cm}<{\centering}p{1cm}<{\centering}p{1.5cm}<{\centering}p{1.5cm}<{\centering}p{1.5cm}<{\centering}p{1.5cm}<{\centering}p{1.5cm}<{\centering}p{1.5cm}<{\centering}p{1.5cm}<{\centering}}
    \caption{The light curves fitting results with FRED model and derived parameters of 15 GRBs.} 
    \label{Table3} \\
    \hline
    \hline
    \multicolumn{1}{c}{GRB} & 
    \multicolumn{1}{c}{$\rm t_1\sim t_2$} & 
    \multicolumn{1}{c}{Channel} & 
    \multicolumn{1}{c}{$\tau_{1}$} & 
    \multicolumn{1}{c}{$\tau_{2}$} & 
    \multicolumn{1}{c}{$\rm t_{s}$} & 
    \multicolumn{1}{c}{$w$} & 
    \multicolumn{1}{c}{$\rm t_{p}$}&
    \multicolumn{1}{c}{$\rm t_{r}$}&
    \multicolumn{1}{c}{$\rm t_{d}$}
    \\
    \multicolumn{1}{c}{$\quad$} & 
    \multicolumn{1}{c}{(s)} & 
    \multicolumn{1}{c}{(keV)} & 
    \multicolumn{1}{c}{(s)} & 
    \multicolumn{1}{c}{(s)} & 
    \multicolumn{1}{c}{(s)} & 
    \multicolumn{1}{c}{(s)} & 
    \multicolumn{1}{c}{(s)} & 
    \multicolumn{1}{c}{(s)} & 
    \multicolumn{1}{c}{(s)} 
   \\
    \hline
    \endfirsthead
    \multicolumn{10}{c}{{\bfseries \tablename\ \thetable{} -- continued from previous page}} \\
    \hline
    \multicolumn{1}{c}{GRB} & 
    \multicolumn{1}{c}{$\rm t_1\sim t_2$} & 
    \multicolumn{1}{c}{Channel} & 
    \multicolumn{1}{c}{$\tau_{1}$} & 
    \multicolumn{1}{c}{$\tau_{2}$} & 
    \multicolumn{1}{c}{$\rm t_{s}$} & 
    \multicolumn{1}{c}{$w$} & 
    \multicolumn{1}{c}{$\rm t_{p}$}&
    \multicolumn{1}{c}{$\rm t_{r}$}&
    \multicolumn{1}{c}{$\rm t_{d}$}
    \\
    \multicolumn{1}{c}{$\quad$} & 
    \multicolumn{1}{c}{(s)} & 
    \multicolumn{1}{c}{(keV)} & 
    \multicolumn{1}{c}{(s)} & 
    \multicolumn{1}{c}{(s)} & 
    \multicolumn{1}{c}{(s)} & 
    \multicolumn{1}{c}{(s)} & 
    \multicolumn{1}{c}{(s)} & 
    \multicolumn{1}{c}{(s)} & 
    \multicolumn{1}{c}{(s)} \\
    \hline
    \endhead
    \hline
    \endfoot
    \hline
    \endlastfoot
080916C	&	-1 	$\sim$	100 	&	8$\sim$30		&$	1.2 	^{+	0.4 	}_{-	0.2 	}$&$	33.5 	^{+	1.8 	}_{-	1.4 	}$&$	-1.1 	^{+	0.1 	}_{-	0.5 	}$&$	44.4 	^{+	2.6 	}_{-	1.9 	}$&$	5.2 	^{+	1.0 	}_{-	0.8 	}$&$	5.5 	^{+	0.7 	}_{-	0.5 	}$&$	38.9 	^{+	2.1 	}_{-	1.6 	}$	\\
$\quad$	&	-1 	$\sim$	100 	&	30$\sim$70		&$	0.7 	^{+	0.1 	}_{-	0.2 	}$&$	30.3 	^{+	1.0 	}_{-	1.0 	}$&$	-1.0 	^{+	0.1 	}_{-	0.1 	}$&$	38.7 	^{+	1.3 	}_{-	1.2 	}$&$	3.8 	^{+	0.3 	}_{-	0.3 	}$&$	4.2 	^{+	0.3 	}_{-	0.2 	}$&$	34.5 	^{+	1.1 	}_{-	1.1 	}$	\\
$\quad$	&	-1 	$\sim$	100 	&	70$\sim$100		&$	0.6 	^{+	0.1 	}_{-	0.2 	}$&$	27.3 	^{+	1.0 	}_{-	1.1 	}$&$	-0.9 	^{+	0.1 	}_{-	0.2 	}$&$	34.5 	^{+	1.3 	}_{-	1.4 	}$&$	3.2 	^{+	0.4 	}_{-	0.3 	}$&$	3.6 	^{+	0.3 	}_{-	0.3 	}$&$	30.9 	^{+	1.2 	}_{-	1.2 	}$	\\
$\quad$	&	-1 	$\sim$	100 	&	100$\sim$150		&$	0.8 	^{+	0.2 	}_{-	0.2 	}$&$	24.2 	^{+	0.8 	}_{-	0.8 	}$&$	-1.0 	^{+	0.1 	}_{-	0.1 	}$&$	31.6 	^{+	1.0 	}_{-	1.0 	}$&$	3.3 	^{+	0.3 	}_{-	0.2 	}$&$	3.7 	^{+	0.2 	}_{-	0.2 	}$&$	27.9 	^{+	0.9 	}_{-	0.9 	}$	\\
$\quad$	&	-1 	$\sim$	100 	&	150$\sim$200		&$	0.9 	^{+	0.2 	}_{-	0.2 	}$&$	21.3 	^{+	1.2 	}_{-	1.2 	}$&$	-1.0 	^{+	0.1 	}_{-	0.2 	}$&$	28.9 	^{+	1.6 	}_{-	1.6 	}$&$	3.4 	^{+	0.5 	}_{-	0.4 	}$&$	3.8 	^{+	0.4 	}_{-	0.3 	}$&$	25.1 	^{+	1.4 	}_{-	1.4 	}$	\\
$\quad$	&	-1 	$\sim$	100 	&	200$\sim$300		&$	0.8 	^{+	0.2 	}_{-	0.2 	}$&$	20.8 	^{+	1.7 	}_{-	1.7 	}$&$	-1.0 	^{+	0.1 	}_{-	0.1 	}$&$	27.7 	^{+	2.2 	}_{-	2.2 	}$&$	3.0 	^{+	0.6 	}_{-	0.5 	}$&$	3.4 	^{+	0.4 	}_{-	0.4 	}$&$	24.2 	^{+	1.9 	}_{-	1.9 	}$	\\
$\quad$	&	-1 	$\sim$	100 	&	300$\sim$500		&$	8.0 	^{+	8.5 	}_{-	2.8 	}$&$	2.6 	^{+	0.7 	}_{-	0.6 	}$&$	-1.5 	^{+	0.4 	}_{-	1.1 	}$&$	7.3 	^{+	2.3 	}_{-	1.5 	}$&$	3.1 	^{+	2.5 	}_{-	1.4 	}$&$	2.4 	^{+	1.0 	}_{-	0.5 	}$&$	4.9 	^{+	1.4 	}_{-	1.0 	}$	\\
$\quad$	&	-1 	$\sim$	100 	&	500$\sim$900		&	——					&	——					&	——					&	——					&	——					&	——					&	——						\\
\end{longtable}

\clearpage

\begin{table*}
\centering
\renewcommand\tabcolsep{5pt}
\renewcommand\arraystretch{1}
\caption{The value of power-law index $k$ and corresponding to Spearman correlation coefficients ($\rho_p$ for the point estimation and $\rho_{\rm s}$ for the error consideration) of 15 GRBs.}
\begin{tabular}{ccccc} 
\hline\hline
GRB&$\rm t_{1}\sim t_{2}$($s$)&$k$& $\rho_p$&$\rho_{\rm s}$\\

\hline
080916C	&	-1	$\sim$	100	&$	-0.19 	^{+	0.04 	}_{-	0.05 	}$&	-1.00 	&$	-0.99 	^{+	0.06 	}_{-	0.01 	}	$	\\
	&	-1	$\sim$	16	&$	-0.27 	^{+	0.07 	}_{-	0.09 	}$&	-1.00 	&$	-0.96 	^{+	0.08 	}_{-	0.04 	}	$	\\
100414A	&	-7	$\sim$	38	&$	-0.10 	^{+	0.09 	}_{-	0.10 	}$&	-0.86 	&$	-0.39 	^{+	0.39 	}_{-	0.32 	}	$	\\
120624B	&	0	$\sim$	28	&$	-0.35 	^{+	0.17 	}_{-	0.21 	}$&	-1.00 	&$	-0.50 	^{+	0.39 	}_{-	0.28 	}	$	\\
	&	10	$\sim$	28	&$	-0.32 	^{+	0.08 	}_{-	0.11 	}$&	-0.96 	&$	-0.96 	^{+	0.07 	}_{-	0.04 	}	$	\\
120711A	&	60	$\sim$	80	&$	-0.16 	^{+	0.09 	}_{-	0.11 	}$&	-0.96 	&$	-0.54 	^{+	0.39 	}_{-	0.29 	}	$	\\
	&	80	$\sim$	114	&$	-0.19 	^{+	0.09 	}_{-	0.10 	}$&	-1.00 	&$	-0.71 	^{+	0.39 	}_{-	0.18 	}	$	\\
\hline
\end{tabular}
\label{Table4}
\end{table*}


\appendix

\section{The fitting results of other six GRBs (Figure A1-A6) in group (I), five GRBs (Figure A7-A11) in group (II), and GRB 200524A (Figure A12) in group (III)}\label{App_a}
\begin{figure} 
    \centering %
    \begin{minipage}{0.45\textwidth} %
        \centering %
        \includegraphics[width=\textwidth]{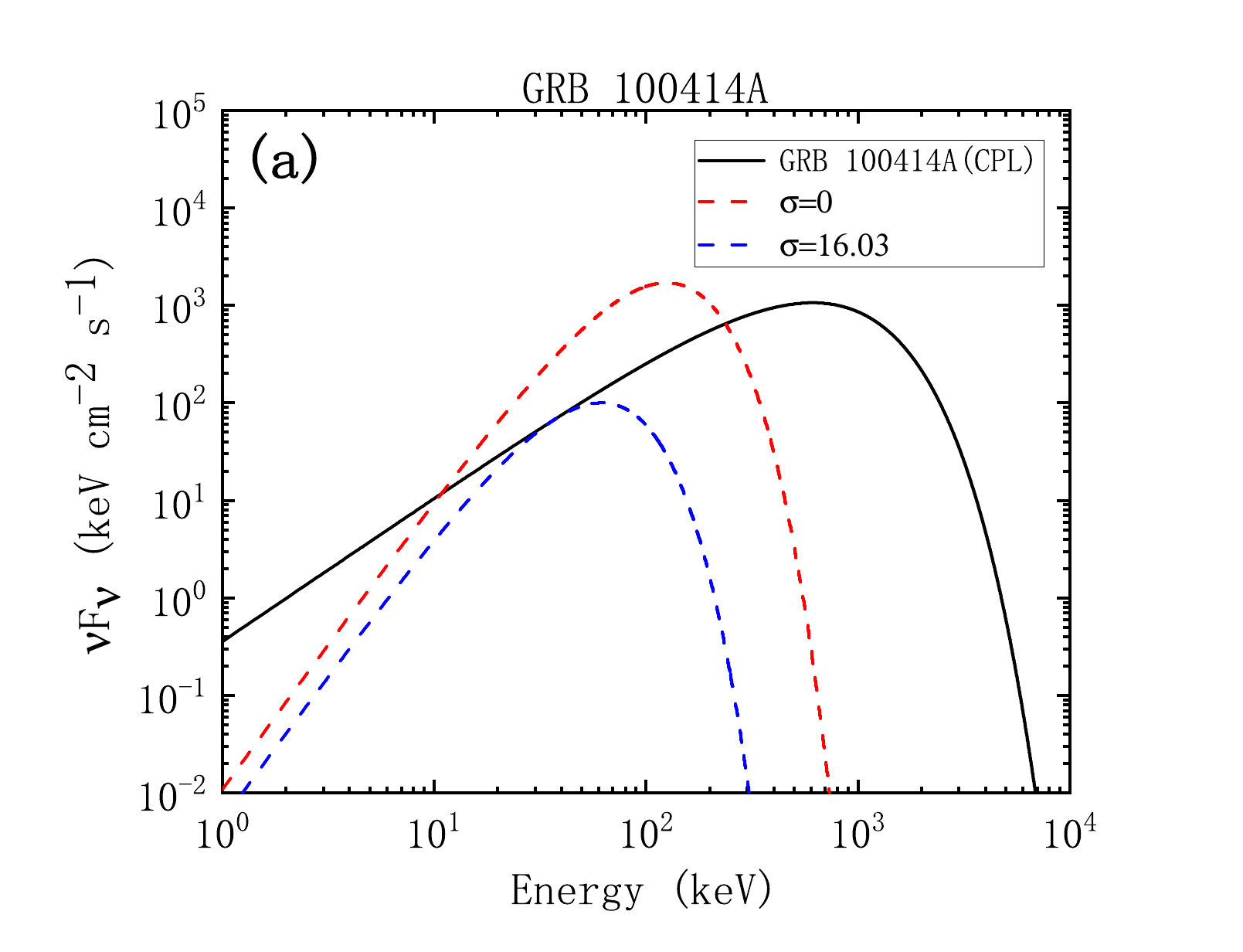} %
        \vskip\baselineskip %
        \includegraphics[width=\textwidth]{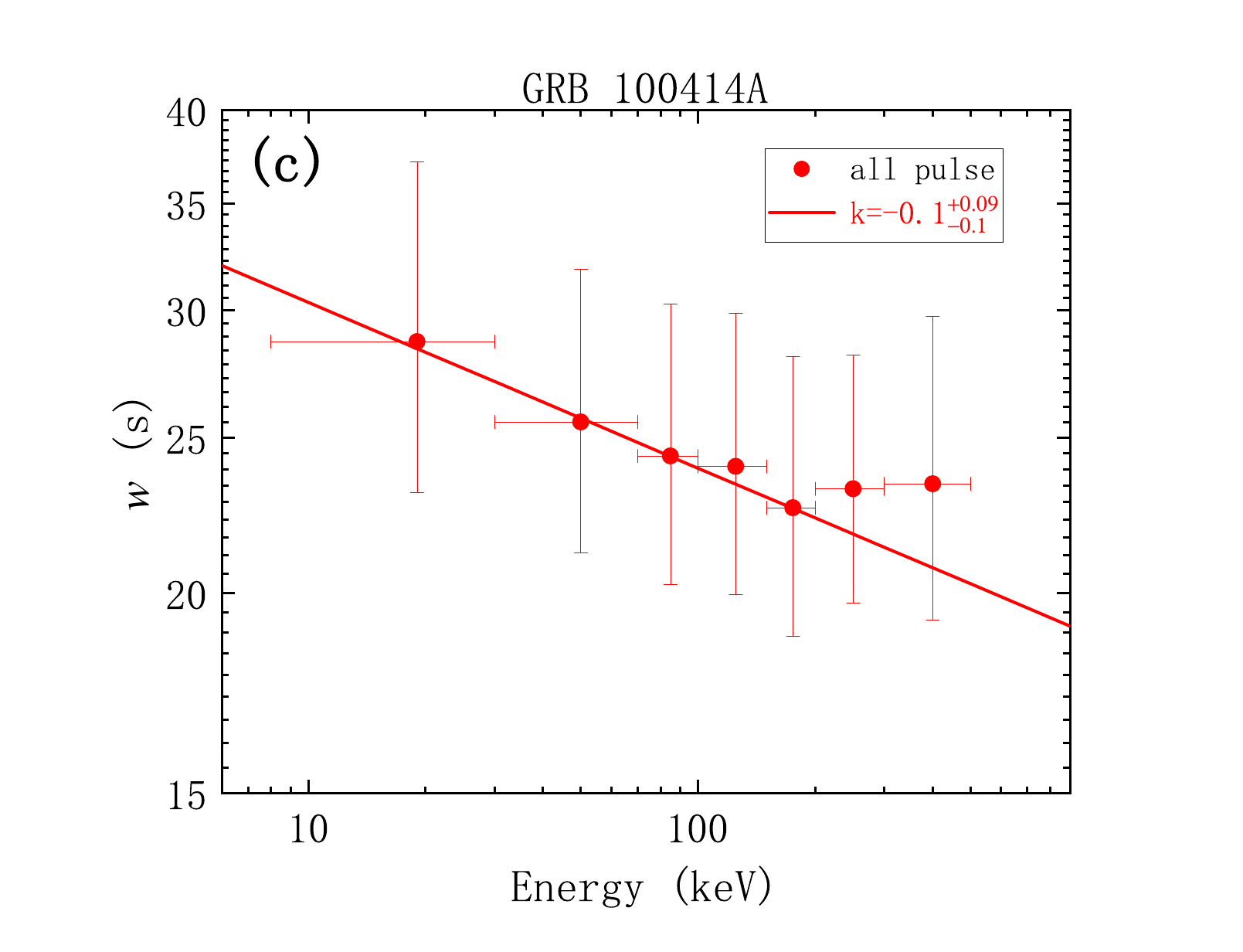} %
    \end{minipage}
    \hfill %
    \begin{minipage}{0.45\textwidth} %
        \centering %
        \includegraphics[width=\textwidth]{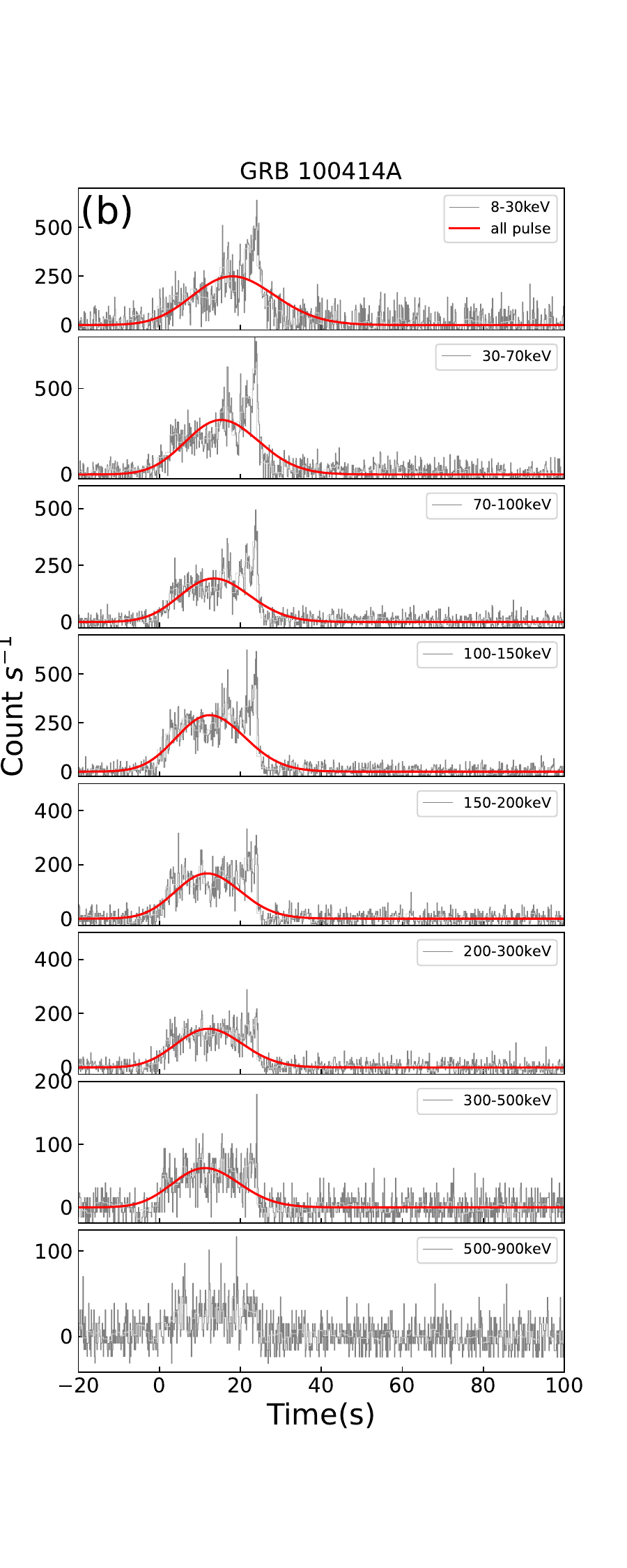} %
    \end{minipage}
    \caption{Spectral and temporal analysis of GRB 100414A for group (I). (a‌). The predicted lower limits of the photosphere spectra (dashed
    lines) with $R_{0}=10^{10}$ cm and observed non-thermal spectrum (solid line). (b). The light curves of the prompt emission of GRB 100414A (gray) in the different energy ranges and the FRED model fitting (red solid lines). (c). The pulse width ($w$) is derived from FRED model fitting as a function of energy and the power-law fitting (solid red line).}  
    \label{fig:A1}
\end{figure}

\begin{figure} 
    \centering %
    \begin{minipage}{0.45\textwidth} %
        \centering %
        \includegraphics[width=\textwidth]{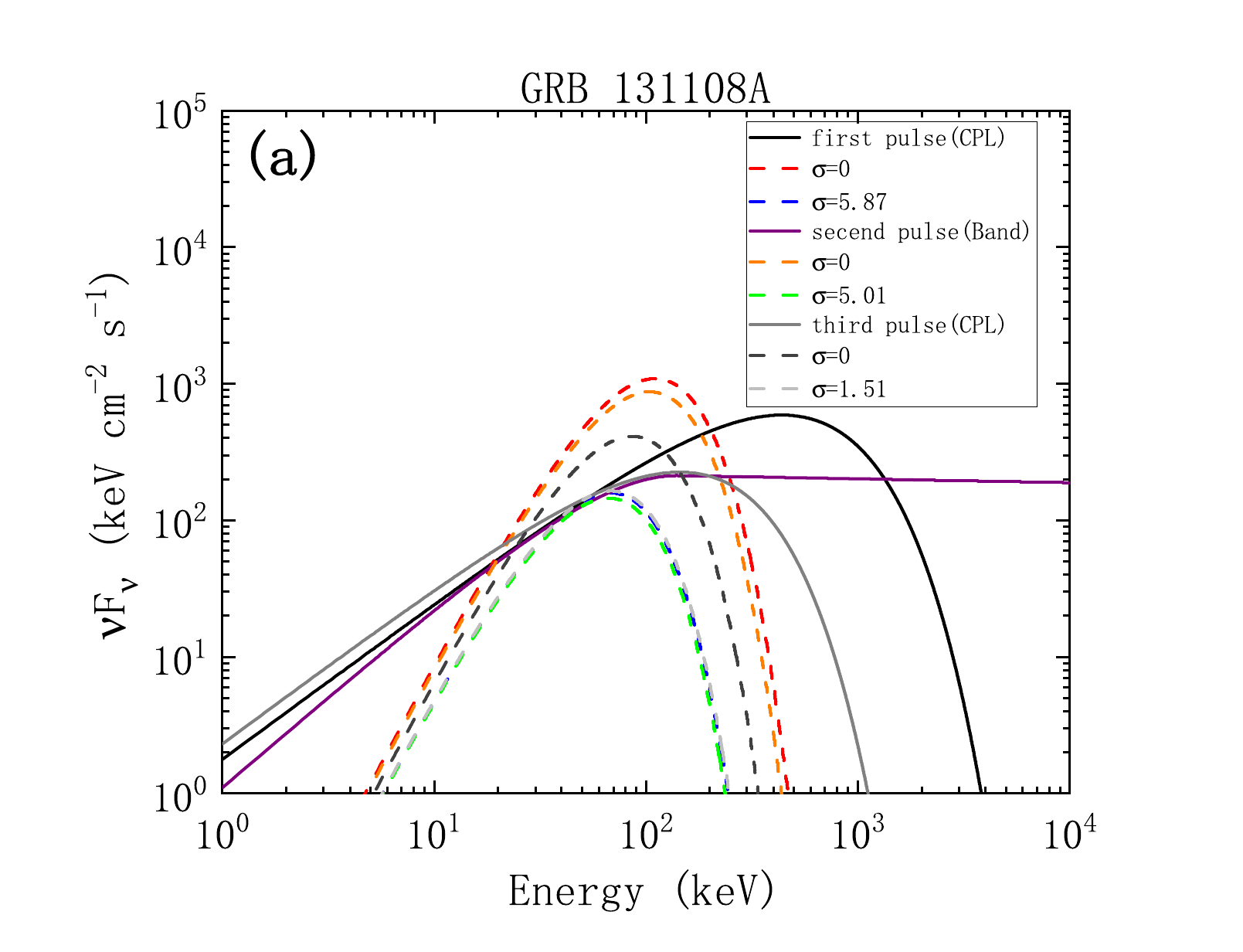} %
        \vskip\baselineskip %
        \includegraphics[width=\textwidth]{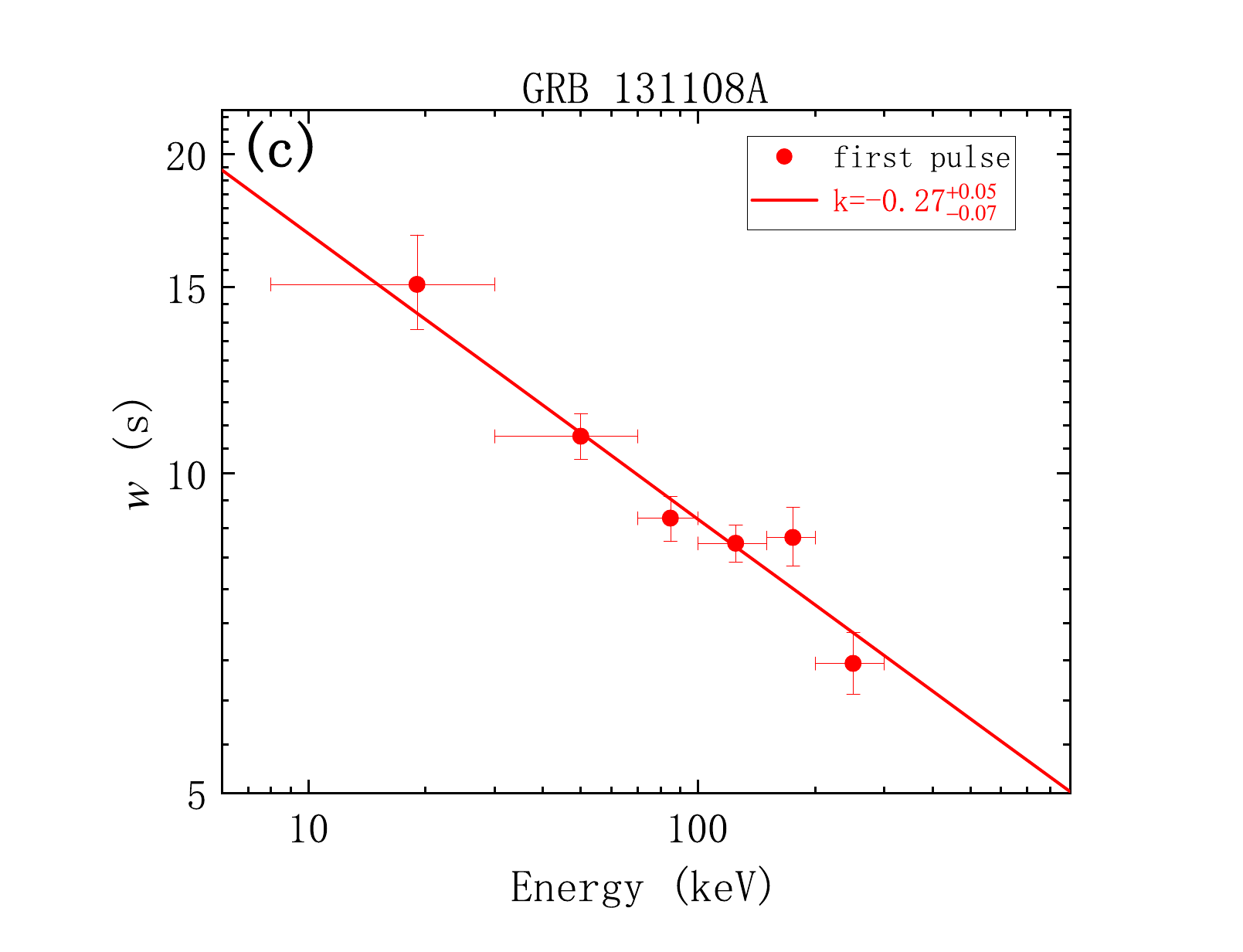} %
    \end{minipage}
    \hfill %
    \begin{minipage}{0.45\textwidth} %
        \centering %
        \includegraphics[width=\textwidth]{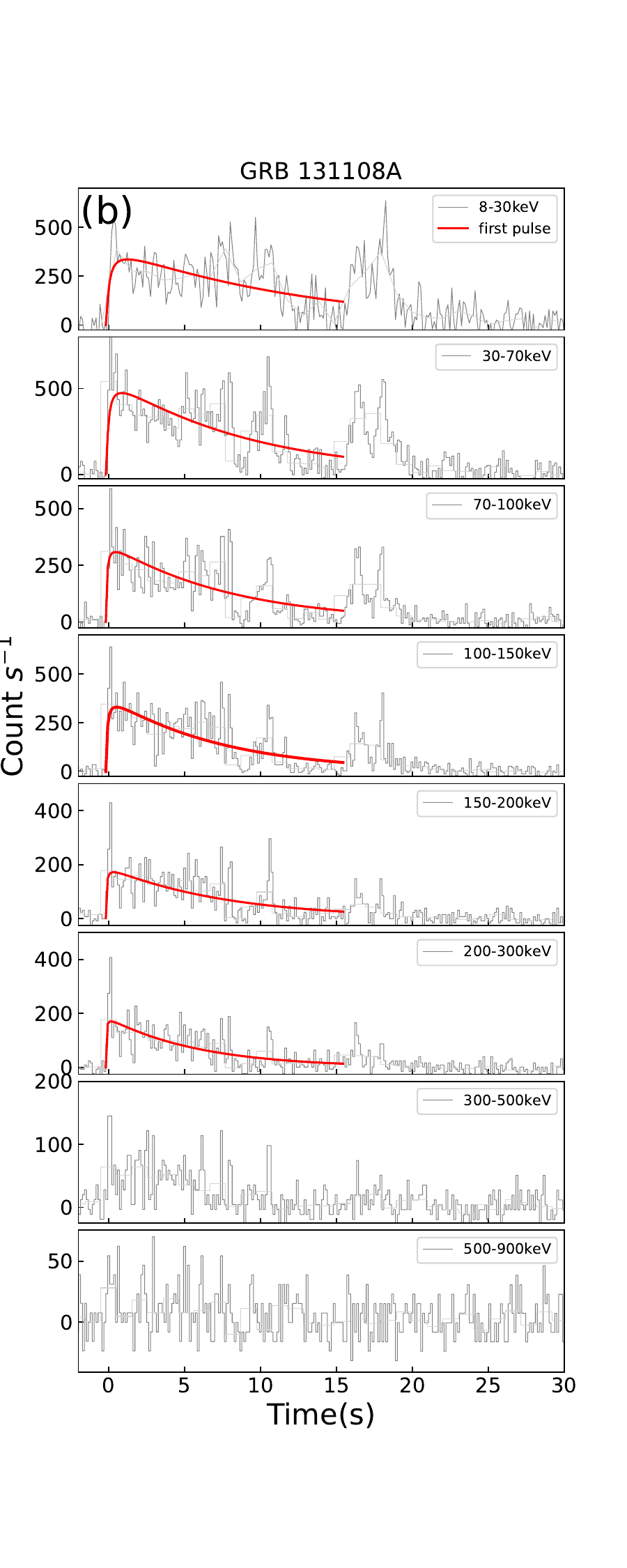} %
    \end{minipage}
    \caption{‌Similar to Figure \ref{fig:A1}, but for GRB 131108A in group (I).} %
     \label{fig:A2}
\end{figure}

\begin{figure} 
    \centering %
    \begin{minipage}{0.45\textwidth} %
        \centering %
        \includegraphics[width=\textwidth]{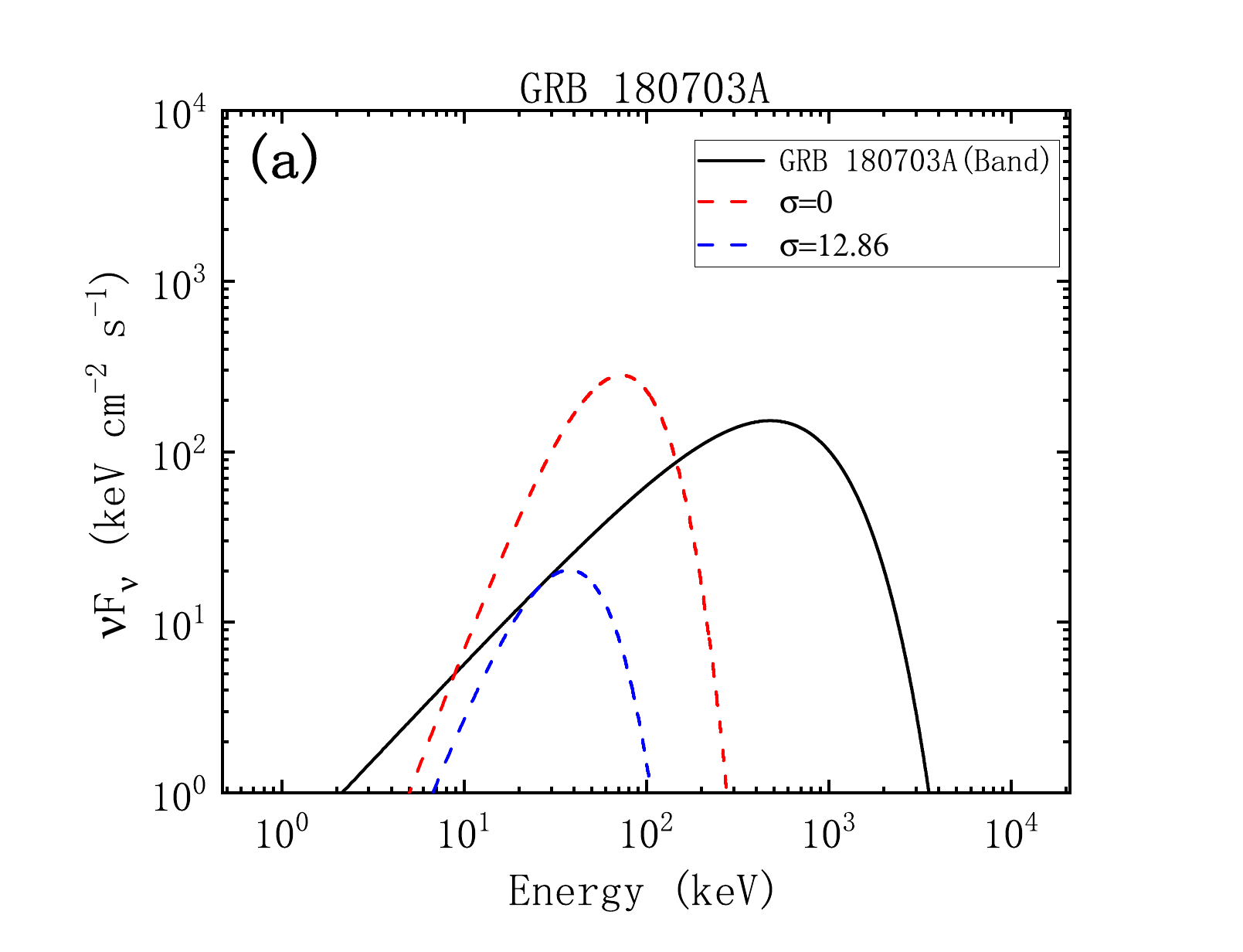} %
        \vskip\baselineskip %
        \includegraphics[width=\textwidth]{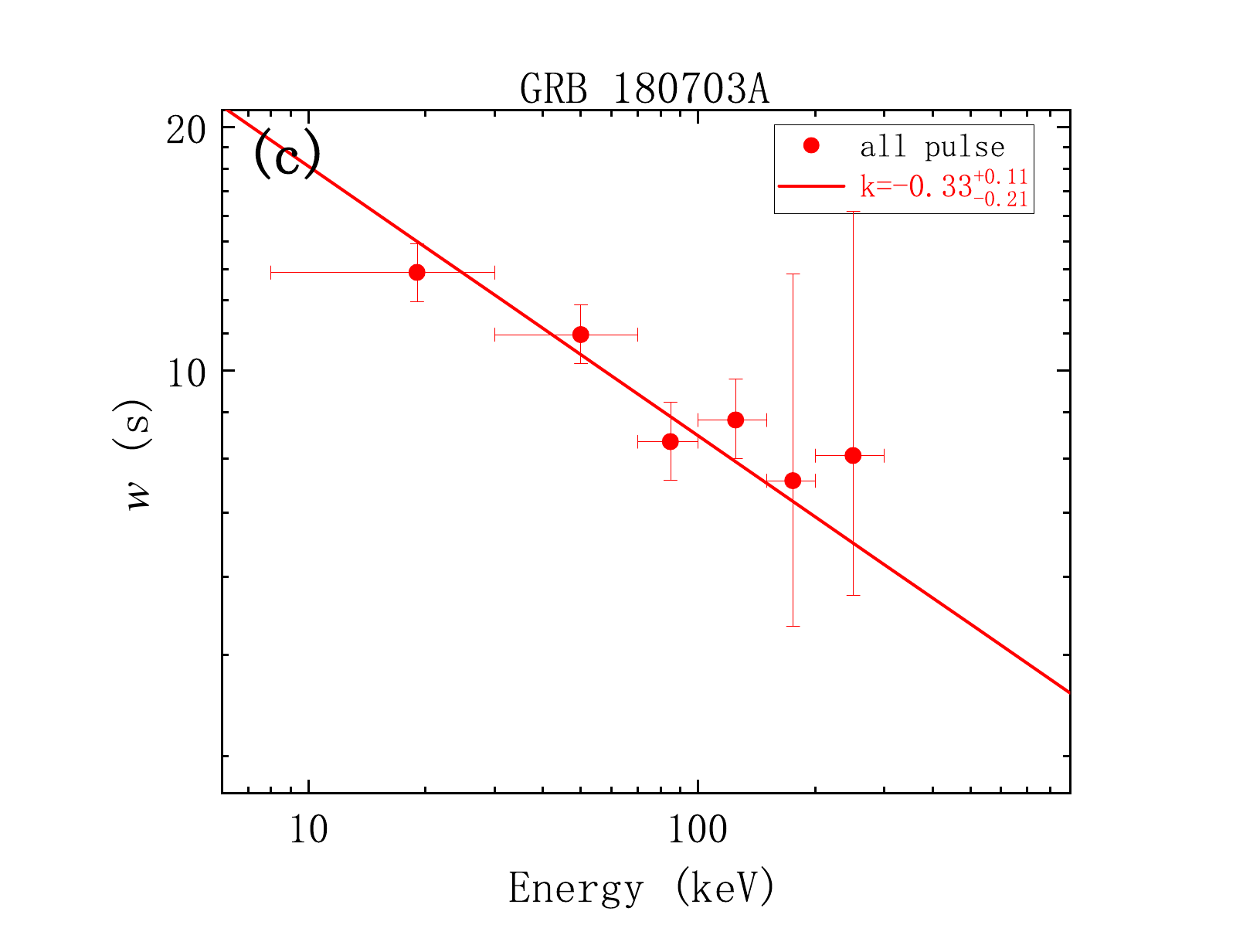} %
    \end{minipage}
    \hfill %
    \begin{minipage}{0.45\textwidth} %
        \centering %
        \includegraphics[width=\textwidth]{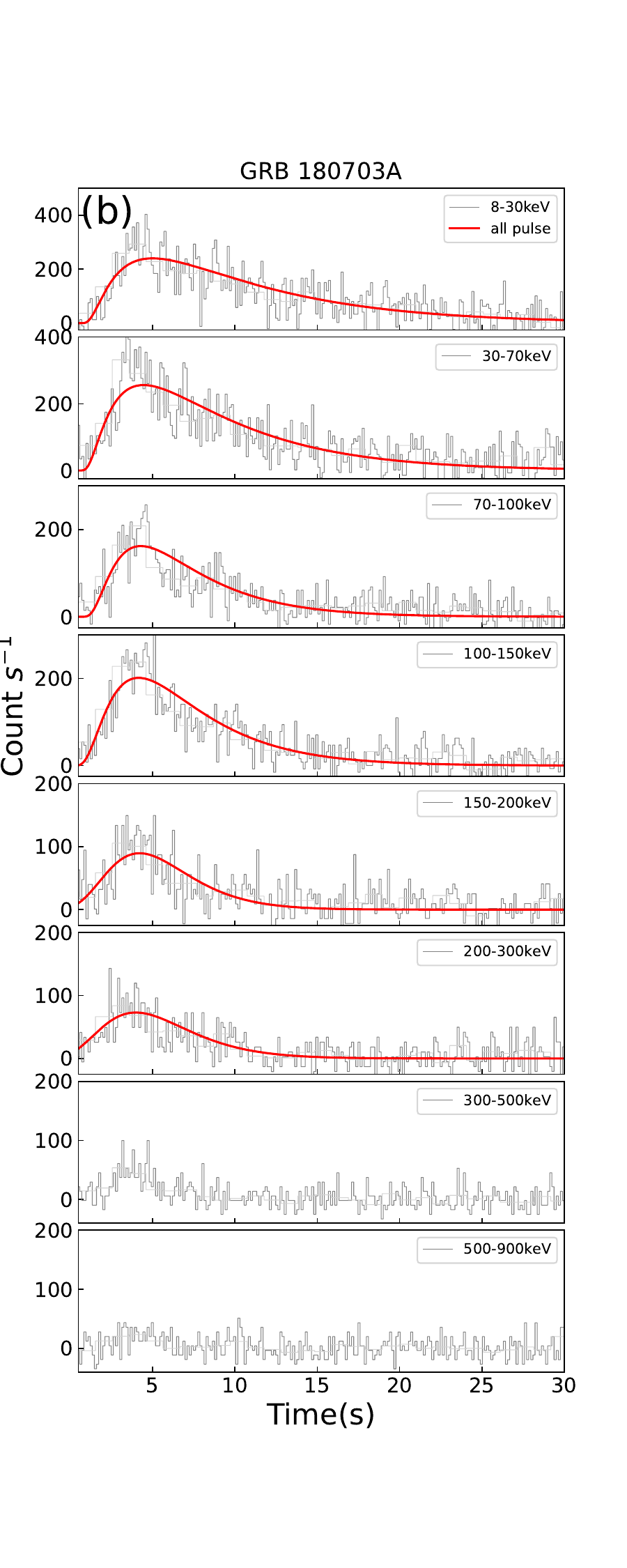} %
    \end{minipage}
    \caption{‌Similar to Figure \ref{fig:A1}, but for GRB 180703A in group (I).} %
     \label{fig:A3}
\end{figure}

\begin{figure} 
    \centering %
    \begin{minipage}{0.45\textwidth} %
        \centering %
        \includegraphics[width=\textwidth]{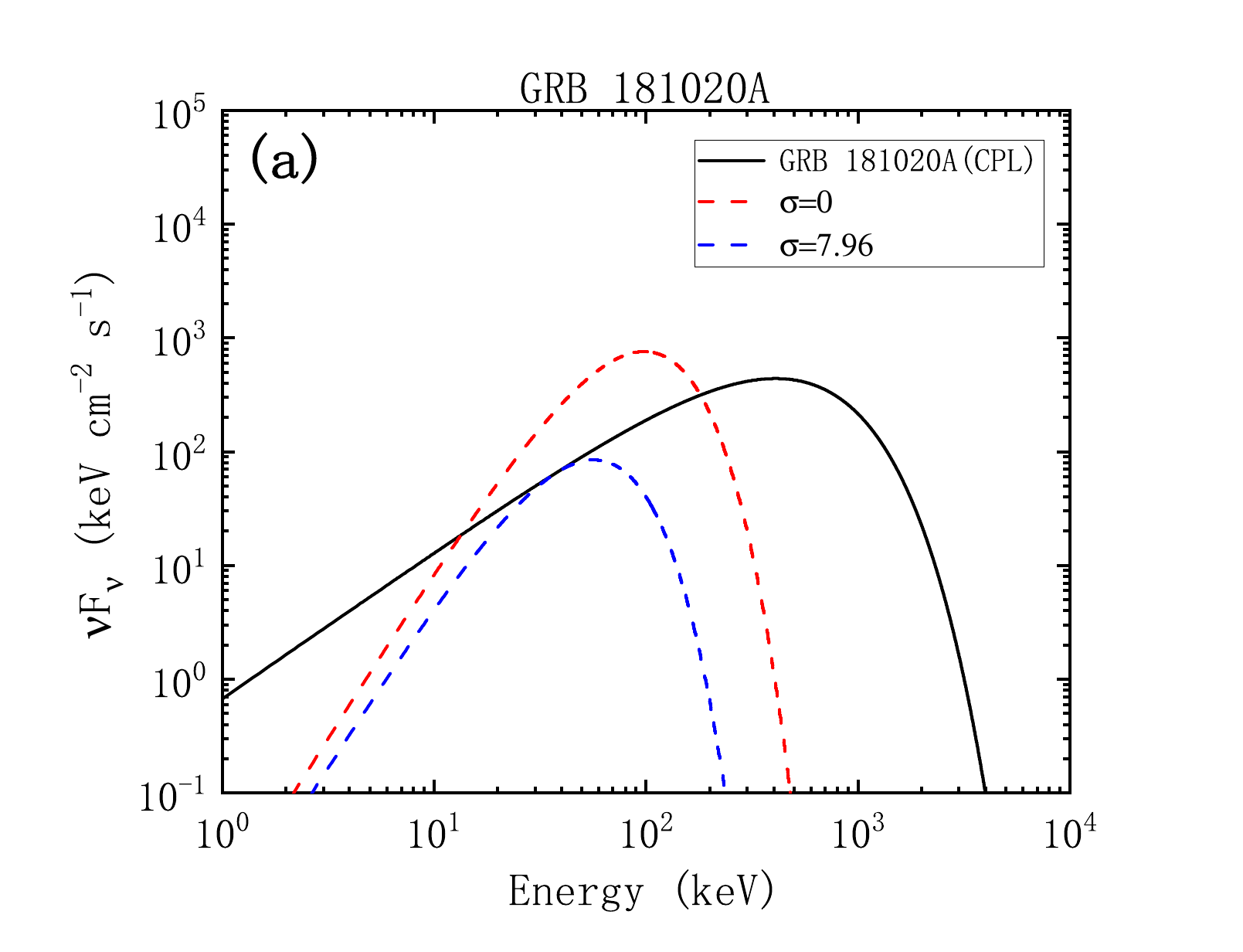} %
        \vskip\baselineskip %
        \includegraphics[width=\textwidth]{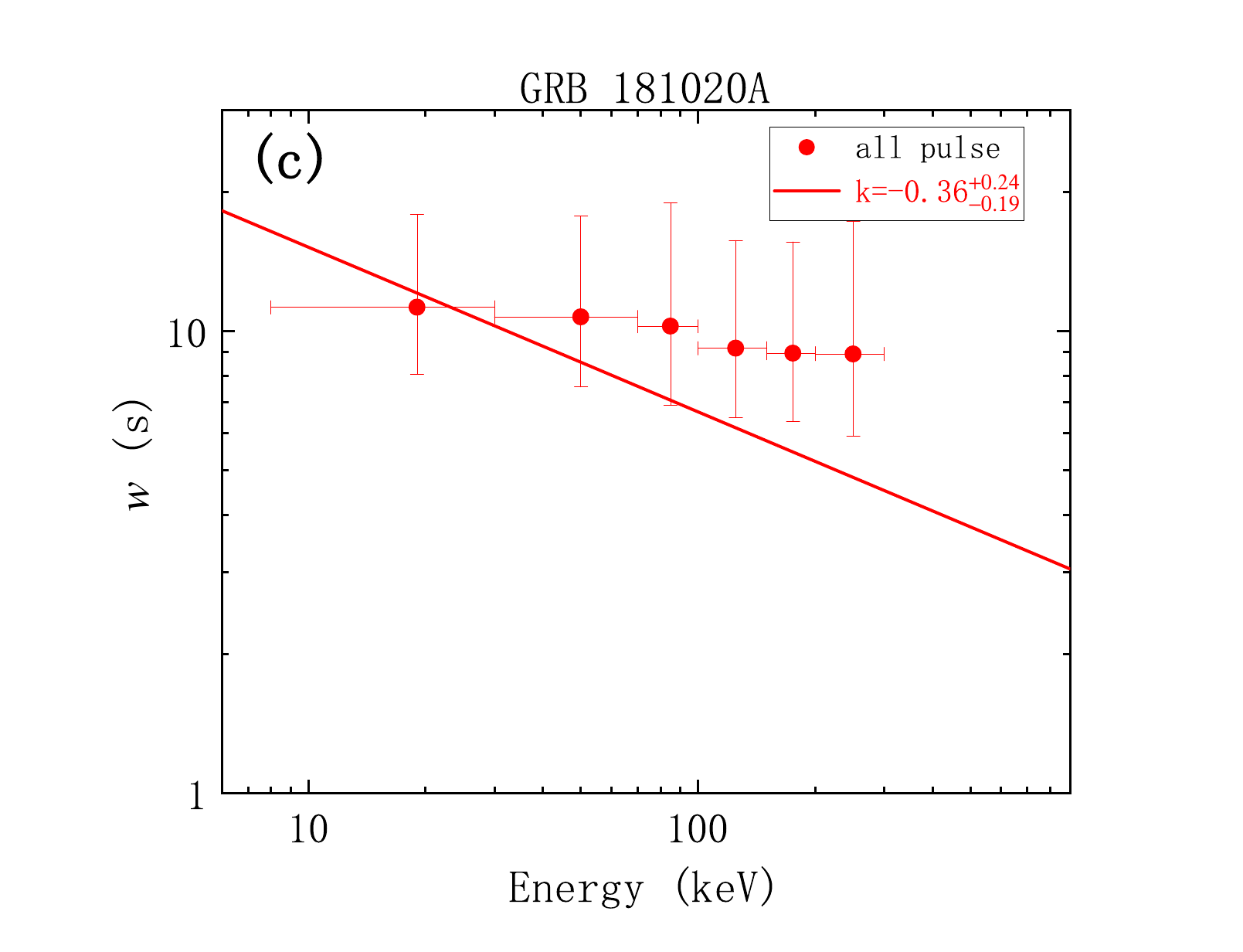} %
    \end{minipage}
    \hfill %
    \begin{minipage}{0.45\textwidth} %
        \centering %
        \includegraphics[width=\textwidth]{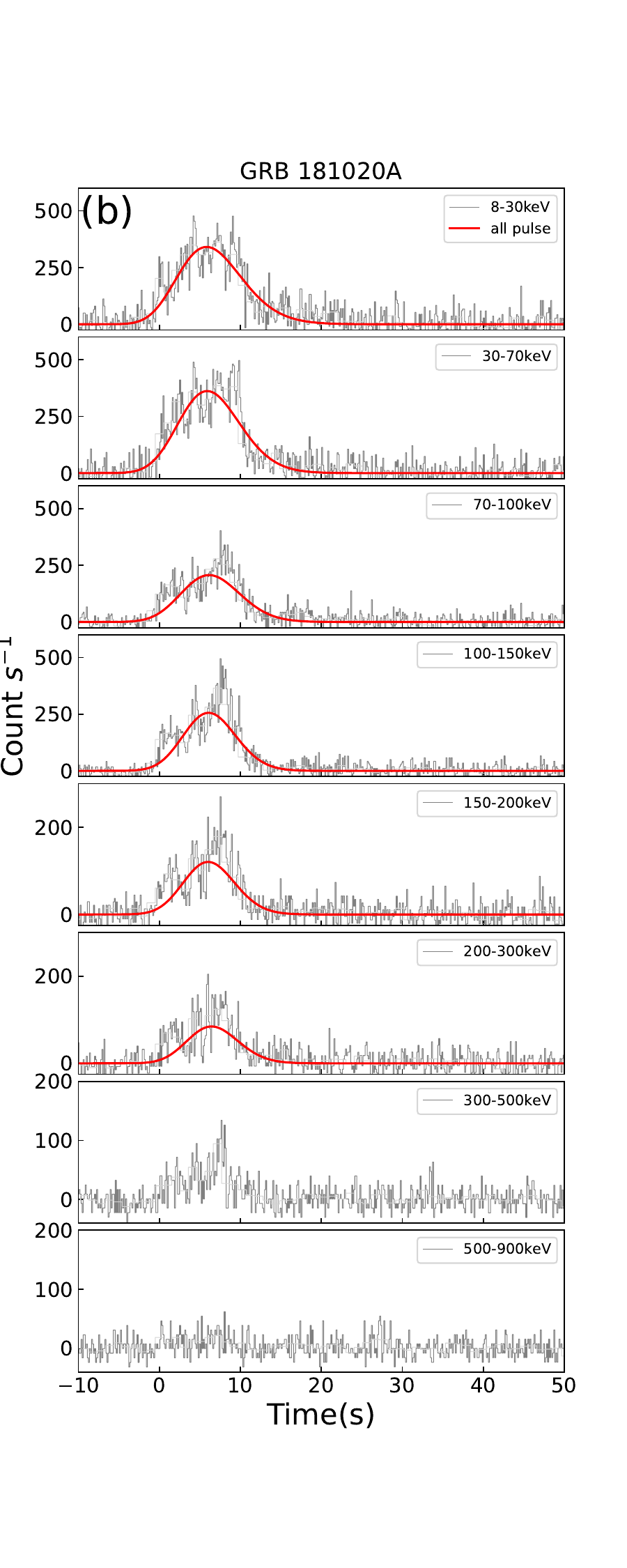} %
    \end{minipage}
    \caption{Similar to Figure \ref{fig:A1}, but for GRB 181020A in group (I).‌} %
     \label{fig:A4}
\end{figure}

\begin{figure} 
    \centering %
    \begin{minipage}{0.45\textwidth} %
        \centering %
        \includegraphics[width=\textwidth]{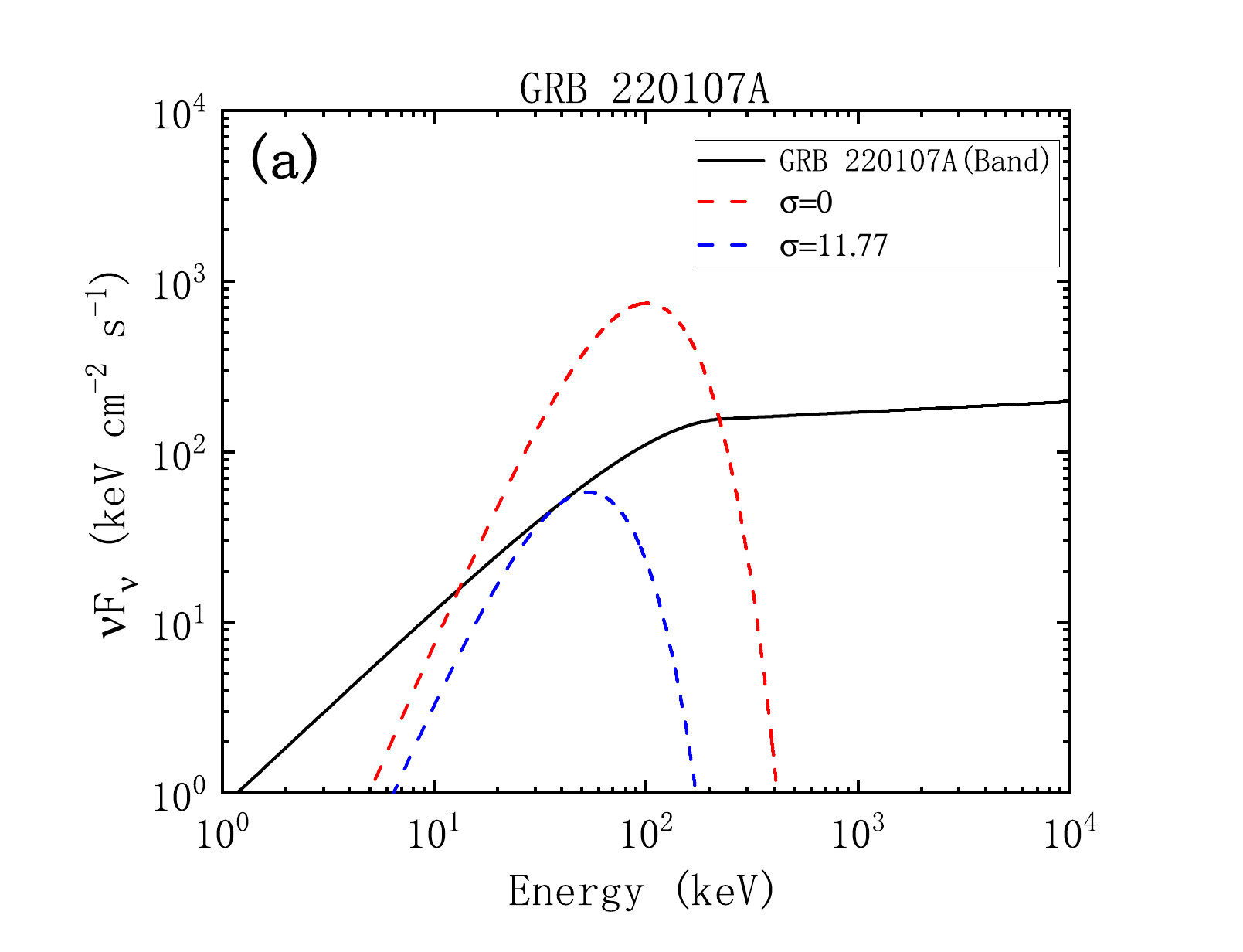} %
        \vskip\baselineskip %
        \includegraphics[width=\textwidth]{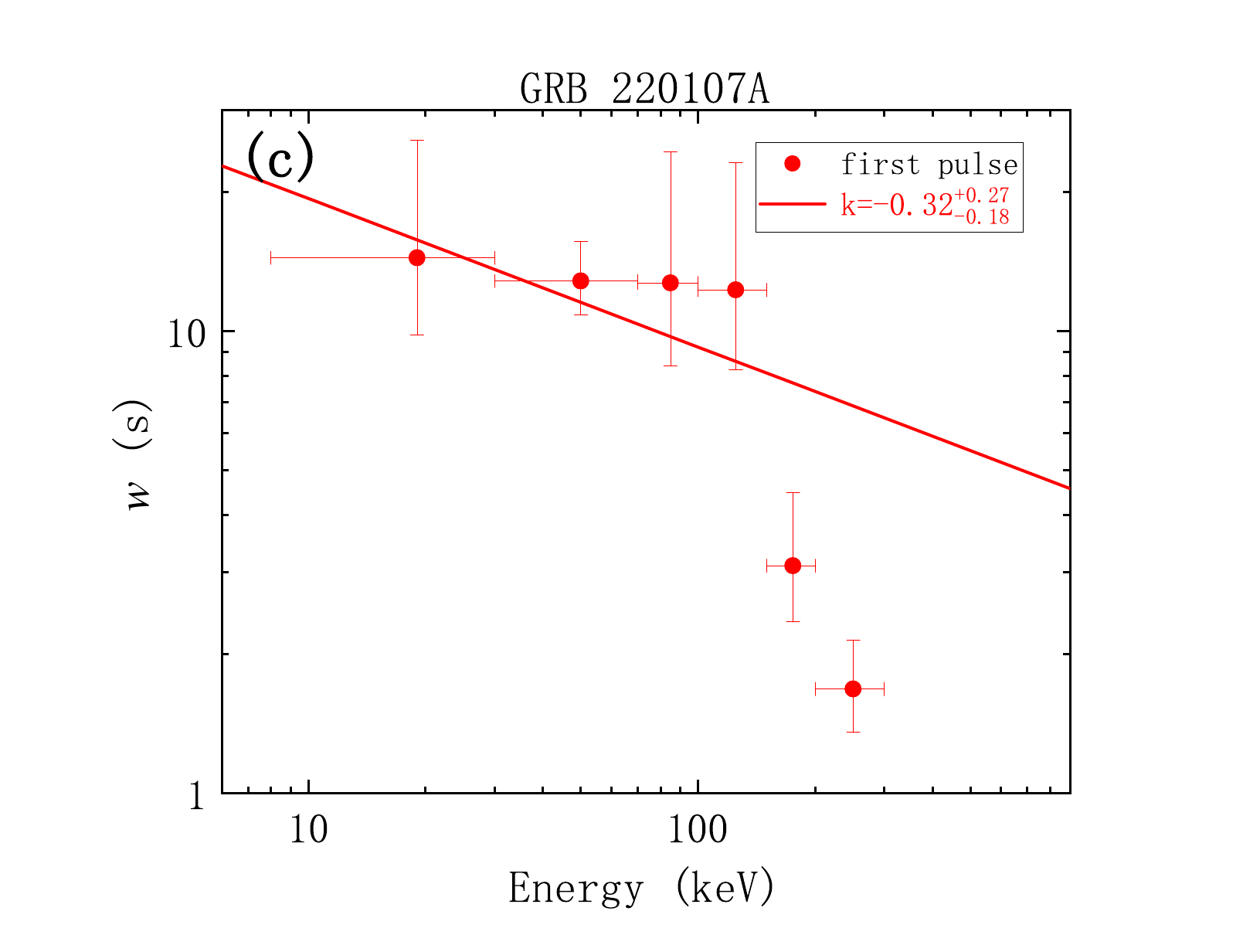} %
    \end{minipage}
    \hfill %
    \begin{minipage}{0.45\textwidth} %
        \centering %
        \includegraphics[width=\textwidth]{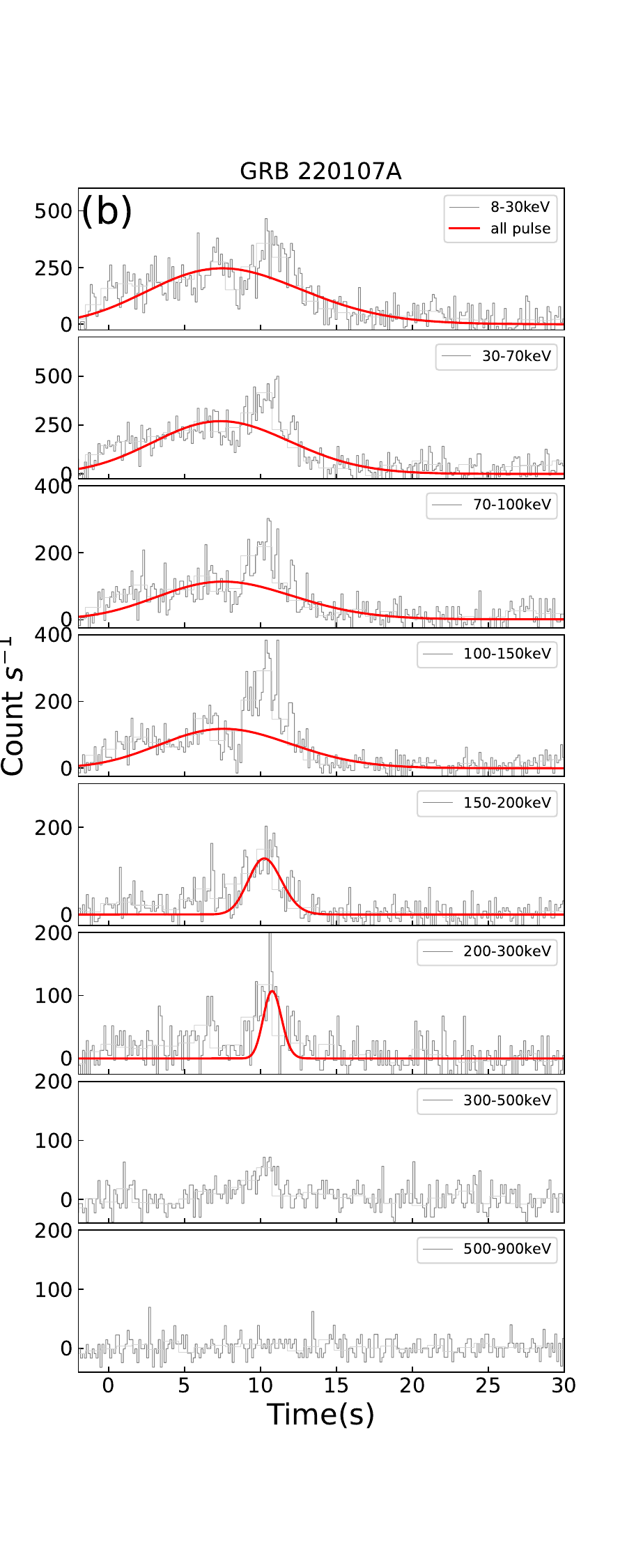} %
    \end{minipage}
    \caption{‌‌Similar to Figure \ref{fig:A1}, but for GRB 220107A in group (I).‌‌} %
     \label{fig:A5}
\end{figure}

\begin{figure} 
    \centering %
    \begin{minipage}{0.45\textwidth} %
        \centering %
        \includegraphics[width=\textwidth]{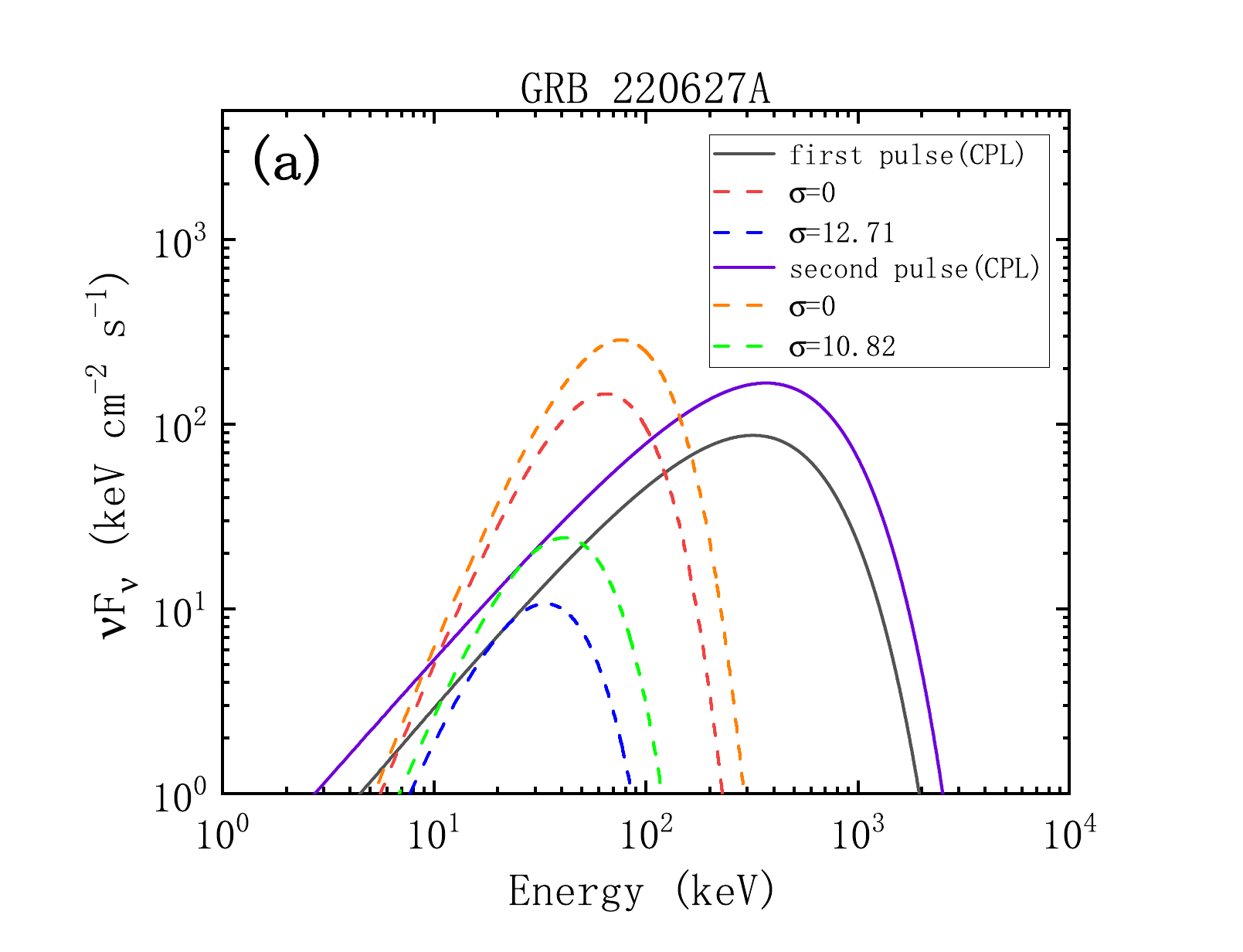} %
        \vskip\baselineskip %
        \includegraphics[width=\textwidth]{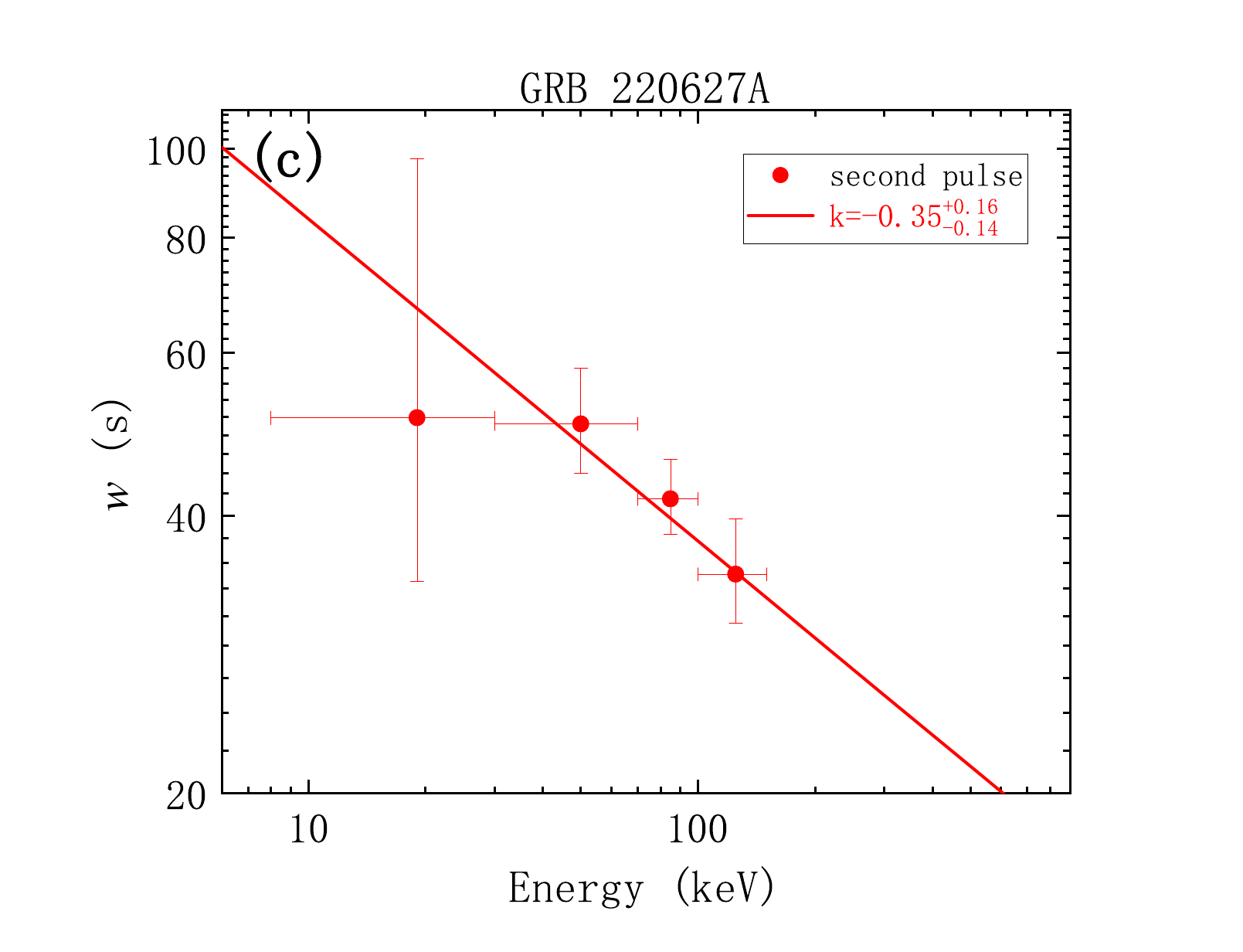} %
    \end{minipage}
    \hfill %
    \begin{minipage}{0.45\textwidth} %
        \centering %
        \includegraphics[width=\textwidth]{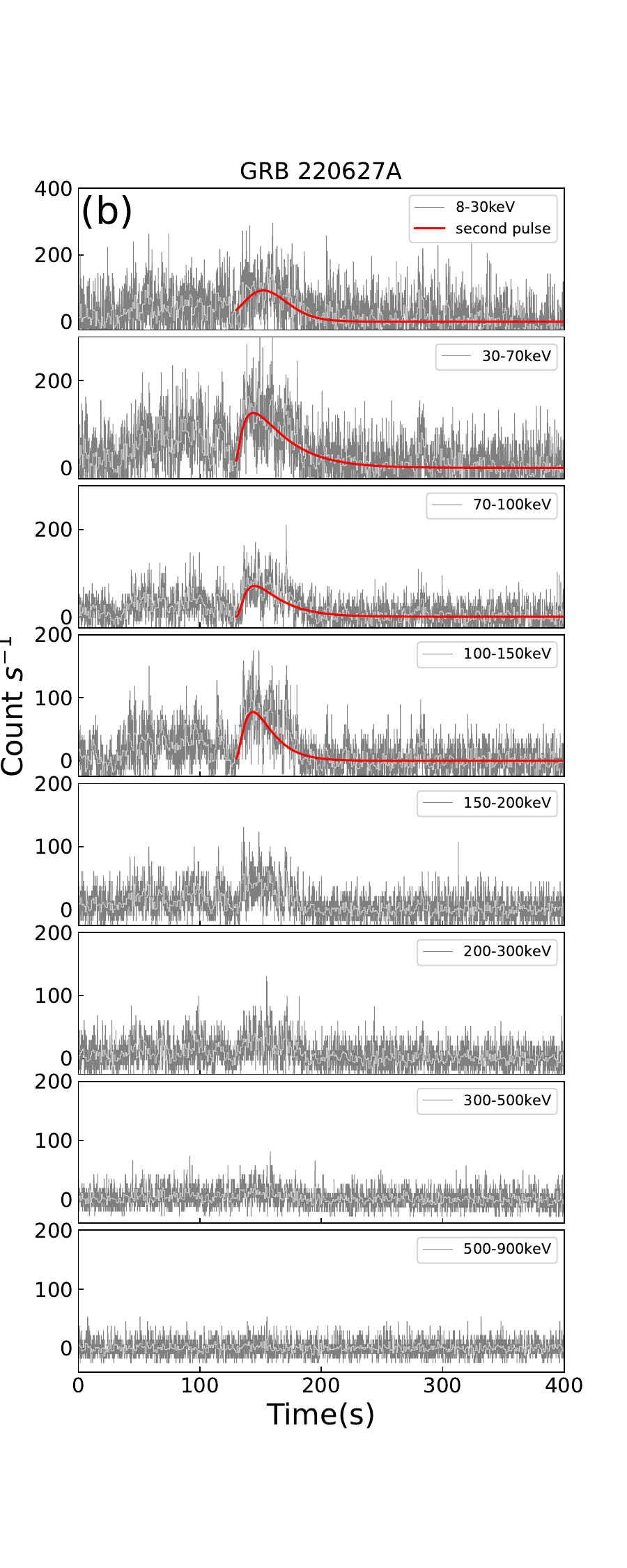} %
    \end{minipage}
    \caption{‌Similar to Figure \ref{fig:A1}, but for GRB 220627A in group (I).‌‌} %
     \label{fig:A6}
\end{figure}

\begin{figure} 
    \centering %
    \begin{minipage}{0.45\textwidth} %
        \centering %
        \includegraphics[width=\textwidth]{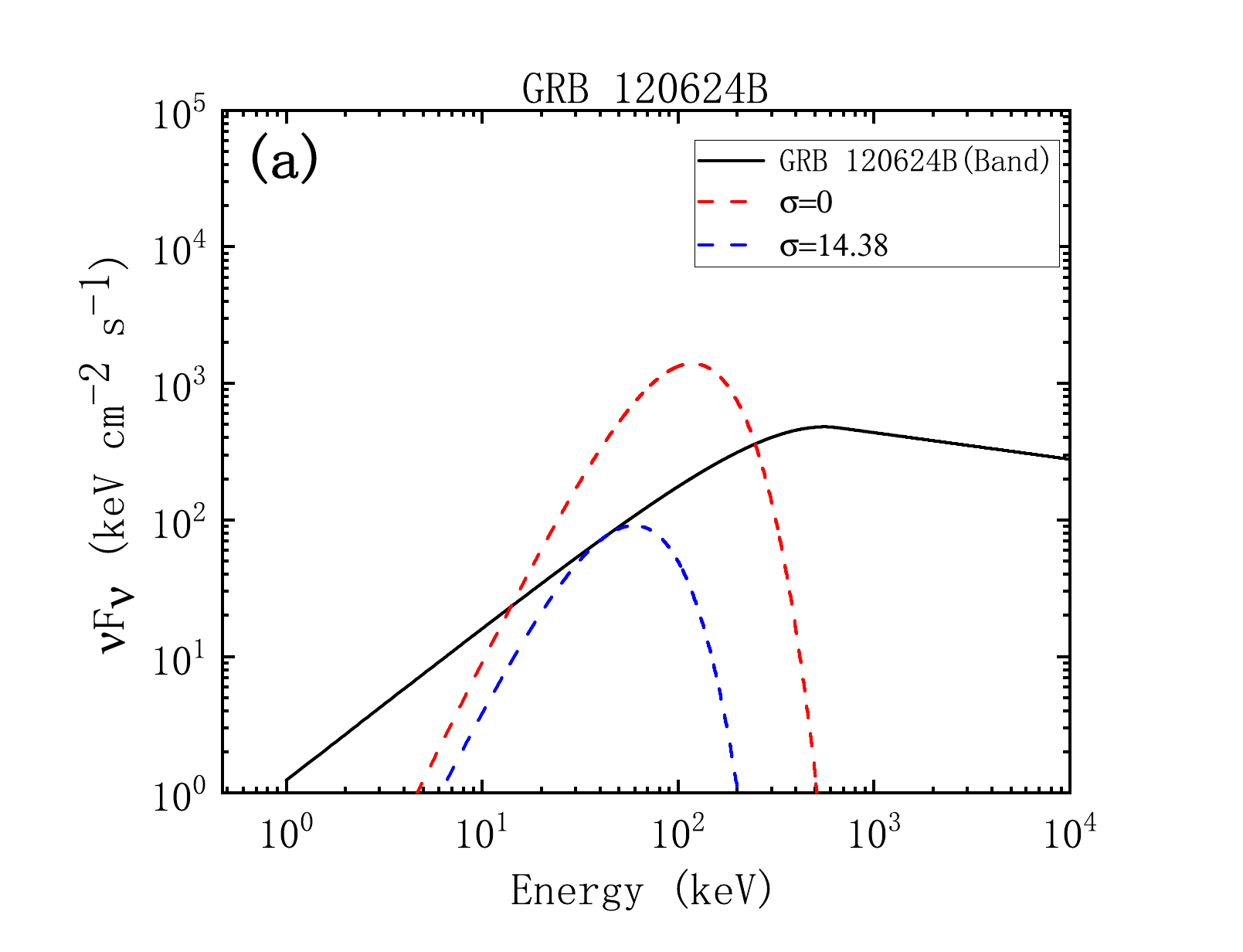} %
        \vskip\baselineskip %
        \includegraphics[width=\textwidth]{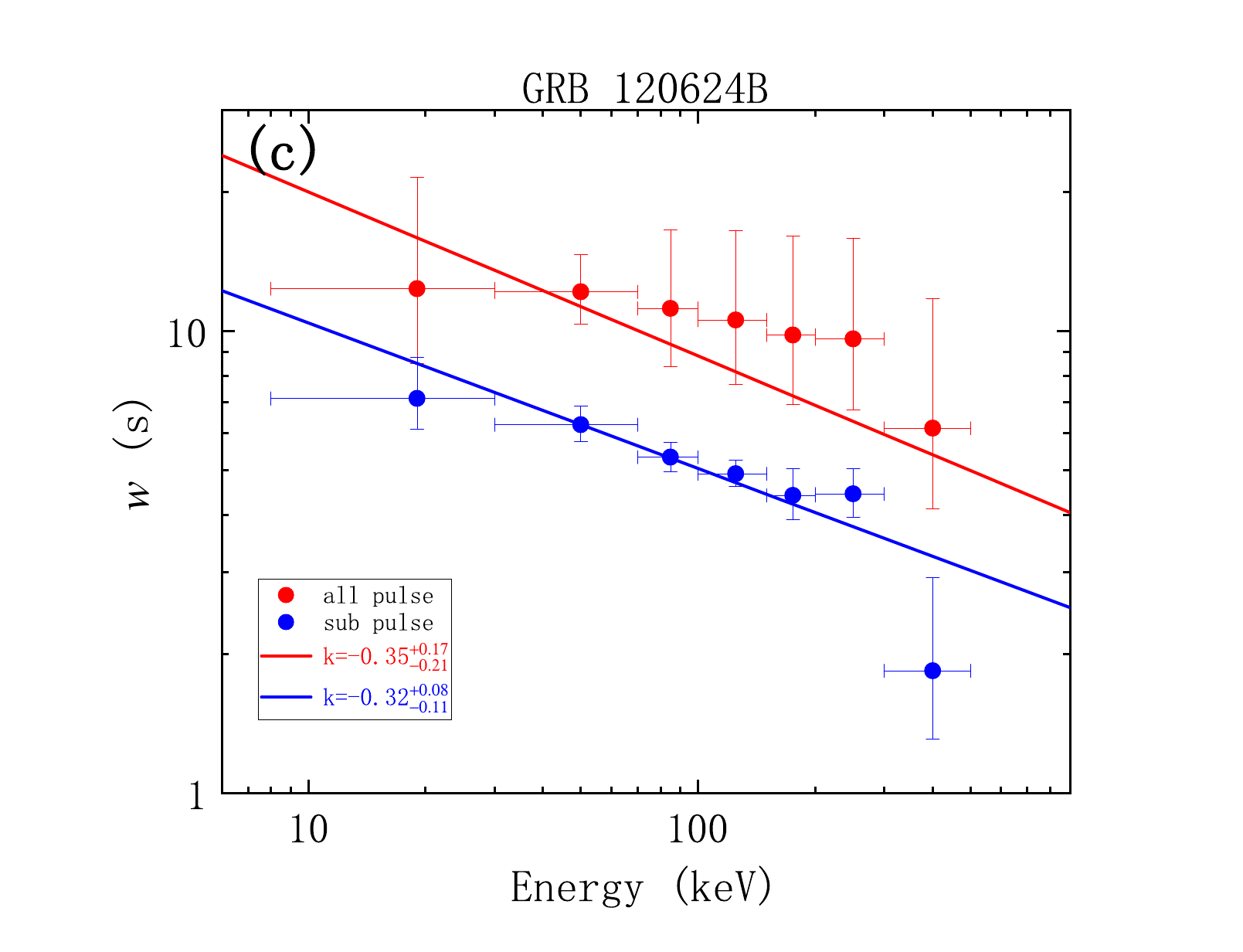} %
    \end{minipage}
    \hfill %
    \begin{minipage}{0.45\textwidth} %
        \centering %
        \includegraphics[width=\textwidth]{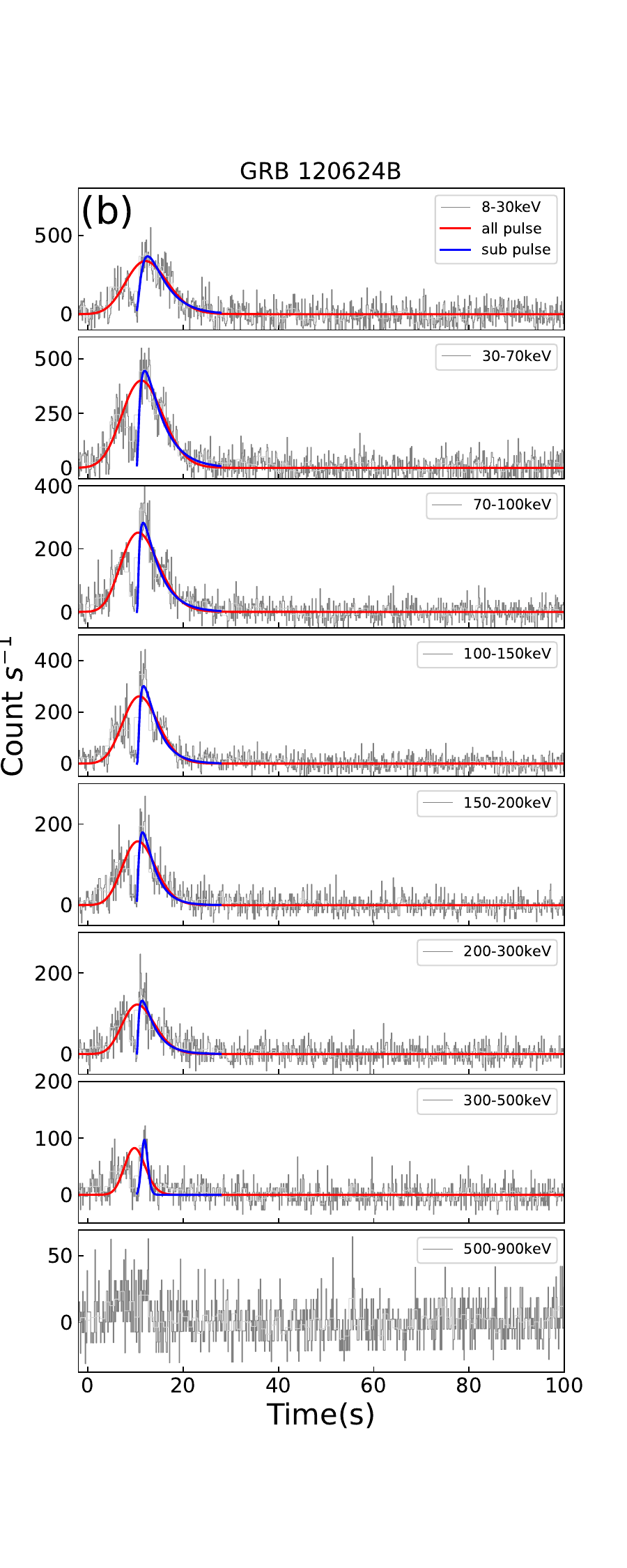} %
    \end{minipage}
    \caption{Spectral and temporal analysis of GRB 120624B for group (II). (a‌). The predicted lower limits of the photosphere spectra (dashed
    lines) with $R_{0}=10^{10}$ cm and observed non-thermal spectrum (solid line). (b). The light curves of the prompt emission of GRB 120624B (gray) in the different energy ranges and the FRED model fitting (red and blue solid lines). (c). The pulse width ($w$) is derived from the FRED model fitting for sub-pulse and whole pulse as a function of energy and the power-law fitting (red and blue solid lines).}  
     \label{fig:B1}
\end{figure}

\begin{figure} 
    \centering %
    \begin{minipage}{0.45\textwidth} %
        \centering %
        \includegraphics[width=\textwidth]{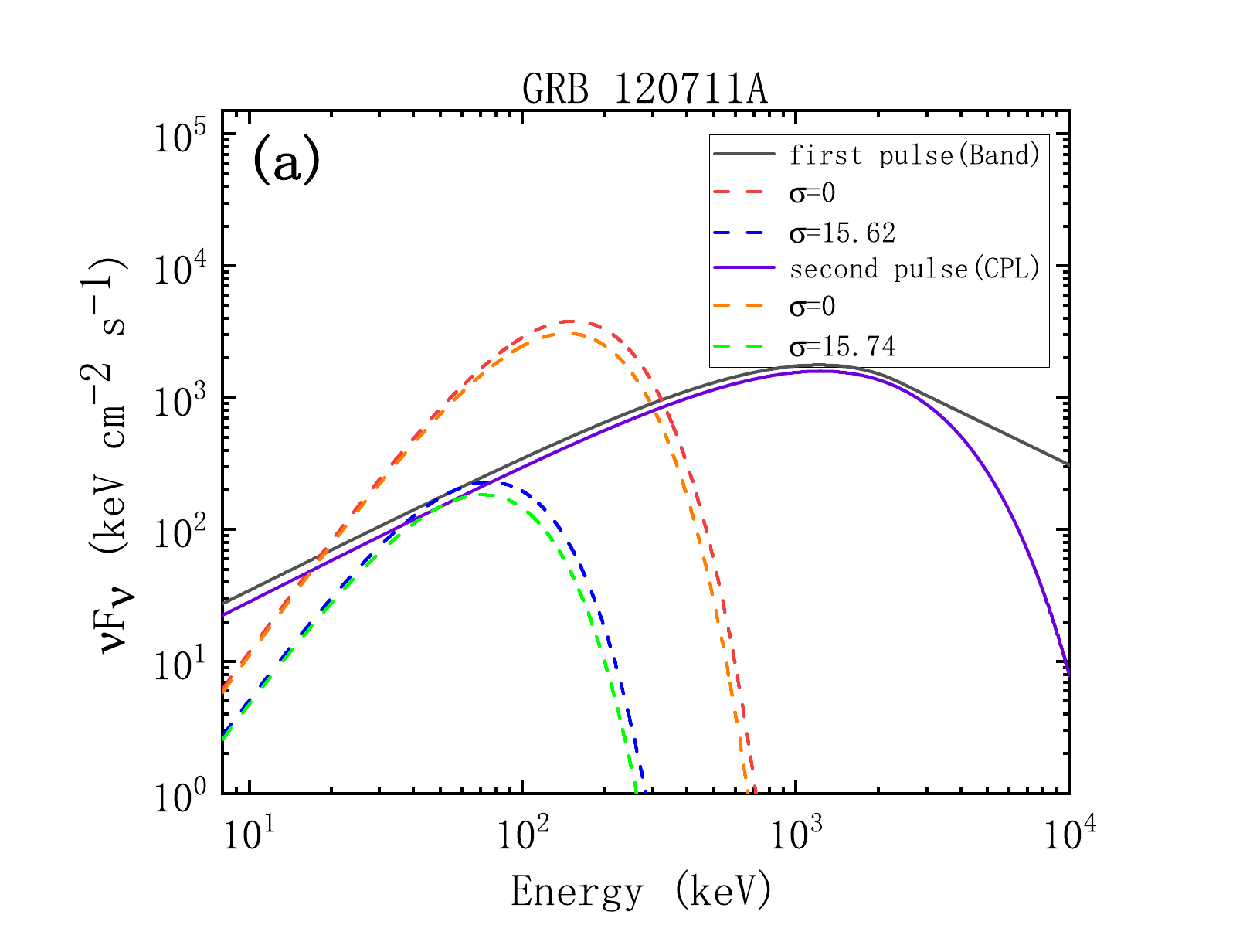} %
        \vskip\baselineskip %
        \includegraphics[width=\textwidth]{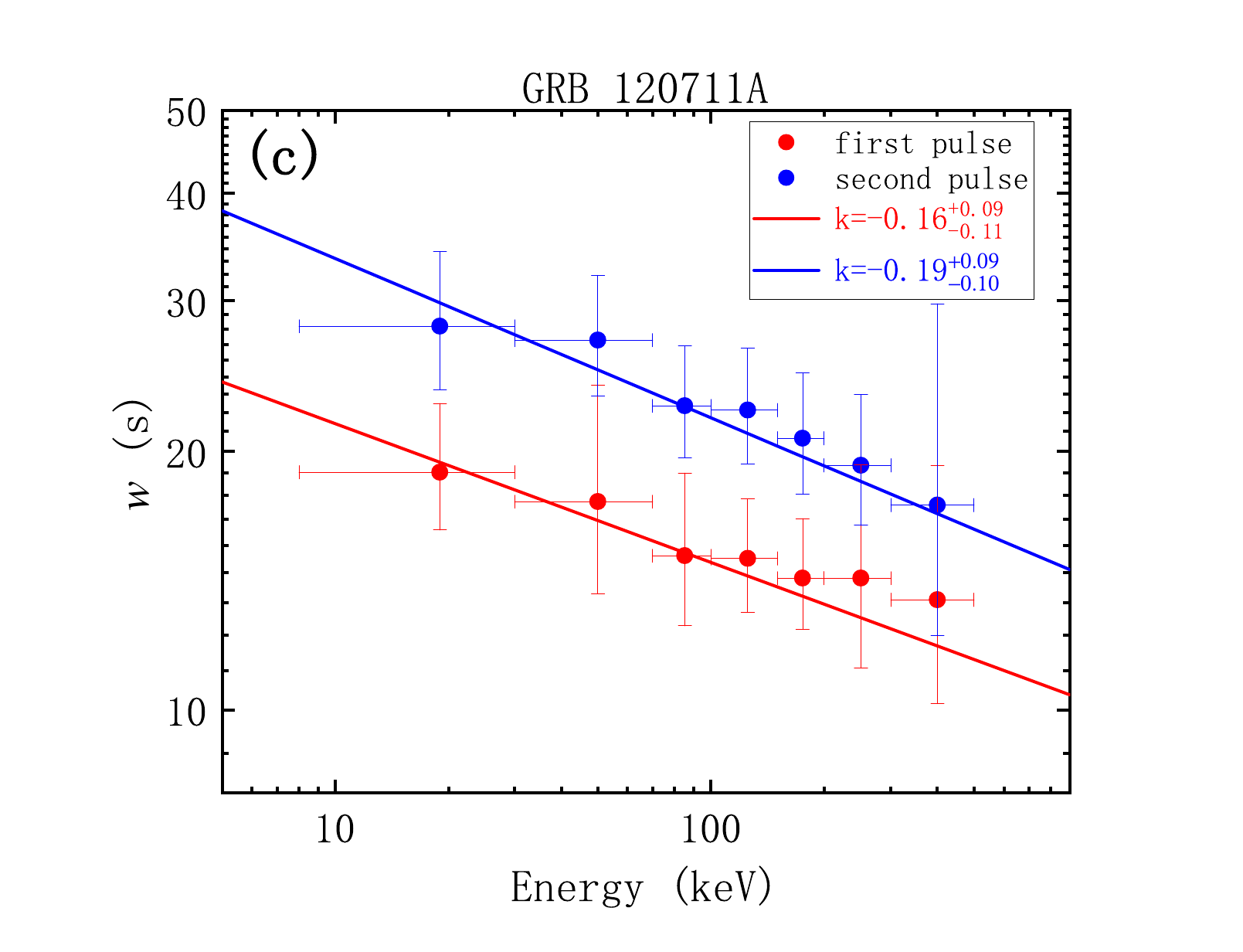} %
    \end{minipage}
    \hfill %
    \begin{minipage}{0.45\textwidth} %
        \centering %
        \includegraphics[width=\textwidth]{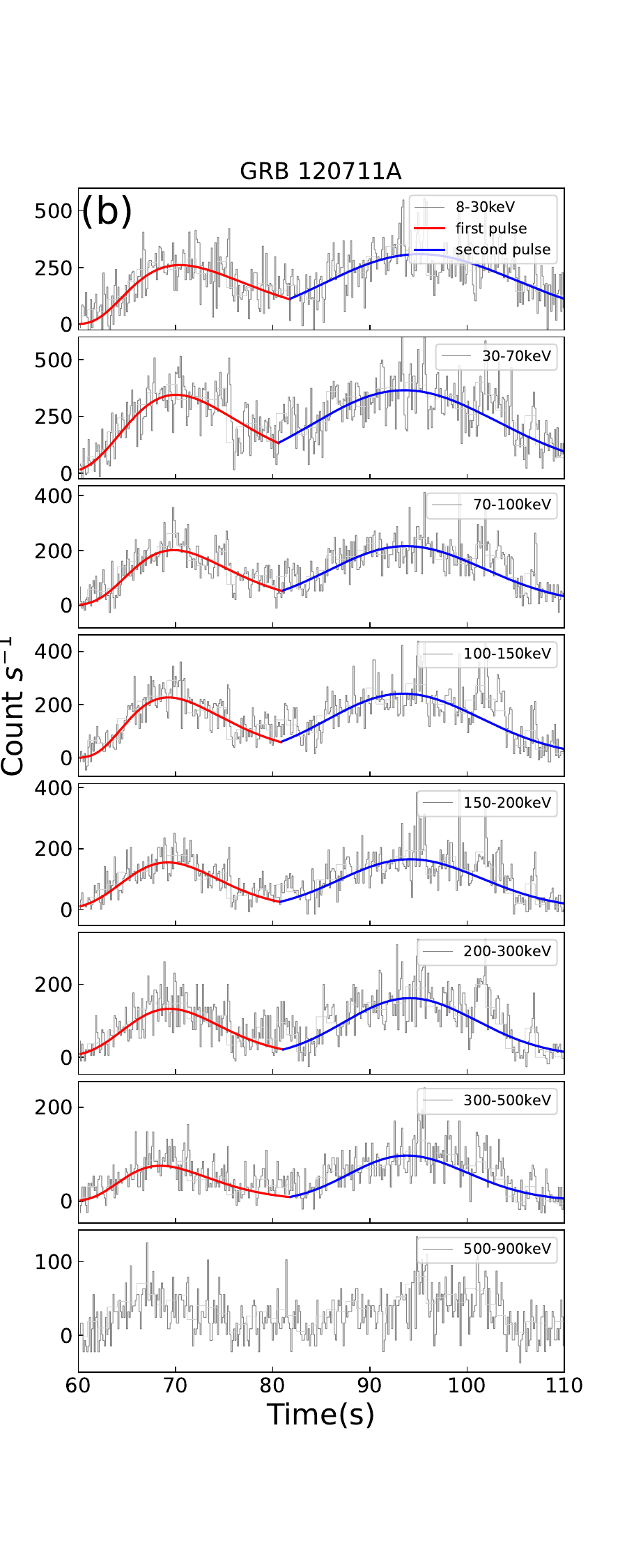} %
    \end{minipage}
    \caption{Spectral and temporal analysis of GRB 120711A for group (II). (a‌). The predicted lower limits of the photosphere spectra (dashed
    lines) with $R_{0}=10^{10}$ cm and observed non-thermal spectrum (solid lines). (b). The light curves of prompt emission of GRB 120711A (gray) in the different energy ranges and the FRED model fitting (red and blue solid lines). (c). The pulse width ($w$) is derived from the FRED model fitting for two sub-pulses as a function of energy and the power-law fitting (red and blue solid lines).} %
     \label{fig:B2}
\end{figure}

\begin{figure} 
    \centering %
    \begin{minipage}{0.45\textwidth} %
        \centering %
        \includegraphics[width=\textwidth]{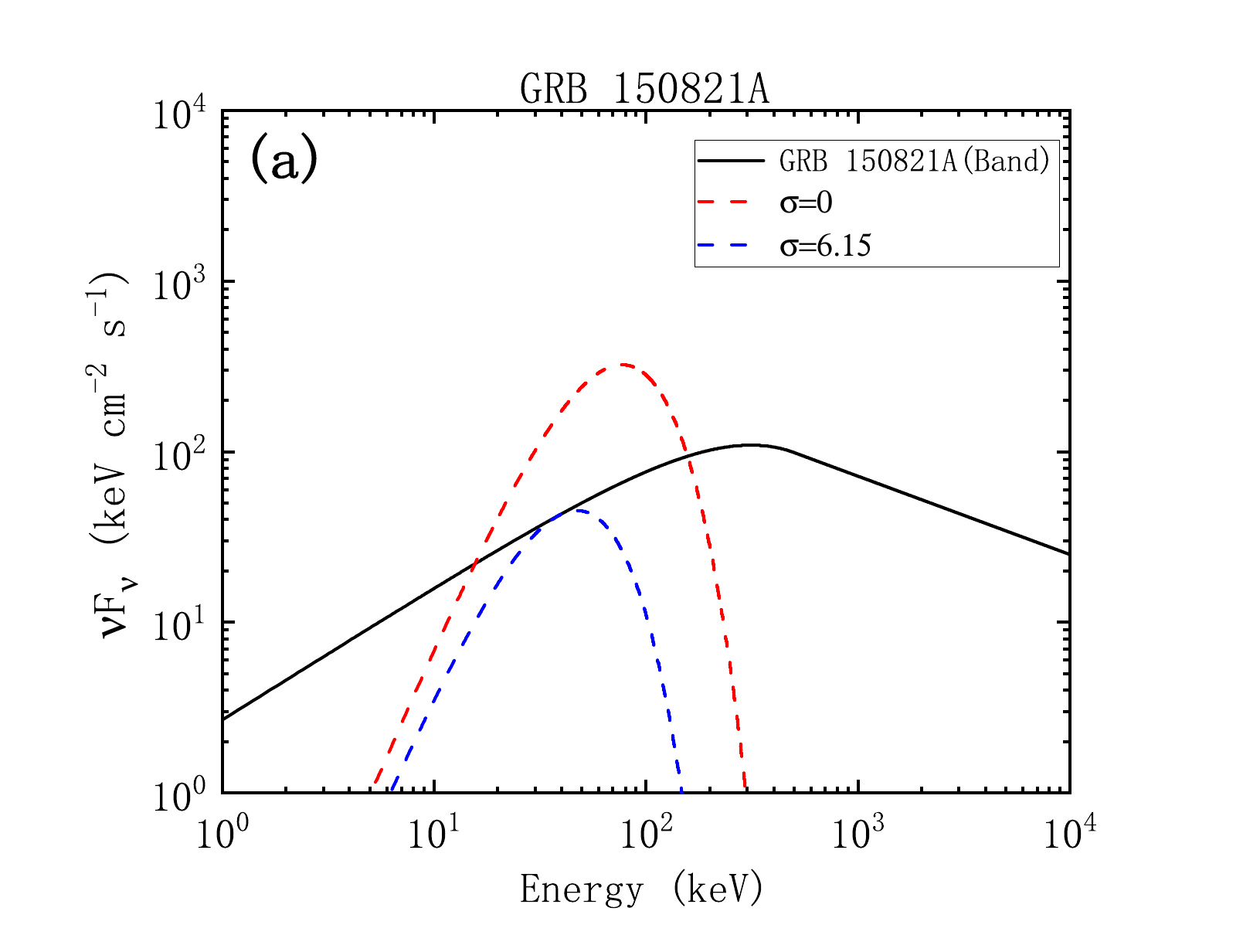} %
        \vskip\baselineskip %
        \includegraphics[width=\textwidth]{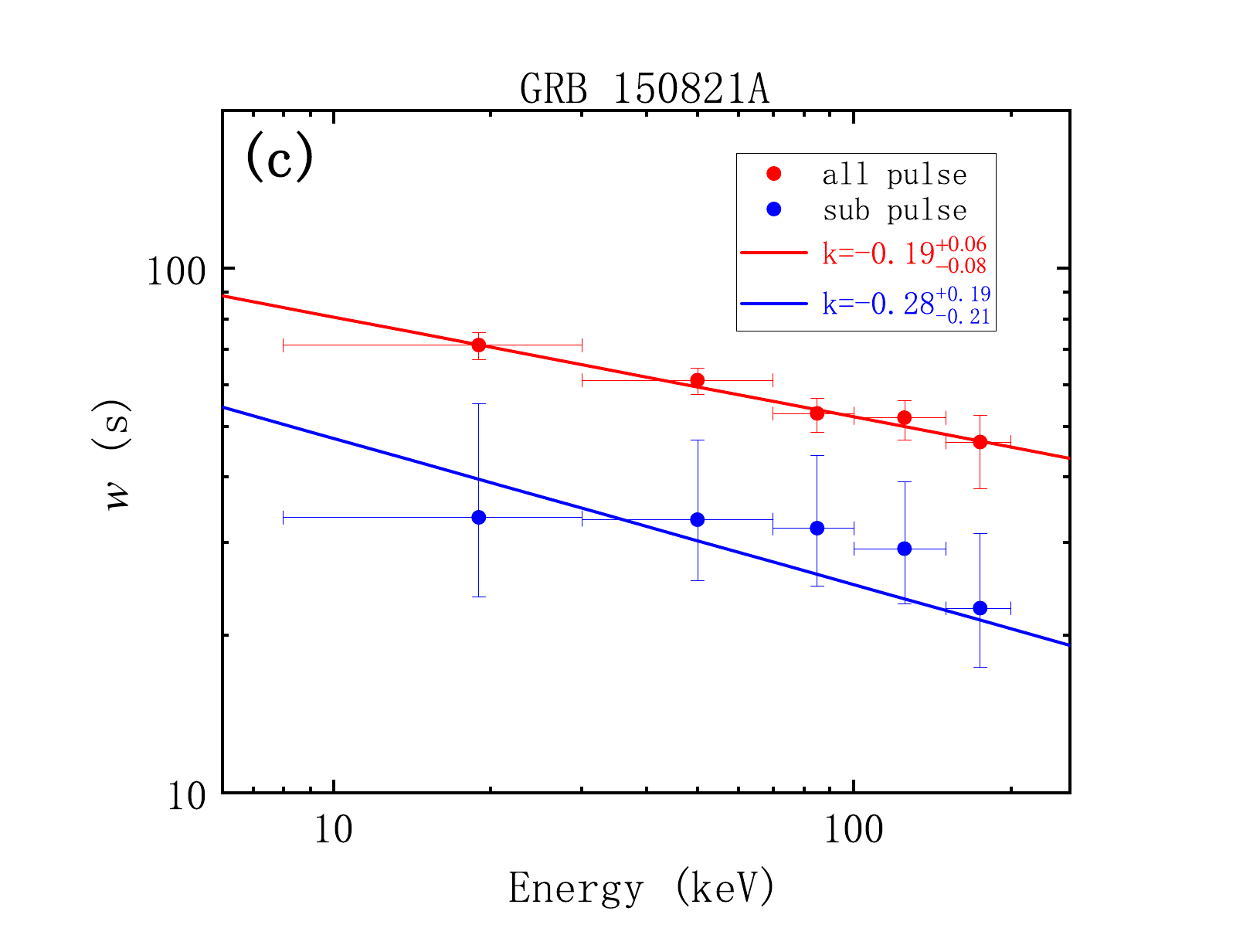} %
    \end{minipage}
    \hfill %
    \begin{minipage}{0.45\textwidth} %
        \centering %
        \includegraphics[width=\textwidth]{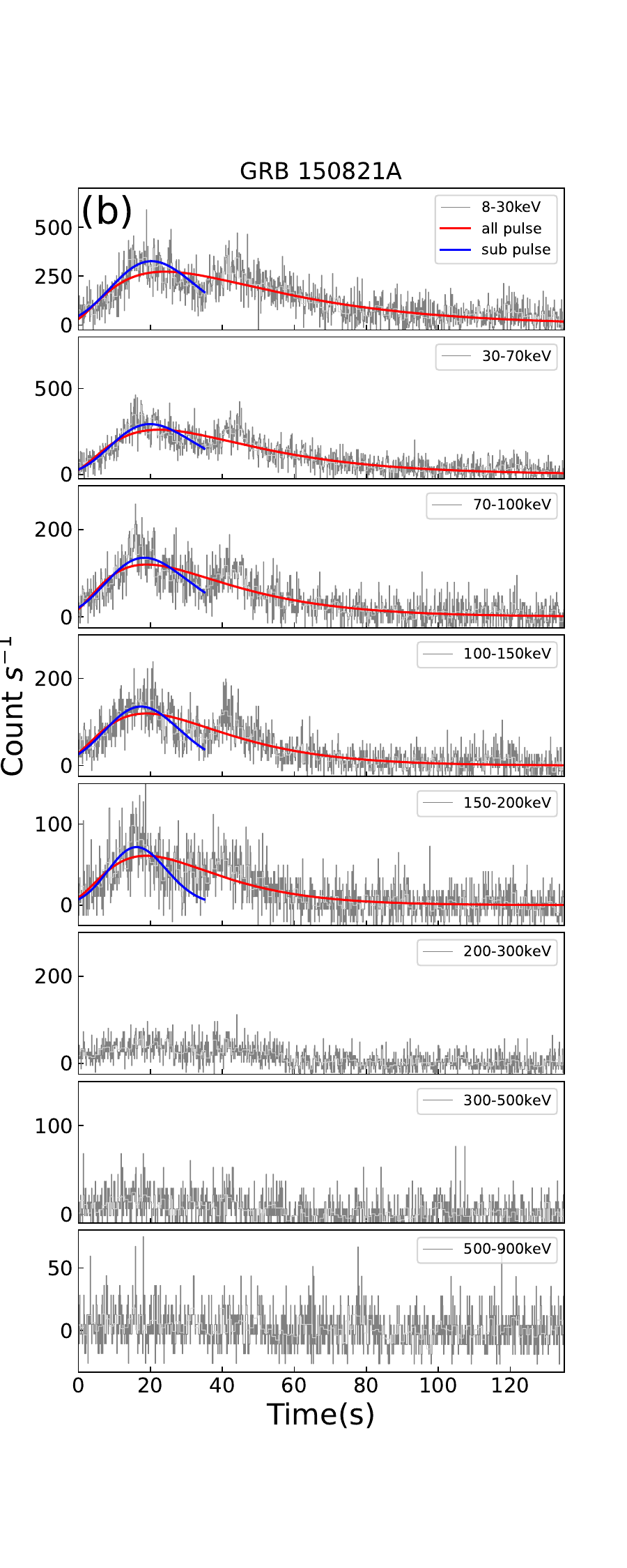} %
    \end{minipage}
    \caption{‌Similar to Figure \ref{fig:B1}, but for GRB 150821A in group (II).‌} 
     \label{fig:B3}
\end{figure}

\begin{figure} 
    \centering %
    \begin{minipage}{0.45\textwidth} %
        \centering %
        \includegraphics[width=\textwidth]{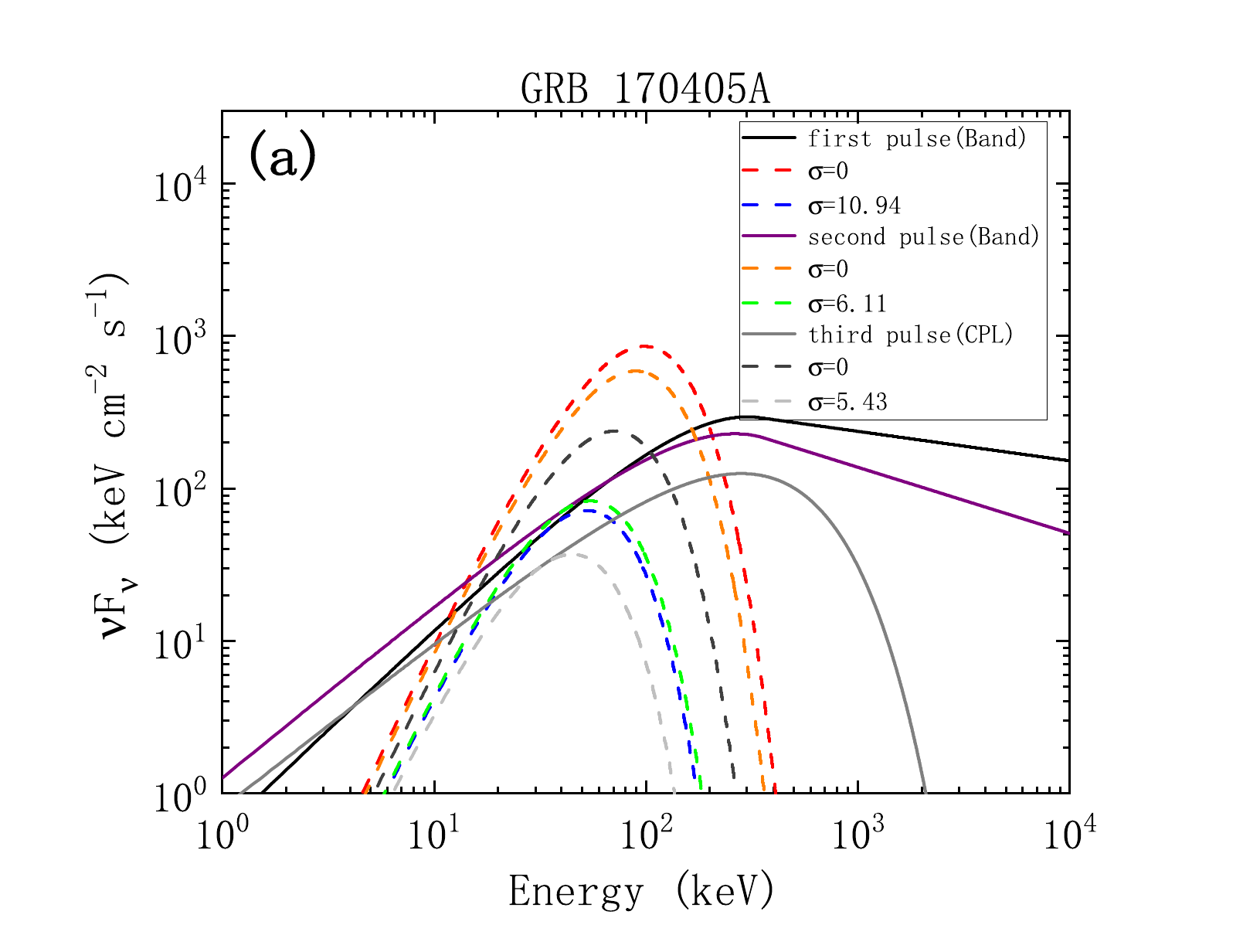} %
        \vskip\baselineskip %
        \includegraphics[width=\textwidth]{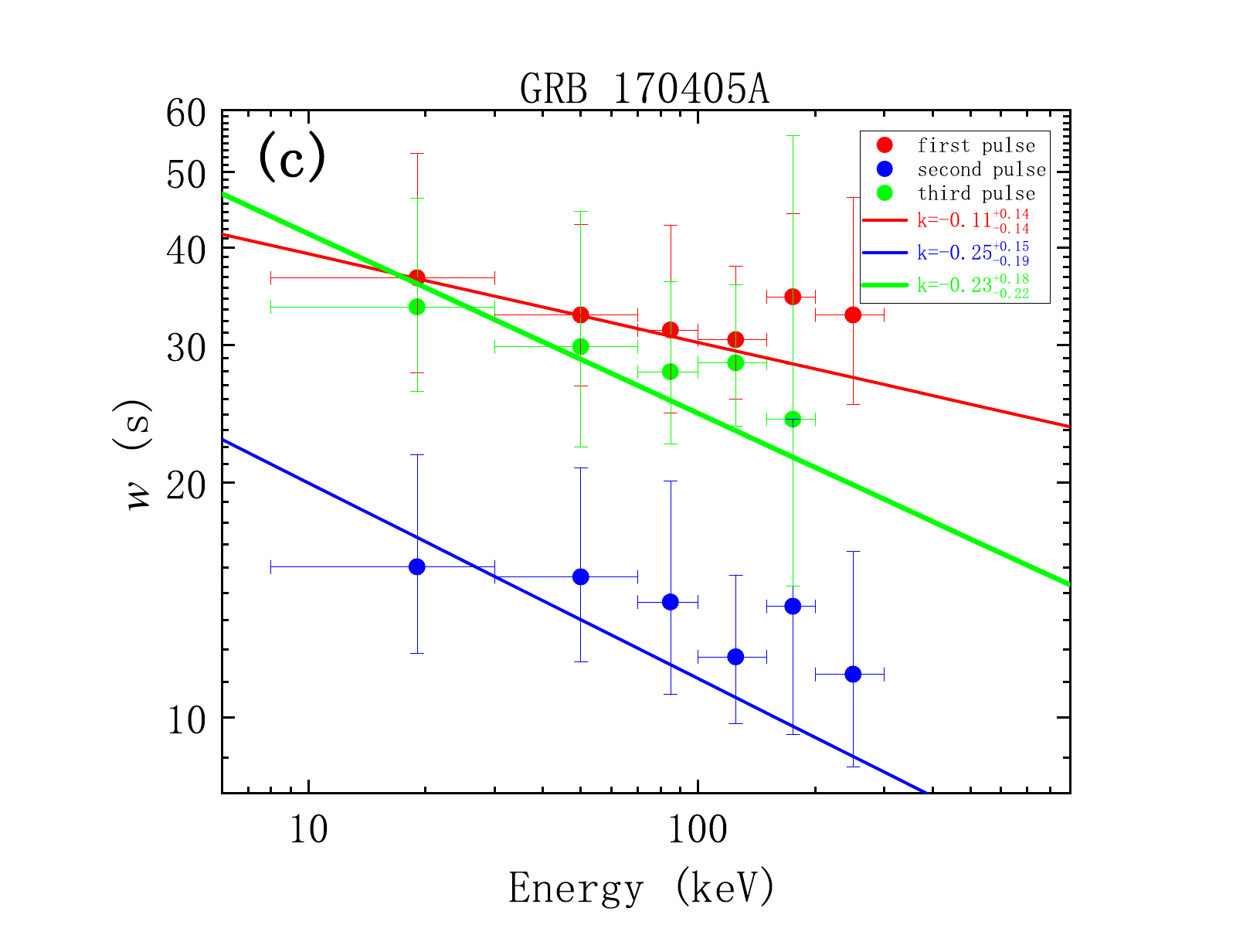} %
    \end{minipage}
    \hfill %
    \begin{minipage}{0.45\textwidth} %
        \centering %
        \includegraphics[width=\textwidth]{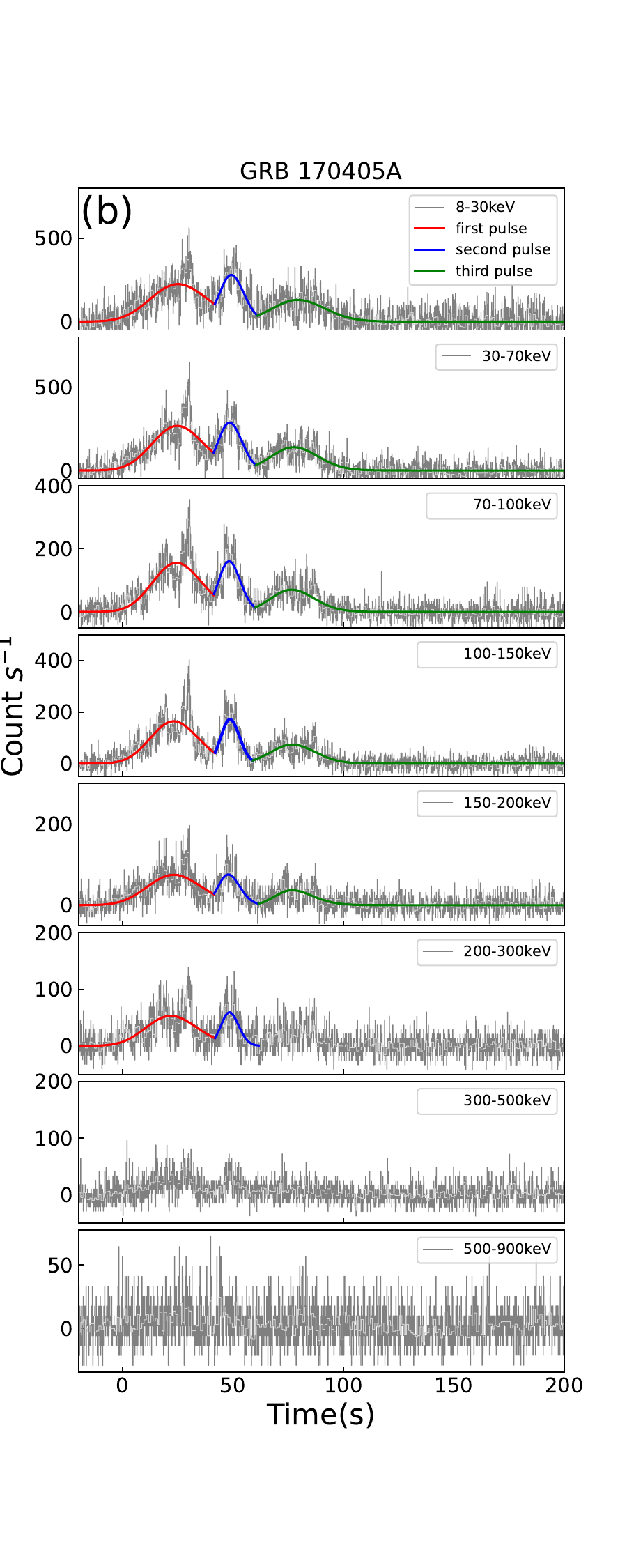} %
    \end{minipage}
    \caption{‌‌‌Similar to Figure \ref{fig:B2}, but for GRB 170405A in group (II).} %
     \label{fig:B4}
\end{figure}

\begin{figure} 
    \centering %
    \begin{minipage}{0.45\textwidth} %
        \centering %
        \includegraphics[width=\textwidth]{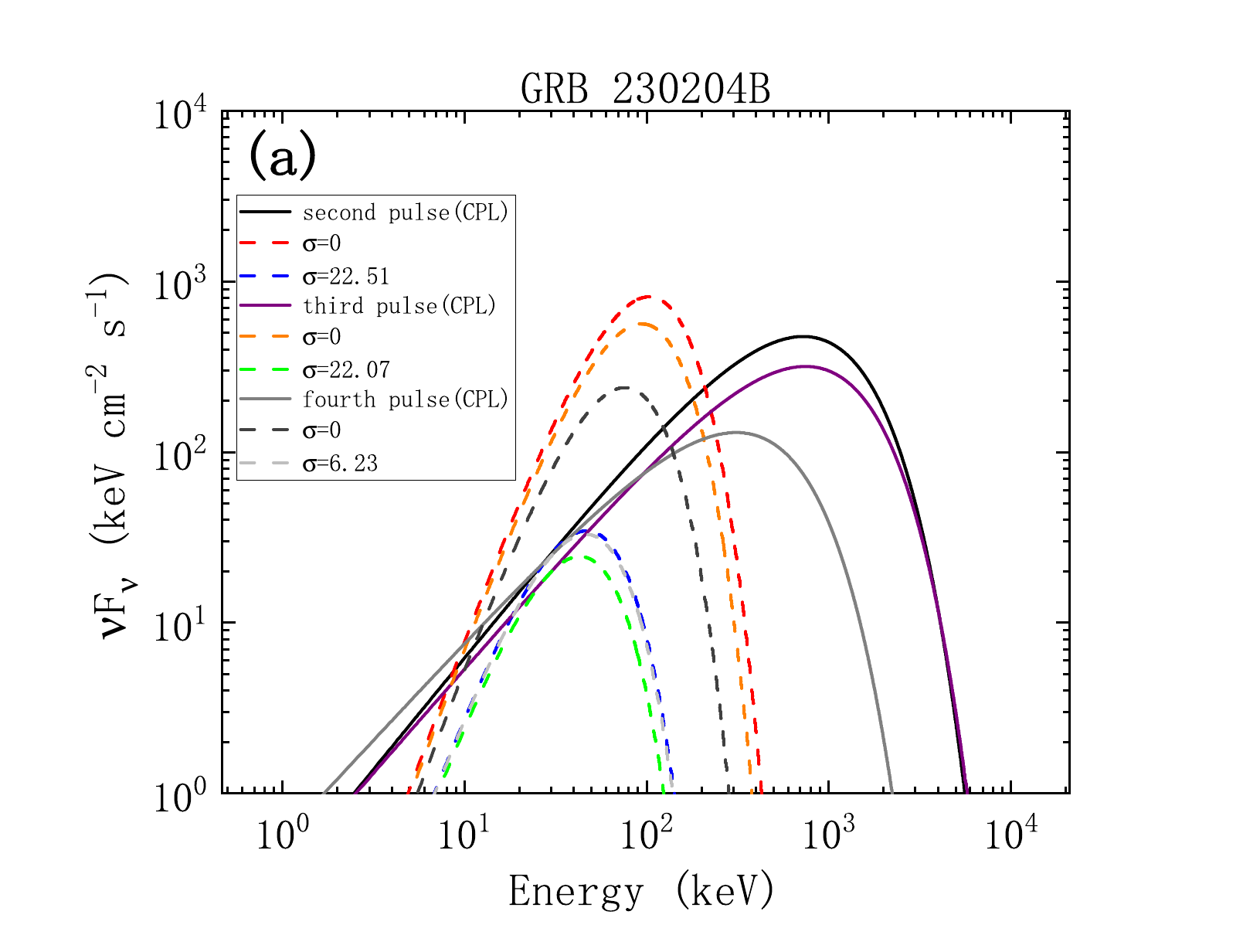} %
        \vskip\baselineskip %
        \includegraphics[width=\textwidth]{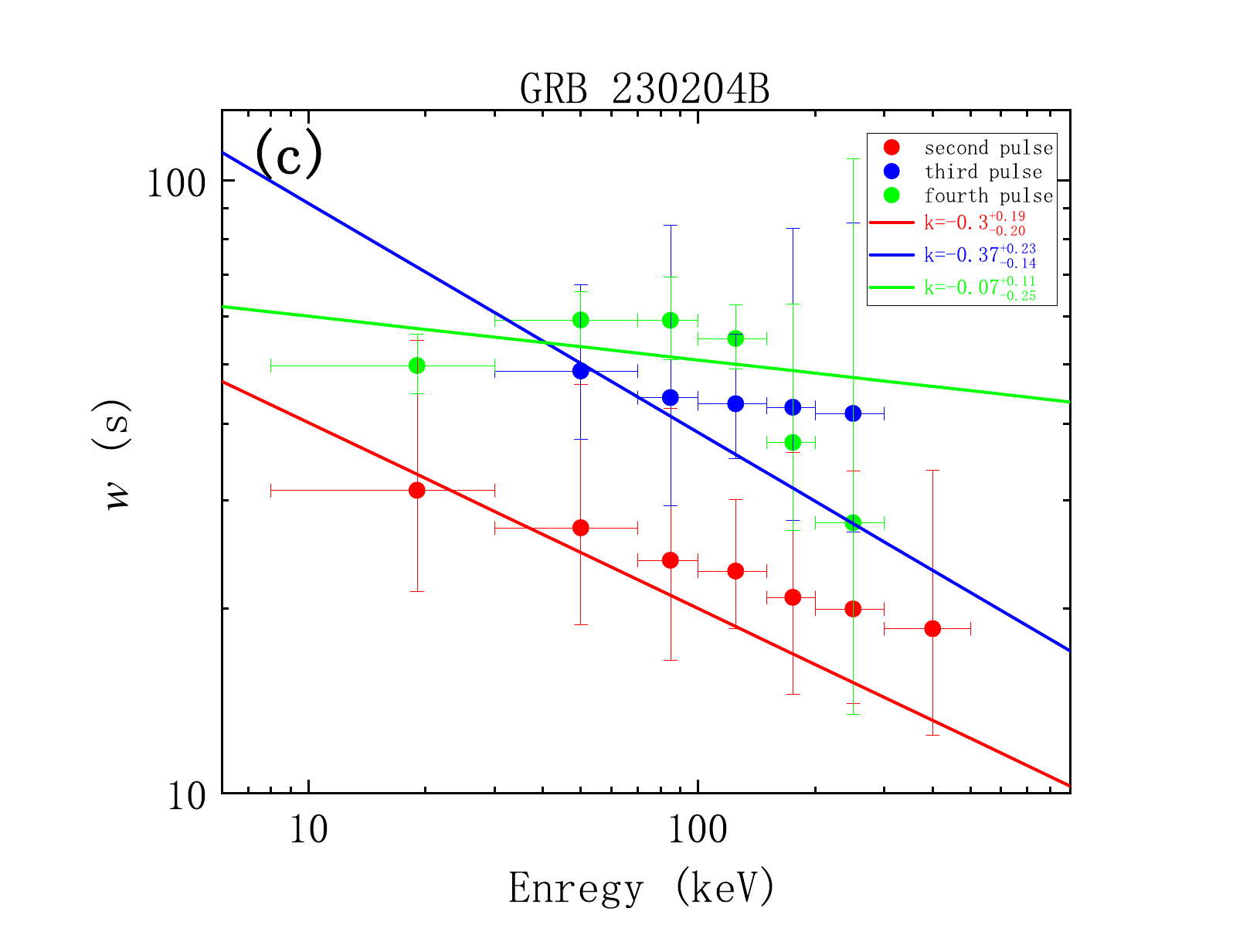} %
    \end{minipage}
    \hfill %
    \begin{minipage}{0.45\textwidth} %
        \centering %
        \includegraphics[width=\textwidth]{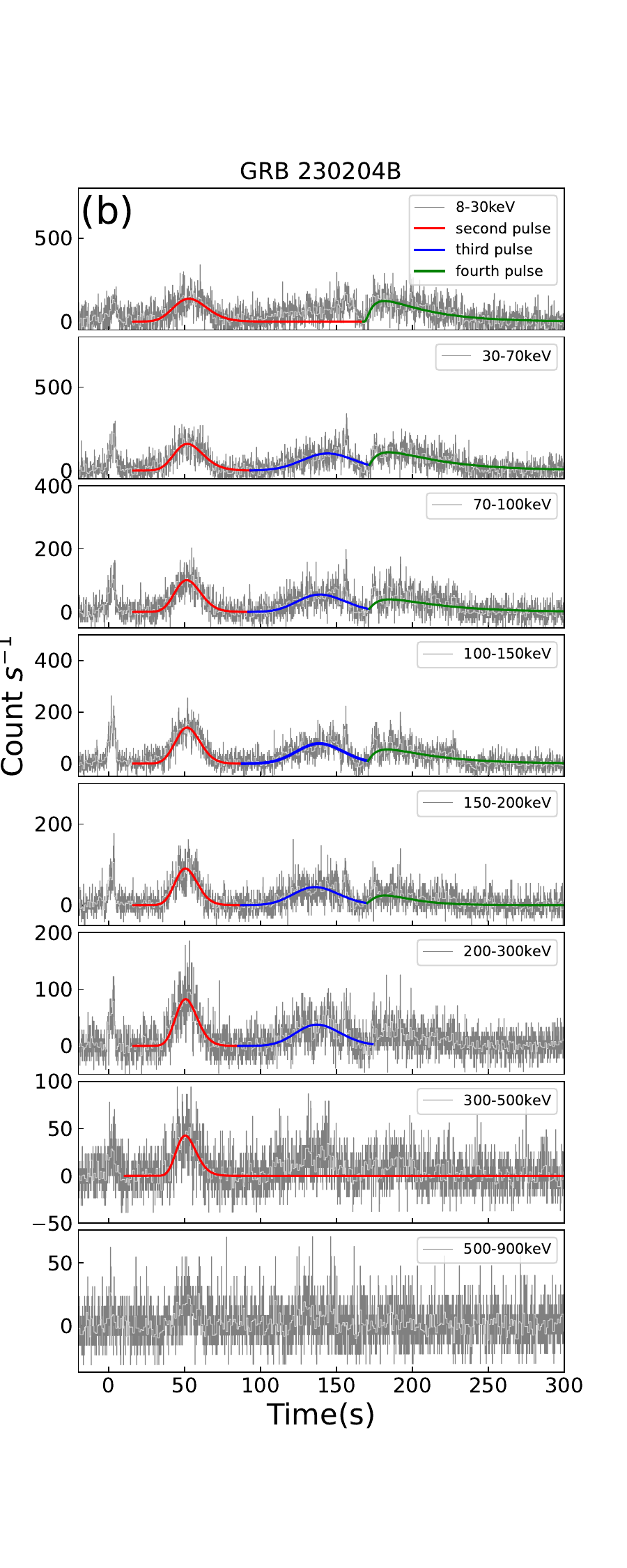} %
    \end{minipage}
    \caption{Similar to Figure \ref{fig:B2}, but for GRB 230204B in group (II).} %
     \label{fig:B5}
\end{figure}

\begin{figure} 
    \centering %
    \begin{minipage}{0.45\textwidth} %
        \centering %
        \includegraphics[width=\textwidth]{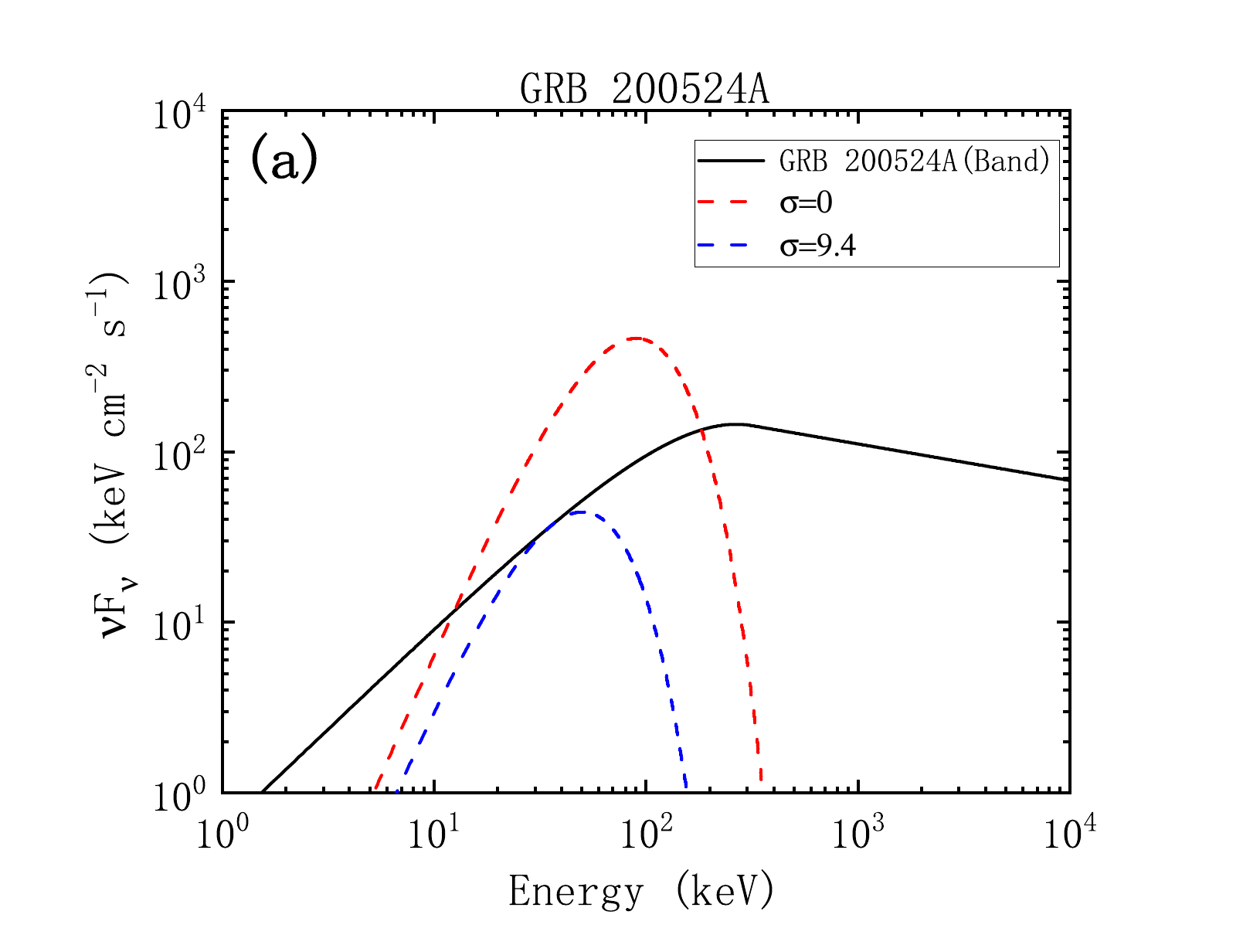} %
        \vskip\baselineskip %
        \includegraphics[width=\textwidth]{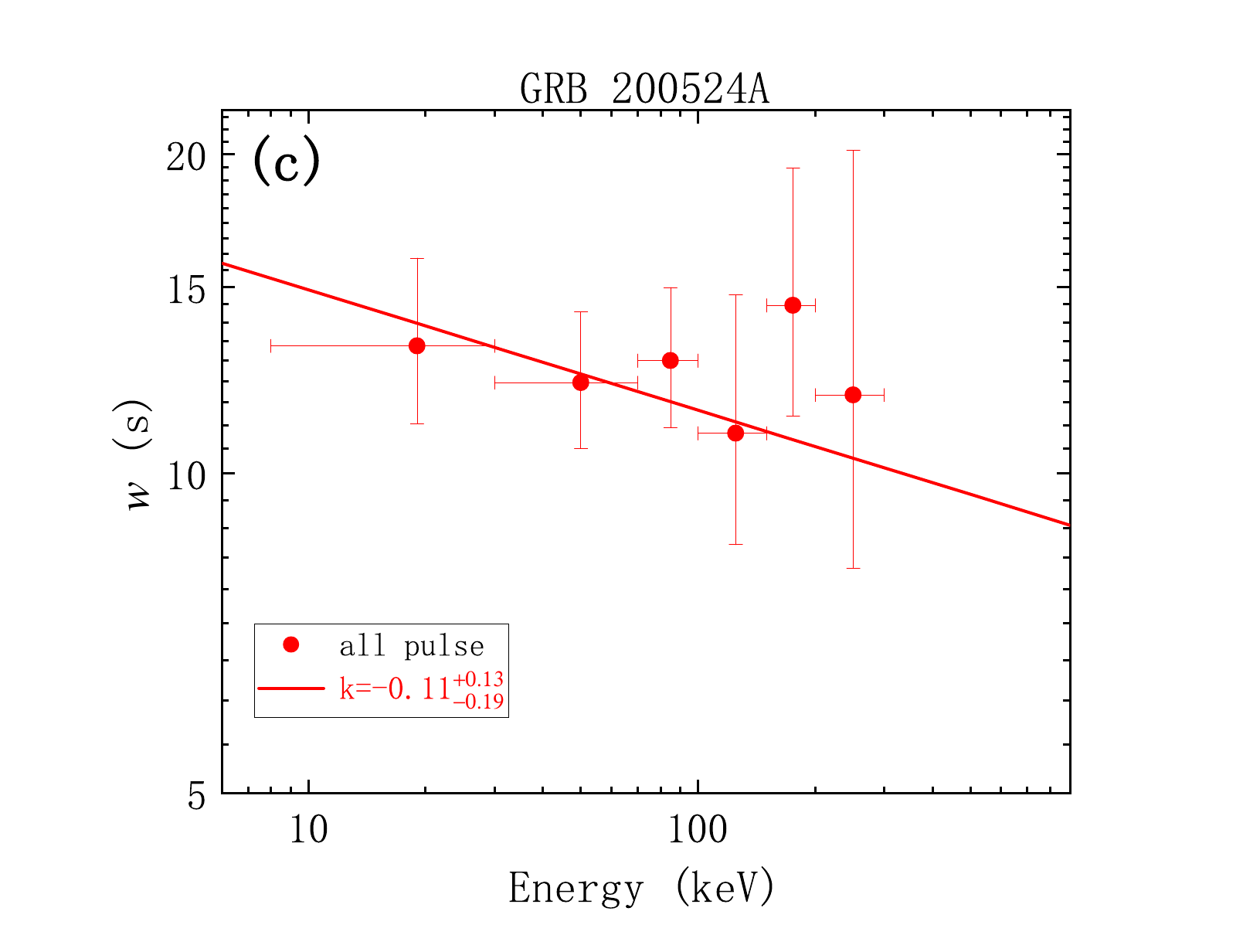} %
    \end{minipage}
    \hfill %
    \begin{minipage}{0.45\textwidth} %
        \centering %
        \includegraphics[width=\textwidth]{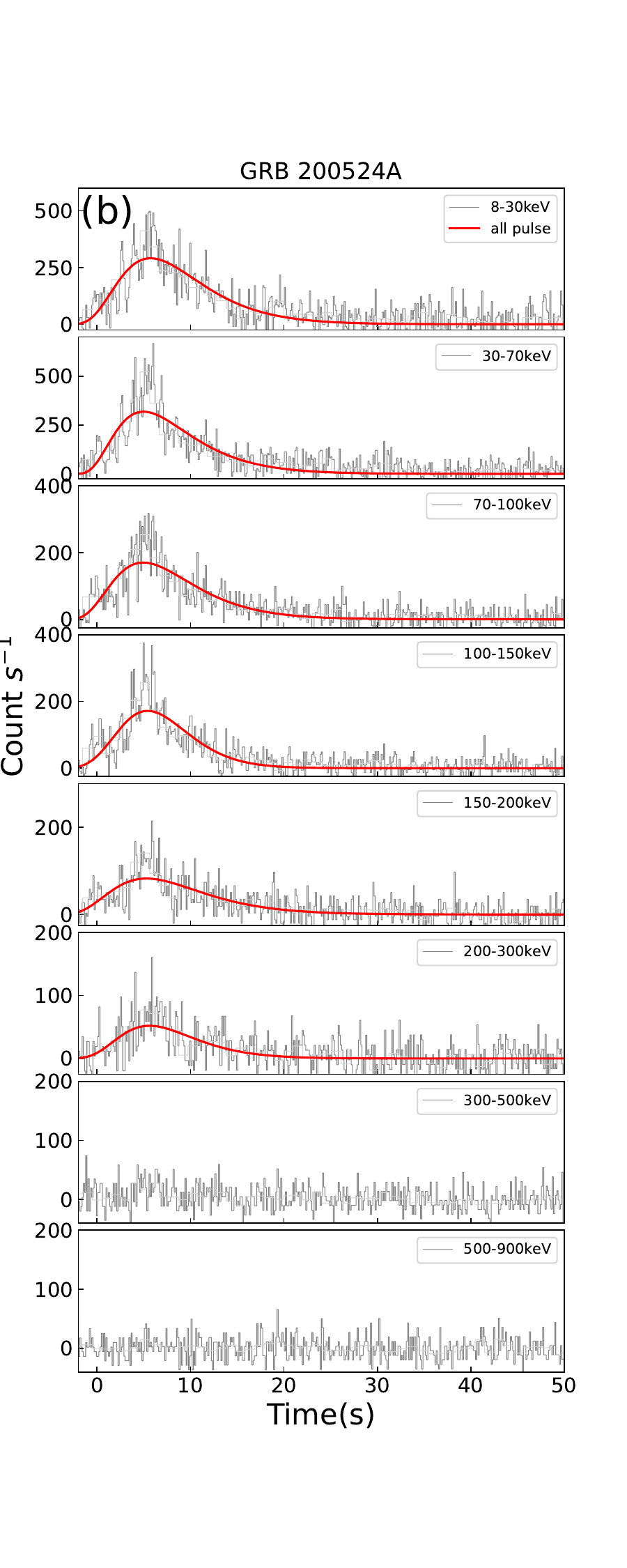} %
    \end{minipage}
    \caption{Spectral and temporal analysis of GRB 200524A for group (III). (a‌). The predicted lower limits of the photosphere spectra (dashed
    lines) with $R_{0}=10^{10}$ cm and observed non-thermal spectrum (solid lines). (b). The light curves of prompt emission of GRB 200524A (gray) in the different energy ranges and the FRED model fitting (red solid lines). (c). The pulse width ($w$) is derived from the FRED model fitting for the bright pulse as a function of energy and the power-law fitting (red solid lines).‌‌‌} %
     \label{fig:C1}
\end{figure}

\bsp	
\label{lastpage}
\end{document}